\documentclass[
 reprint,
 amssymb, amsmath,
 aip,jcp                                                                                                                                                
]{revtex4-1}
\usepackage[USenglish]{babel}

\usepackage{soul}
\usepackage{amsmath,amssymb,bm}
\usepackage{comment}
\usepackage{mathtools} 
\usepackage{graphicx}
\usepackage{xcolor}
\usepackage{physics}
\usepackage[colorlinks=true, allcolors=blue]{hyperref}
\usepackage{wrapfig}
\usepackage{lipsum,booktabs}
\usepackage{bm}
\usepackage{rotating}
\usepackage{multirow}
\usepackage{float}
\usepackage{adjustbox}
\usepackage[normalem]{ulem}

\newcommand{\omegamf}[0]{\Omega_{\text{MF}}}

\newcommand{\gc}[0]{{\text{GC}}}
\newcommand{\pgc}[0]{p^{\gc}}

\newcommand{\gcano}[0]{{\text{GC}}}

\newcommand{\Trac}[0]{\text{Tr}}
\newcommand{\mm}[0]{{\scriptscriptstyle \text{MM}}}
\newcommand{\opt}[0]{{\scriptscriptstyle \text{opt}}}
\newcommand{\qm}[0]{{\scriptscriptstyle \text{QM}}}
\newcommand{\nesmall}[0]{{\nsmall\esmall}}
\newcommand{\nnsmall}[0]{{\scriptscriptstyle \text{NN}}}
\newcommand{\eesmall}[0]{{\scriptstyle \text{ee}}}
\newcommand{\esmall}[0]{{\scriptstyle \text{e}}}
\newcommand{\elsmall}[0]{{\scriptstyle \text{el}}}
\newcommand{\elsmalll}[0]{{\scriptscriptstyle \text{el}}}
\newcommand{\nsmall}[0]{{\scriptscriptstyle \text{N}}}

\newcommand{\BO}[0]{{\scriptscriptstyle \text{BO}}}

\newcommand{\calF}[1]{\mathcal{F}_{{#1}}}

\newcommand{\brnucl}[1]{{\bf R}_{#1}}
\newcommand{\brnucltot}[0]{{\bf R}^{\NN}}
\newcommand{\bri}[1]{\textbf{r}_{#1}}
\newcommand{\bq}[0]{{\textbf{Q}}}
\newcommand{\bqtot}[0]{{\tilde{\textbf{Q}}^{\NMM}}}
\newcommand{\bqtotone}[0]{{\tilde{\textbf{Q}}}}
\newcommand{\bomega}[0]{{{\bf\Omega}}}
\newcommand{\br}[0]{\textbf{r}}

\newcommand{\Ham}[1]{ \hat{{\bf H}}_{{#1}}}
\newcommand{\hkine}[0]{\hat{T}_{\scriptscriptstyle \text{e}}}
\newcommand{\hkinn}[0]{\hat{{T}}_{\nsmall}}
\newcommand{\hkinMM}[0]{T_{N_\text{c}}}
\newcommand{\vnn}[0]{\hat{{W}}_{\nnsmall}}
\newcommand{\vtot}[0]{\hat{\mathcal{W}}_{\nnsmall}^{\text{tot}}}
\newcommand{\hvne}[0]{\hat{W}_\nesmall}
\newcommand{\vne}[0]{{v}_\nesmall}
\newcommand{\hwee}[0]{\hat{W}_\eesmall}
\newcommand{\VMM}[0]{W_{N_\text{c}}}

\newcommand{\dertwoel}[0]{\bold{h}_\text{Ne}}
\newcommand{\dertwomm}[0]{\bold{h}_\text{N}^\text{MM}}
\newcommand{\dertwommel}[0]{\bold{h}_\text{e}^\text{MM}}
\newcommand{\rhomm}[0]{n}
\newcommand{\rhommc}[0]{n_c}
\newcommand{\rhommopt}[0]{\rhomm_{\opt}}

\newcommand{\rhoel}[0]{\rho}
\newcommand{\rhoelopt}[0]{\rho_{\opt}}

\newcommand{\Prho}[1]{\hat{\mathcal{P}}_{ {\text{#1}}}}
\newcommand{\Prhob}[2]{\hat{\mathcal{P}}_{{\text{#1}}}^{\text{#2}}}
\newcommand{\Prhoa}[1]{\hat{\mathcal{P}}^{\text{#1}}}
\newcommand{\Prhoc}[0]{\hat{\mathcal{P}}}
\newcommand{\Orho}[1]{\hat{{\bm \rho}}_{ {{#1}}}}

\newcommand{\Peq}[0]{p_{\text{eq}}^{\gcano}}
\newcommand{\req}{{\bf R}^{\NN}_\text{eq}}
\newcommand{\Uint}[0]{\mathcal{U}_{\text{int}}}
\newcommand{\Unucl}[0]{\mathcal{U}_{\text{N}}}
\newcommand{\Sint}[0]{\mathcal{S}_{\text{int}}}
\newcommand{\Sqm}[0]{\mathcal{S}_{\text{N}}}
\newcommand{\Smm}[0]{\mathcal{S}_{\mm}}

\newcommand{\MystFunc}[0]{\mathcal{W}^{\text{eff}}_{\nsmall}}
\newcommand{\mystFunc}[0]{{w}^{\text{eff}}_{\nsmall}}
\newcommand{\PartF}[0]{\mathcal{Z}}
\newcommand{\minu}[1]{\underset{#1}{\text{min}}}

\newcommand{\cpF}[1]{{\mathcal{W}}_{{#1}}^{\mm}}
\newcommand{\cpf}[1]{{v}_{{#1}}^{\mm}}
\newcommand{\cpFbis}[1]{\hat{\mathcal{W}}_{{#1}}}

\newcommand{\eigvec}[0]{\Phi^{\alpha}_n}
\newcommand{\eigval}[0]{\mathcal{E}^{\alpha}_{n}}
\newcommand{\eigvalz}[0]{\mathcal{E}^{0}_{n}}
\newcommand{\Ne}{\text{N}_\text{e}}
\newcommand{\NN}{\text{N}_\text{N}}

\newcommand{\NMM}{\text{N}_\text{c}}

\begin{document}
\title{Prediction of the aqueous redox properties of functionalized quinones using a new QM/MM variational formulation.}
\author{Maxime Labat}
\affiliation{Sorbonne Université, CNRS, Physico-Chimie des Électrolytes et Nanosystèmes
Interfaciaux, PHENIX, F-75005 Paris, France}

\author{Guillaume Jeanmairet$^*$}
\email{guillaume.jeanmairet@sorbonne-universite.fr}
\affiliation{Sorbonne Université, CNRS, Physico-Chimie des Électrolytes et Nanosystèmes Interfaciaux, PHENIX, F-75005 Paris, France}
\affiliation{Réseau sur le Stockage Électrochimique de l'Énergie (RS2E), FR CNRS3459, 80039 Amiens Cedex, France}

\author{Emmanuel Giner$^*$}
\email{emmanuel.giner@lct.jussieu.fr}
\affiliation{Sorbonne Université, CNRS, Laboratoire de Chimie Théorique, Sorbonne Université, F-75005 Paris, France}

\begin{abstract}
\textbf{Abstract}\\
We recently proposed to couple quantum mechanics (QM) methods with molecular density functional theory (MDFT) to describe mixed quantum-classical systems [J. Chem. Phys. 161, 014113 (2024)]. This approach is particularly appropriate to account for solvent effect into QM calculations. Motivated by the growing interest in quinones as potential electrolytes for aqueous redox-flow batteries, we apply the QM/MDFT framework  to compute the two-electrons  redox potentials of a series of Benzoquinone/Hydroquinone couples in aqueous solution. However, since  these molecules are made of several dozens of atoms, their geometries are not trivial. This motivates the development of a geometry optimization procedure within the QM/MDFT framework. 

To this end, we introduce a new variational formulation for the grand potential of a mixed quantum-classical system.  Within the  Born-Oppenheimer approximation, and neglecting electronic entropy, the quantum solute is described by a product of electronic and nuclear density matrices, both depending parametrically  on   coordinates of the classical solvent. It can then be shown that a functional of the total density matrix  satisfies a variational principle for the grand potential. Using a mean-field approximation, we express  the grand potential of the mixed quantum-classical system as a variational problem which depends only on the nuclear density matrix. The nuclei  experience an external field generated by the electronic and classical one-particle densities.

In practice, the computation of the grand potential is reduced to a sequence of density optimizations. First, the classical solvent density and the solute electronic density are optimized for a fixed solute nuclear geometry, using the previously reported mixed quantum mechanics/classical procedure. Subsequently, the solute geometry is optimized for a fixed solvent configuration.  Finally, the redox potentials of a selection of Benzoquinone/Hydroquinone couples are computed after geometry optimizations. The  predictions are in good agreement with QM calculation using a  continuum solvent model and with experimental data. 
\end{abstract}
 
\maketitle

\section{Introduction\label{sec:Intro}}

Recent advancements in redox flow batteries (RFBs) highlight their potential for large-scale energy storage, particularly for intermittent renewable sources like solar and wind\cite{DinZhaZhaZhoYu-CSR-18, SinKimKanByon-CSR-19, KwaJuAzi-CR-20}. These batteries utilize external storage tanks for liquid-phase electrolytes. This design facilitates  maintenance and allows for scalable energy accumulation\cite{SinKimKanByon-CSR-19}. Traditional RFBs primarily employ metal-based active materials such as vanadium, iron, and zinc \cite{NoaRozHerFis-ACIE-15}. However, concerns over their cost and  toxicity have hindered their widespread utilization \cite{KeaShaWal-IJER-12, WinHagJanHagMarSch-ACIE-17}. Organic redox-active materials, particularly quinones, offer a promising alternative due to their abundance and highly tunable electrochemical windows\cite{Huskandal-N-14, DinZhaZhaZhoYu-CSR-18,Qingandal-SA-18, LeeKoKwoLeeKuKimKan-J-18, SinKimKanByon-CSR-19}.

Quinones are particularly noted for their rapid redox kinetics, making them suitable for advanced RFB technologies \cite{Huskandal-N-14, Gerhardtandal-AEM-17}. 
During their reduction process, each quinone can accept up to two electrons due to their characteristic two carbonyl groups. The precise mechanism of this reduction process depends strongly on the nature of the solvent. In organic aprotic solvents, where quinones are highly soluble, the reduction is a sequence of two single-electron transfers \cite{SasKasOhuOhsOht-JES-90, ShamSalGolShaMou-JSSE-01,WanXuCasChoiLiYan-CC-12,RoyGuin-JES-14}. Alternatively, it has been highlighted that quinone derivatives undergo a simultaneous double-electron reduction and double proton transfer in acidic aqueous solutions \cite{GuinDasMan-IJE-11, Huskandal-N-14, HuyAnsCavStaHam-JACS-16}. These proton-coupled electron transfers (PCET) have encouraged the development of aqueous RFBs (ARFBs). However, practical applications of quinones in ARFBs are often limited by their low solubility and suboptimal half-cell potentials\cite{YanHooWanPraNar-JES-14,WedDraKonBen-SR-16}.

To increase the redox potential window of quinones, a common strategy is to create  new quinone derivatives by functionalizing them with electron-withdrawing or donating groups. Due to gigantic chemical space of possible quinones,  experimental screening would be too time-consuming and costly.  To circumvent this problem, the  idea of high-throughput computational screening has emerged\cite{khetan_high-throughput_2023, ErSuhMarAsp-CS-15, HuyAnsCavStaHam-JACS-16, KimGooZimm-JPCC-16, DinZhaZhaZhoYu-CSR-18, SinKimKanByon-CSR-19, ZhaKheEr-SR-20,elhajj_first_2024}. In most studies, the chemical properties of each compound are computed using electronic Density Functional Theory (DFT), while the solvent is represented with a continuum solvent model, among which  the polarizable continuum model (PCMs) is very popular. 

Continuum solvent models allow accounting for solvation effects at a very moderate computational cost but suffer from several limitations. The major one is that they replace a series of interactions between molecules with a macroscopic electrostatic picture.  Classical density functional theory (cDFT) is a promising alternative approach that can account for solvation effects at a competitive cost while retaining a molecular description.

In short, cDFT allows computing the grand-potential by minimizing a functional of the liquid density \cite{Mermin-PhysRev-65,Evans-nature-79,dwandaru_variational_2011}. This minimum is reached for the equilibrium density.  Similarly to its electronic counterpart, cDFT is theoretically well grounded but approximations in the choice of the functional  describing the interactions between particles, here solvent molecules, are required to make its utilisation practical. Several attempts to use cDFT as a solver for QM/MM calculations have been proposed\cite{LiscAria-PRB-13, SunLetQchGunOzhAri-SX-17,TanCaiZhaLiu-JPCC-18,Troudbal-ChemRxiv-25}. 

We recently proposed to use  molecular density functional theory (MDFT), a flavour of cDFT that is designed to study solvation of classical solutes in rigid, classical solvents to account for solvent effects in QM calculations. Formally, one can show  that there is a variational principle for the free energy in terms of the full QM/MM density matrix. The computation of the free energy  can be reformulated using constrained-search methods \`a la Levy-Lieb into a variational problem involving only the quantum and classical one-body densities \cite{JeanLabGiner-arXiv-24}.  In practice, the QM/MDFT approach can be used with any QM techniques such as electronic DFT\cite{JeaLevBor-JCTC-20} or selected CI\cite{LabGinJean-JCP-24}, and has been used to predict ground- and excited-states properties of organic molecules, and chemical reactivity.  

In this paper, electronic DFT and MDFT are used to compute the aqueous redox potentials of a series of quinones. In particular, we focus on the simultaneous double proton charge electron transfer reduction. While our previous applications of the QM/MDFT approach were limited to  small molecules such as water or formaldehyde, the considered quinones are composed of several dozens of atoms, and some functional groups are quite flexible. This makes the choice of the geometry of solute to be used in the calculation less straightforward. Thus, we propose,  in Sec.~\ref{sec:Theory}, a theoretical framework to carry on the geometry optimization of the solute within the QM/MDFT formalism. This is done by formulating a variational principle for the grand potential as depending only on the nuclear variables, \textit{i.e.} the nuclear density matrix. The nuclei, however, experience an effective potential that is due to the electrons and the solvent molecules. We then propose a series of practical approximations that allow performing the computation of the grand potential as a series of quantum and classical densities optimizations followed by geometry optimization. The harmonic approximation is used to account for the vibrational motion of the nuclei and requires the computation of the Hessian matrix of the nuclei-nuclei interaction energy. In Sec.~\ref{sec:Methology}, we describe some theoretical and computational details before focusing on the redox properties of quinones in Sec.~\ref{sec:Applications}. The quality of our results is assessed by a thorough comparison with previous predictions obtained by Huynh  \textit{et al.} \cite{HuyAnsCavStaHam-JACS-16} using eDFT in a conductor-like PCM (C-PCM) and to experimental data reported by the same group.

\section{Theory: variational framework for QM/MM calculations \label{sec:Theory}}

This section is dedicated to the description of the grand potential of a QM/MM system where 
both the electrons and nuclei of the solute are treated quantum mechanically, while the solvent is treated at the MM level. Starting from a very general QM/MM framework for a nuclei-electron/classical solvent system and thanks to well-defined approximations, we obtain a formalism similar to that used in "thermodynamical corrections" in standard QM calculations using a continuous solvent model.  We therefore derive an approximation for the grand potential of an electron-nuclei QM system 
solvated by classical molecules, based on geometry optimization and frequency calculations. 

We first present in Sec.~\ref{subsec:ExGSEDMat} the Hamiltonian of the QM/MM system 
and establish the exact equilibrium density matrix (EDM) in the semi-grand canonical ensemble, 
together with the corresponding grand potential. Then, in Sec.~\ref{sec:bo_zee} we show how the exact EDM can be simplified within the Born Oppenheimer (BO) and zero electronic entropy (ZEE) approximations, 
which correspond to an adiabatic electronic ground state approximation while keeping the entropic term for both the QM nuclei and classical solvent. 
{ Then, inspired by a previous work\cite{JeanLabGiner-arXiv-24}, in Sec.~\ref{subsec:VarForm}  we propose a variational formulation of the grand potential in terms of high-dimensional objects. In Sec.~\ref{subsec:VarForm}~\&~\ref{subsec:NuclEffectPot}, by using  a mean mean-field approximation, which ignore the   explicit dependency of the quantum objects on the classical degrees of freedom, it becomes possible to reuse the recently proposed eDFT/cDFT mean-field coupling. Eventually, by approximating the nuclear effective potential as a harmonic potential, we obtain a practical computational scheme to approximate the Helmholtz free energy of the QM/MM system, which is presented in Sec.~\ref{subsec:harmonic}. }

\subsection{Exact QM/MM Equilibrium Density Matrix in the semi-grand canonical ensemble and grand potential\label{subsec:ExGSEDMat}}

We consider a solute/solvent QM/MM system at finite temperature $T$ in a fixed volume $V$.
The solute is described at the QM level and is constituted of $\Ne$ electrons, whose positions are denoted by
\mbox{$\br^{\Ne} = \{  \bri{1}, \dots, \bri{\Ne}  \} $} and  $\NN$ nuclei 
identified by their positions \mbox{$ \brnucltot = \{  \brnucl{1}, \dots, \brnucl{\NN} \} $}, 
 charges \mbox{$\textbf{Z}^{\NN} = \{  Z_{1}, \dots, Z_{\NN}  \} $}, 
and  masses \mbox{$ \{  M_{1}, \dots, M_{\NN} \} $}.
On the other hand, the solvent environment is described by classical rigid molecules, whose number $\NMM$ is allowed to fluctuate due to exchange with a reservoir at fixed chemical potential $\mu$. The appropriate statistical ensemble to describe the current system is thus the semi-grand canonical ensemble $(\Ne,\NN,\mu,V,T)$ .

For a given number $\NMM$ of solvent molecules, a solvent configuration is described by the molecules center of mass  
\mbox{$ \bq^{\NMM} = \{  \bq_{1}, \dots, \bq_{\NMM}  \} $}, 
and their orientations \mbox{${\bf \bomega}^{\NMM} = \{  {\bf \bomega}_{1}, \dots, {\bf \bomega}_{\NMM}  \} $}, where ${\bf \bomega}_i$ is a shorthand  for the three Euler angles $(\theta_i, \phi_i,\Psi_i)$ of the $i^\text{th}$ molecule. For conciseness, these positional and orientational variables are collectively denoted as  $\bqtot=\{\bq^{\NMM},\bomega^{\NMM}\}$. The superscript $\NMM$ indicates the number of particles, which,  as a reminder, is allowed to fluctuate. Additionally, the solvent molecules possess linear momenta \mbox{$ \textbf{P}^{\NMM} = \{  \textbf{P}_{1}, \dots, \textbf{P}_{\NMM}  \} $}. From this point on, atomic units are used.

The total Hamiltonian of this QM/MM system with $\NMM$ classical particles is defined as:
\begin{equation}
    \Ham{\text{tot}}(\bqtot) = \Ham{\qm}(\bqtot) +  \hkinMM(\textbf{P}^{\NMM}) + \VMM(\bqtot),
    \label{eq:Htot1}
\end{equation}
where $\hkinMM$ is the kinetic energy of $\NMM$ solvent molecules
and $\VMM$ is the potential energy of interaction.
{Note that, since the solvent is constituted of rigid molecules, their angular momenta also contribute to Hamiltonian through a rotational kinetic energy term. However, similarly to the linear momenta, these angular momenta are not coupled to the other degrees of freedom and can therefore be integrated analytically in the classical (grand-)partition function. To avoid unnecessarily cumbersome notation,  the rotational kinetic energy term is ignored in Eq.~\eqref{eq:Htot1} and in the following derivation, without any loss of generality.  }

In Eq.~\eqref{eq:Htot1}, $\Ham{\qm}(\bqtot)$ is the quantum Hamiltonian describing the electrons and nuclei.
Since it also includes the QM/MM interaction, it depends parametrically on the coordinates $\bqtot$ of the $\NMM$ classical particles. The QM Hamiltonian appearing in Eq.~\eqref{eq:Htot1} is defined as the sum of nuclear and electronic contributions:
\begin{equation}
    \Ham{\qm}(\bqtot) = \hkinn + \cpF{\nsmall} (\bqtot) + \Ham{\BO}(\bqtot, \brnucltot) ,
    \label{eq:HQM}
\end{equation}
where $\hkinn$ is the  kinetic energy operator of the nuclei and $\cpF{\nsmall}$ is their potential energy of interaction  with $\NMM$ molecules of solvent  at configuration $ \bqtot$. The electronic Hamiltonian $\Ham{\BO}(\bqtot)$ is the sum of the usual electronic Born-Oppenheimer (BO) Hamiltonian and the interaction between the electrons and $\NMM$ classical solvent molecules at configuration $ \bqtot$, labelled $\cpF{\esmall} (\bqtot)$, 
which therefore reads:
\begin{equation}
\label{eq:Hel}
 \begin{aligned}
    \Ham{\BO}(\bqtot, \brnucltot)  =& \hkine + \hwee + \hvne (\brnucltot) +\vnn(\brnucltot)  \\
    &+ \cpF{\esmall} (\bqtot).
 \end{aligned}
\end{equation}
In Eq.~\eqref{eq:Hel}, the first four terms are the usual contributions to the electronic BO Hamiltonian, \textit{i.e.} 
the electronic kinetic operator $\hkine$, mutual Coulomb repulsion $\hwee$, Coulomb attraction with the nuclei $\hvne$, and nuclear-nuclear Coulomb repulsion $\vnn$, respectively.

The QM/MM terms $\cpF{\nsmall}$ and $\cpF{\esmall}$ appearing in Eqs.~\eqref{eq:HQM} and \eqref{eq:Hel}
lack a unique, practical, definition and several propositions for these terms have been made in the literature. {Therefore, we will carry the general derivations without explicitly defining $\cpF{\nsmall}$ and $\cpF{\esmall}$, and simply assume the following generic forms,}
\begin{equation}
\label{eq:small_wnmm}
 \cpF{\nsmall} = \sum_{i=1}^{\NMM}\sum_{j=1}^{\NN} w_{\text{N}}^{\mm}(\mathbf{R}_j,\tilde{\mathbf{Q}}_i), 
\end{equation}
\begin{equation}
 \cpF{\esmall} = \sum_{i=1}^{\NMM}\sum_{j=1}^{\Ne} w_{\text{e}}^{\mm}(\mathbf{r}_j,\tilde{\mathbf{Q}}_i),
\end{equation}
where $w_{\text{N}}^{\mm}(\mathbf{R},\tilde{\mathbf{Q}})$ and $w_{\text{e}}^{\mm}(\mathbf{r},\tilde{\mathbf{Q}})$ are functions describing the 
pairwise nuclear-solvent and electron-solvent interactions, respectively. 

The only assumption is that the QM/MM interaction is pairwise additive, both for electrons and nuclei. We further emphasize that, while we chose to include these coupling terms into the Quantum Hamiltonian, they also act on the classical particles.

Within these definitions, the equilibrium density matrix (EDM)  for the QM/MM system with $\NMM$ classical solvent molecules in the canonical ensemble 
can be expressed as a quantum density matrix parametrized by the classical degrees of freedom  $(\bqtot,\textbf{P}^{\NMM})$, namely:  
\begin{equation}
 \Prho{eq}^{\NMM}(\bqtot,\textbf{P}^{\NMM}) = \frac{e^{-\beta (\Ham{\qm}(\bqtot)+\VMM(\bqtot))} 
  \text{e}^{- \beta \hkinMM(\textbf{P}^{\NMM}) }}{\PartF_{\NMM}},
      \label{eq:EDMatq}
\end{equation}
where the superscript $"\NMM"$ emphasizes that it is the canonical EDM when there are $"\NMM"$ classical particles, and $\PartF_{\NMM}$ is the partition function of the QM/MM system with $\NMM$ classical solvent molecules,
\begin{equation}
 \label{eq:z_n}
\begin{aligned}
  \PartF_{\NMM}\!\!&=\!\!\iint \!\!\Trac_\text{QM}[e^{-\beta \Ham{\qm}(\bqtot)}] \text{e}^{- \beta \big(\VMM(\bqtot)+\hkinMM\big)}\!d\textbf{P}^{\NMM}d\bqtot\!\!. 
\end{aligned}
\end{equation}
The "$\Trac_\text{QM}$" operator in Eq.~\eqref{eq:z_n} denotes the usual  trace over 
the quantum degrees of freedom.
The form of the EDM as in Eq.\eqref{eq:EDMatq} is  intuitive, as it follows the standard $\exp(-\beta H)$ form of statistical physics. It is worth emphasizing that this expression can be derived from first principles using Wigner transformation techniques and semi-classical expansions 
(see Refs. \onlinecite{NieKapCic-JCP-01,JeanLabGiner-arXiv-24}). 

Once the EDM and the partition function have been defined in the canonical ensemble it is rather natural to extend the formalism to the semi-grand canonical ensemble where the number of classical particles, $\NMM$, is allowed to fluctuate. Therefore, the corresponding EDM can be expressed as the direct sum over the canonical density matrices for all possible particle number $\NMM$, weighted by the Boltzman factor associated to the chemical potential $\mu$. 
This construction of the semi-grand canonical ensemble is conceptually similar to the definition of the Fock space in quantum mechanics, where it is formulated as the direct sum of all $N$-particle Hilbert spaces. The grand canonical equilibrium density matrix associated to the number of classical particle $\NMM$ is defined as:
\begin{equation}
	\Prho{eq}^\text{GC}(\bqtot,\textbf{P}^{\NMM}) = \frac{e^{-\beta (\Ham{\qm}(\bqtot)+\VMM(\bqtot)-\mu\NMM )} \text{e}^{- \beta \hkinMM(\textbf{P}^{\NMM}) }}{\Xi},
	\label{eq:gc_edm}
\end{equation}
where  the grand-partition function $\Xi$ is defined as 
\begin{equation}
\label{eq:xi}
 \begin{aligned}
 \Xi  & = \sum_{\NMM=0}^\infty\frac{1}{\NMM!} \PartF_{\NMM} \text{e}^{ \beta\mu\NMM }. \\
 \end{aligned}
\end{equation}
In the following, as we will constantly be dealing with quantities both in the canonical and semi-grand canonical ensembles, it is useful to introduce some conventions. 
When considering a quantity, say $O$, without explicit mention to the number of particles $\NMM$, it will then be the semi-grand canonical extension of the operator $O^{\NMM}$, the latter being defined for each number of classical particles $\NMM$. 
As usually done in quantum mechanics with the notion of Fock space,  
the semi-grand canonical extension of an operator $O^{\NMM}$ can be formally defined in terms of a direct sum, over all particle number sectors $\NMM$, of the canonical operators $O^{\NMM}$. 
Within the present context, it is also useful to introduce the semi-grand canonical classical trace operator, $ \Trac_\text{GC}$, 
of a semi-grand canonical operator $O$ as follows
\begin{equation}
\Trac_\text{GC}[O] = \sum_{\NMM=0}^\infty \frac{1}{\NMM!}\Trac[O^{\NMM}],
\end{equation}
where $\Trac[O^{\NMM}]$ is the canonical trace over the quantum and $\NMM$ classical particles degrees of freedom. 
Within these definitions, the partition function of Eq. \eqref{eq:xi} becomes 
\begin{equation}
\Xi  = \Trac_\text{GC}[\text{e}^{-\beta (\Ham{\qm}+W-\mu N)} \text{e}^{- \beta T }],
\end{equation}
where $\Ham{\qm}$, $W$, $T$ and $N$ are the semi-grand canonical extensions of the quantum Hamiltonian, classical potential energy, classical kinetic energy, and classical number of particles, respectively.

One can then naturally define the grand potential, $\Omega$, that is  the thermodynamic state function of this system 
\begin{equation}
 \Omega = - \frac{1}{\beta} \log(\Xi).
\end{equation}
One of the objectives of this paper is to propose a variational formulation of the grand potential, which extends our previous work in two significant ways. First, the classical solvent particles are now described within a grand-canonical framework, allowing their number to fluctuate.  Second, the nuclei of the solute are no longer treated as clamped, but instead their degrees of freedom are explicitly accounted for in the expression of the grand-canonical functional. We begin by introducing two well-established approximations that simplify the expression of the equilibrium density matrix: the Born-Oppenheimer approximation and the zero electronic entropy (ZEE) approximation.


\subsection{Born Oppenheimer and zero electronic entropy approximations\label{sec:bo_zee}}
For a given number of classical particle $\NMM$, the grand-canonical equilibrium density matrix $\Prhob{eq}{GC}$ can be expressed as the product of a purely classical probability distribution $\Peq(\bqtot,\textbf{P}^{\NMM})$ with  a purely quantum electronic-nuclear
density matrix , $ \Orho{\text{eq}} (\bqtot)$, parametrically depending on the
classical 
variable:
\begin{equation}
 \Prhob{eq}{GC}(\bqtot,\textbf{P}^{\NMM}) = \Peq(\bqtot,\textbf{P}^{\NMM}) \Orho{\text{eq}} (\bqtot),
      \label{eq:EDMateq_prod}
\end{equation}
with 
\begin{equation}
    \Peq(\bqtot,\textbf{P}^{\NMM}) =  \PartF_{\text{QM}} (\bqtot) 
  \frac{\text{e}^{- \beta \big(\VMM (\bqtot)+\hkinMM(\textbf{P}^{\NMM})-\mu\NMM \big) }}{\Xi},
    \label{eq:Pmm_Q}
\end{equation}
\begin{equation}
 \Orho{\text{eq}} (\bqtot) =   \frac{\text{e}^{- \beta \Ham{\qm}(\bqtot)}}{\PartF_{\text{QM}}(\bqtot)}  ~\text{and}~   
      \label{eq:qm_mat}
\end{equation}

 \begin{equation}
	\PartF_{\text{QM}}(\bqtot) = \text{Tr}_\text{QM}   \left\{\text{e}^{- \beta \Ham{\qm}(\bqtot)}\right\}.
	\label{eq:Zqm}
\end{equation}
It is important to note that, although $\Prhob{eq}{GC}$ is written as a product of a QM density matrix and a classical  probability distribution, it explicitly incorporates QM/MM correlation effects, as the quantum density matrix depends parametrically on the classical solvent coordinates $\bqtot$. Furthermore, the classical probability distribution is influenced by the quantum subsystem through the normalization condition (\textit{i.e.} through $\PartF_{\text{QM}} (\bqtot)$), thus preserving the coupling between the two subsystems.

To simplify the quantum density matrix $\Orho{\text{eq}}$, several standard approximations can be employed. First, making the BO approximation allows for rewriting $\Orho{\text{eq}}$ as a weighted sum over electronic-nuclear adiabatic states, labelled by indices $(\alpha,n)$, where $\alpha$ denotes the electronic state and $n$ the nuclear vibrational state, \textit{i.e.}
\begin{equation}
	\Orho{\text{eq}} (\bqtot) =  
	\sum_{\alpha,n} 
	\frac{e^{-\beta \eigval(\bqtot)}}{\PartF_{\text{QM}}(\bqtot)} \ket{\eigvec (\bqtot)} \bra{\eigvec(\bqtot)}.
	\label{eq:QMEDmat}
\end{equation}
In the above expression, $\eigval(\bqtot)$ and $\ket{\eigvec (\bqtot)}$ are the eigenvalues and eigenvectors of the total electronic-nuclear adiabatic quantum Hamiltonian (\textit{i.e.} neglecting the non-adiabatic couplings between electrons and nuclei) for a given solvent configuration $\bqtot$. 
This expression can be further simplified in the case of a weakly correlated system at room temperature, 
where one can neglect the contribution from electronically excited adiabatic states.  Under this assumption, only the ground electronic state $\alpha=0$ is retained in the summation of Eq.~\eqref{eq:QMEDmat}.
This approximation, hereafter referred to as 
\textit{zero electronic entropy} (ZEE), is justified by the fact that the 
relative statistical weights of excited adiabatic electronic states decay exponentially with increasing energy gaps. Consequently, their contributions become negligible at room temperature.

Under the ZEE approximation, and exploiting the product structure of the electron-nuclei 
wave function provided by the BO approximation,  
the quantum electronic-nuclear density matrix can  be written as the product of an electronic density matrix  and a nuclear density matrix 
\begin{equation}
\begin{split}
   {\Orho{\text{eq}}(\bqtot)}   = \Orho{\nsmall}^0(\bqtot)\, \Orho{\elsmall}^0 (\bqtot, \brnucltot),
   \label{eq:QMGSEDMat2}
\end{split}
\end{equation}
where $\Orho{\elsmall}^0 (\bqtot)$ is the ground state electronic adiabatic density matrix depending parametrically 
on both the solvent and nuclear coordinates 
\begin{equation}
 \Orho{\elsmall}^0 (\bqtot, \brnucltot) = \ket{\Psi^{0}(\bqtot, \brnucltot)}\bra{\Psi^{0}(\bqtot, \brnucltot)},
   \label{eq:ground_e}
\end{equation}
and $\Orho{\nsmall}^0(\bqtot)$ is the nuclear density matrix which contains contributions from 
all excited nuclear state associated to the adiabatic electronic ground state 
\begin{equation}
\begin{aligned}
   & \Orho{\nsmall}^0(\bqtot)=\\
    &\sum_{n} \frac{e^{- \beta 
 \eigvalz(\bqtot)}}{\PartF_{\textbf{QM}}(\bqtot)}
    \ket{\chi_\nsmall^{0,n}(\brnucltot;\bqtot)} \bra{\chi_\nsmall^{0,n}(\brnucltot;\bqtot)},
   \label{eq:nucl}
\end{aligned}
\end{equation}
and where the  partition function defined in Eq.~\eqref{eq:Zqm} simplifies into a purely nuclear contribution 
\begin{equation}
 \PartF_{\text{QM}}(\bqtot) = \sum_n e^{-\beta \eigvalz(\bqtot)}.
\end{equation}

\subsection{QM/MM Variational Formulation and mean-field approximation\label{subsec:VarForm}}
In the previous sections, we derived an explicit expression for the equilibrium density matrix of a QM/MM system made of $\Ne$ electrons and $\NN$ nuclei , and  a fluctuating number of classical particles. 
This derivation was performed within the semi-grand canonical ensemble,  employing both the  BO and ZEE approximations. We now propose a variational formulation of the grand potential for this system, extending the approach introduced in our previous work within the canonical ensemble \cite{JeanLabGiner-arXiv-24}.  

More specifically, we consider a general QM/MM electron-nuclei-solvent density matrix, $\Prho{}^{\text{GC}}$, which can differ from the equilibrium one, while conserving the product structure of Eq.~\eqref{eq:EDMateq_prod}. Therefore, for a given number of classical particles $\NMM$, it has the following structure
\begin{equation}
	\label{def_gen_prho}
	\Prho{}^{\text{GC}}(\bqtot,\textbf{P}^{\NMM}) = p^{\text{GC}}(\bqtot,\textbf{P}^{\NMM})  \Orho{}(\bqtot).
\end{equation}
In Eq.~\eqref{def_gen_prho}, $p^{\text{GC}}$ is the probability of having  $\NMM$ classical particles with positions and momenta $(\bqtot,\textbf{P}^{\NMM})$, and $ \Orho{}=\Orho{\nsmall} \Orho{\text{el}}$ is the electron-nuclei density matrix, which has a similar form to Eq. \eqref{eq:QMGSEDMat2}, and therefore depends parametrically 
 on the classical solvent configuration $\bqtot$. 
Therefore, $\Prhoc^{\text{GC}}$ is normalized to unity, \textit{i.e.} $\Trac_\text{GC} [\Prhoc^{\text{GC}}] =1$. 
We can now introduce the grand potential functional for the QM/MM system in the semi-grand canonical ensemble
\begin{equation}
 \omega[\Prhoc^{\text{GC}}] =  \Uint[\Prhoc^{\text{GC}}] - \frac{1}{\beta} \Sint[\Prhoc^{\text{GC}}]\label{eq:omega},
\end{equation}
where $\Uint[\Prhoc^{\text{GC}}]$ is the internal energy of the QM/MM system
 \begin{equation}
	\label{eq:Uint_c}
\begin{aligned}
   \Uint[\Prhoc^{\NMM}] = \Trac_\text{GC} \{ 
   p^{\text{GC}}  \Orho{}&\big[ \Ham{\qm} + W
   + T -\mu N\big]  \},
\end{aligned}
 \end{equation}
and $\Sint[\Prhoc^{\text{GC}}]$ is the dimensionless entropy associated with the grand-canonical density matrix $\Prhoc^{\text{GC}}$
\begin{equation}
   \Sint[\Prhoc^{\text{GC}}] =-\Trac_\text{GC}\left\{ p^{\text{GC}}  \Orho{\nsmall} \,\,\text{log} ( p^{\text{GC}}  \Orho{\nsmall}) \right\}.
\label{eq:Sint1}
 \end{equation}
Notice that the definition of the entropy in Eq.~\eqref{eq:Sint1} does not contain the electronic density matrix, which is a consequence of the ZEE approximation.

%

Following the same steps as in Ref. \onlinecite{JeanLabGiner-arXiv-24} (\textit{i.e.} using Jensen's inequality), it can be shown that the functional $\omega$ of Eq.~\eqref{eq:omega} satisfies the variational principle for the grand potential, \textit{i.e.} 
\begin{equation}
	\label{eq:Ftot_min}
	\Omega = \min_{\Prhoa{GC}} \omega[\Prhoa{GC}], \quad \Prhob{eq}{GC} = \text{argmin}\,\,\omega[\Prhoa{GC}],
\end{equation}
where the minimization is performed over all admissible density matrices $\Prhob{eq}{GC}$ consistent with the assumed product structure and normalization constraints of Eq.~\eqref{def_gen_prho}. The novelty of the present approach  lies in formulating the variational principle  within the semi-grand canonical ensemble, and explicitly accounting for the quantum degrees of freedom of the solute nuclei. 

Although the variational form of Eq.~\eqref{eq:Ftot_min} remains highly complex due to the involvement of high-dimensional objects, it provides a foundation for introducing practical approximations that eventually yield to a one-body density-functional formulation. The first step in this strategy is to neglect the explicit electron-solvent QM/MM correlation terms as proposed in earlier works\cite{JeaLevBor-JCTC-20,LabGinJean-JCP-24}, which corresponds to adopting  a mean-field approximation.

Therefore, we propose to extend this mean-field approximation to the electron-nuclei-solvent EDM $\Prhoa{GC}$ in Eq.~\eqref{def_gen_prho}, 
by  ignoring the explicit dependence on $\bqtot$ in the QM part. 
This leads to the following mean-field canonical QM/MM electron-nuclei-solvent density matrix in the semi grand-canonical ensemble
\begin{equation}
 \label{def_gen_prho_mf}
  \Prhob{MF}{GC}(\bqtot,\textbf{P}^{\NMM
  }) = p^{\text{GC}}(\bqtot,\textbf{P}^{\NMM
  }) \Orho{\nsmall}\, \Orho{\elsmall}(\brnucltot).
\end{equation}
Here, the density matrices $\Orho{\nsmall}$ and $\Orho{\elsmall}(\brnucltot)$ are quantum objects that do not depend explicitly on the classical degrees of freedom, and therefore neglect explicit QM/MM correlation effects.
Inserting this form of $\Prhoa{GC}$ into the variational principle of Eq.~\eqref{eq:Ftot_min} yields  the mean-field approximation $\omegamf$ which provides an upper-bound to the exact grand potential $\Omega$, 
\textit{i.e.} 
\begin{equation}
	\label{eq:Ftot_mf}
\omegamf = \min_{\Prhob{MF}{GC}} \omega[\Prhob{MF}{GC}], \quad \Omega \le \omegamf.
\end{equation}

\subsection{Effective potential for the nuclei and QM/MM density functional formalism \label{subsec:NuclEffectPot}}

The variational formulation of Eq.~\eqref{eq:Ftot_mf}, expressed in terms of electron-nuclei-solvent objects and the product structure of $\Prho{MF}$ allows, after some algebraic manipulations which are explicited in Appendix \ref{app:omegamf}, to approximate $\omegamf$ by a variational formulation involving only the  nuclear density matrix.  Nevertheless, the nuclei now experience an effective potential arising from both the electrons and the solvent. More precisely,  $\omegamf$ can be rewritten as follows  
\begin{equation}
 \label{def_fint_nucl}
 \omegamf = \min_{\Orho{\nsmall}} \big( \Unucl [\Orho{\nsmall} ]  - \frac{1}{\beta} \Sqm[\Orho{\nsmall}] \big),
\end{equation}
where $\Sqm[\Orho{\nsmall}] $ is the purely nuclear quantum entropy
\begin{equation}
 \Sqm[\Orho{\nsmall}] = -\text{Tr}\left\{ \Orho{\nsmall} \,\, \text{log} ( \Orho{\nsmall}) \right\},
\label{eq:Sqm}
\end{equation}
and $\Unucl$ is the nuclear quantum energy 
\begin{equation}
\begin{split}
 \Unucl [\Orho{\nsmall} ] = \text{Tr}\{ \Orho{\nsmall} [\hkinn +\vtot(\brnucltot) ] \}.
\label{eq:Uint2}
\end{split}
\end{equation}
The "Tr" symbol in Eqs.~\eqref{eq:Sqm}-\eqref{eq:Uint2} refers to the trace over the nuclear degrees of freedom.
In Eq.~\eqref{eq:Uint2}, $\vtot(\brnucltot)$ is the total effective potential felt by the nuclei defined as 
\begin{equation}
 \vtot(\brnucltot) = \vnn(\brnucltot) +  \MystFunc (\brnucltot),
\label{eq:Vtot}
\end{equation}
which is the sum of the nucleus-nucleus Coulomb interaction and of $\MystFunc$, 
 the effective potential experienced by the nuclei due to the presence of the electrons and solvent molecules. 
The latter is defined as 
\begin{equation}
 \begin{aligned}
 \MystFunc (\brnucltot)  = \minu{\Orho{\elsmalll}, \pgc} &\big( f_{\elsmall}[\Orho{\elsmall}] 
         + f_{\mm}^{\gc}[\pgc] + E_{\qm}^{\mm}[\Orho{\elsmall}, \pgc] \\
         & + V_{\elsmall}^{\nsmall}[\Orho{\elsmall};\brnucltot] + V_{\mm}^{\nsmall}[\pgc;\brnucltot]
 \big),
\label{eq:MystFunc_full}
 \end{aligned}
\end{equation}
with $f_{\elsmall}[\Orho{\elsmall}] $ and $f_{\mm}[\pgc]$ being the intrinsic electronic and classical energy functionals, respectively, defined as 
\begin{equation}
 f_{\elsmall}[\Orho{\elsmall}] = \text{Tr}_\text{QM}\big\{ \Orho{\elsmall}\big( \hkine + \hwee\big)\big\},\label{eq:fel}
\end{equation}
\begin{equation}
 \begin{aligned}
 & f_{\mm}^{\gc}[\pgc]  = \Trac_\text{GC}[\pgc(W+T-\mu N + \frac{1}{\beta}\log(\pgc))]\\
 &  =\sum_{\NMM=0}^{\infty} \iint \pgc (\VMM + \hkinMM -\mu \NMM  + \frac{1}{\beta} \log(\pgc)  \big)d\bqtot d\textbf{P}^{\NMM}.\label{eq:fmm}
 \end{aligned}
\end{equation}
On the other hand, $E_{\qm}^{\mm}[\Orho{\elsmall}, \pgc]$ is the electron-solvent QM/MM interaction
\begin{equation}
 \begin{aligned}
 &E_{\qm}^{\mm}[\Orho{\elsmall}, \pgc] = \Trac_\text{GC}[\pgc \text{Tr}_{\text{QM}}\big[ \Orho{\elsmall} \cpF{\esmall} \big]\big] \\ 
  & = \sum_{\NMM=0}^{\infty}\iint \pgc(\bqtot, \textbf{P}^{\NMM}) \text{Tr}_{\text{QM}}\big[ \Orho{\elsmall} \cpF{\esmall} \big]d\bqtot d\textbf{P}^{\NMM},
 \end{aligned}
\end{equation}
$ V_{\elsmall}^{\nsmall}[\Orho{\elsmall};\brnucltot]$ and $V_{\mm}^{\nsmall}[\pgc;\brnucltot]$ 
are the electron and classical interactions with the nuclei, respectively, defined as 
\begin{equation}
 \label{eq:ven}
 V_{\elsmall}^{\nsmall}[\Orho{\elsmall};\brnucltot] = \text{Tr}_\text{QM} \big[ \Orho{\elsmall} \hvne (\brnucltot) \big],
\end{equation}
\begin{align}
 \label{eq:vmmn}
 V_{\mm}^{\nsmall}&[\pgc,\brnucltot] = \Trac_\text{GC}[p^\text{GC}\cpF{\nsmall}] \nonumber \\ 
&=\sum_{\NMM=0}^{\infty}\iint  p^\text{GC}(\bqtot,\textbf{P}^{\NMM})\cpF{\nsmall} (\bqtot,\brnucltot)d\bqtot d\textbf{P}^{\NMM}.
\end{align}
An important remark is that both the electron-nuclear and the solvent-nuclear potentials are expressed as a sum of pairwise terms (\textit{cf.} Eq.~\eqref{eq:small_wnmm} for the latter). Therefore, for a given nuclear coordinate $\brnucltot$, $V_{\elsmall}^{\nsmall}$ and $V_{\mm}^{\nsmall}$ act on the electrons and solvent molecules as external local potentials expressed as sums of one-body terms. Consequently, $\MystFunc (\brnucltot)$ is obtained as a minimization of a functional describing a QM/MM system made of interacting electrons and classical particles, both embedded in local external potentials. We can therefore identify $\MystFunc (\brnucltot)$ as the mean-field approximation of the grand potential for the electron-solvent system which was formalized in Ref. \onlinecite{JeanLabGiner-arXiv-24}. As shown in  the latter, the Levy–Lieb constrained search formalism\cite{Levy-PNAS-79,Levy-PRA-82, Lieb-IJQC-83} can be used to express 
$\MystFunc (\brnucltot)$  as a functional of the one-body electronic and classical densities. More precisely, $\MystFunc (\brnucltot)$ can be written as 
\begin{equation}
\begin{aligned}
\MystFunc (\brnucltot) = \minu{\rhoel,\rhomm} \,\,\mystFunc[\rhoel,\rhomm,\brnucltot],
\label{eq:MystFunc}
\end{aligned}
\end{equation}
where,  $\rhoel$ and $\rhomm$ are respectively the one-body electronic and classical densities. In \eqref{eq:MystFunc}, $\mystFunc[\rhoel,\rhomm,\brnucltot]$ is the one-body density grand potential functional defined as 
\begin{equation}
 \begin{aligned} 
 \mystFunc[\rhoel,\rhomm,\brnucltot] = 
 & \calF{\elsmall} \left[\rhoel\right]+ (\vne(\brnucltot)|\rhoel) +\calF{\mm} \left[\rhomm\right]      \\
 +&  (\cpf{\nsmall}(\brnucltot)|\rhomm)+ (\rhomm|\cpf{\esmall}|\rhoel).
 \label{eq:MystFunc_small}
	\end{aligned}
\end{equation}
In Eq.~\eqref{eq:MystFunc_small}, $\calF{\elsmall} \left[\rhoel\right]$ is the usual electronic Levy-Lieb\cite{Levy-PNAS-79,Levy-PRA-82, Lieb-IJQC-83} 
universal functional, $(\vne(\brnucltot)|\rhoel)$ is the one-body nuclear-electron interaction, $\calF{\mm} \left[\rhomm\right]$ is  the classical universal functional, as defined by Evans\cite{Evans-nature-79}, $(\cpf{\nsmall}(\brnucltot)|\rhomm)$ is the one-body nuclear-solvent interaction defined as
\begin{equation}
 (\cpf{\nsmall}(\brnucltot)|\rhomm) = \int d\tilde{\mathbf{Q}} \cpf{\nsmall}(\tilde{\mathbf{Q}};\brnucltot) \rhomm(\tilde{\mathbf{Q}}),
\end{equation}
with $\cpf{\nsmall}(\tilde{\mathbf{Q}};\brnucltot) = \sum_{i=1}^{\NN}w_{\text{N}}^{\mm}(\mathbf{R}_i,\tilde{\mathbf{Q}})$,
and $(\rhomm|\cpf{\esmall}|\rhoel)$
 is the mean-field electron-solvent interaction 
\begin{equation}
 \label{eq:mf_qmmm}
 (\rhomm|\cpf{\esmall}|\rhoel) = \int {d}\br d\bqtotone\,\rhoel(\br)\cpf{\elsmall}(\br,\bqtotone)\rhomm(\bqtotone).
\end{equation}

Therefore, by obtaining $\MystFunc (\brnucltot)$ through a coupled functional minimization 
over electrons and solvent one-body densities, we can rigorously approximate the grand potential of a QM/MM  system composed of quantum electrons and nuclei  solely as a functional of the nuclear coordinates. To the best of our knowledge, this represents the first rigorous derivation of such a grand potential formulation explicitly accounting for quantum nuclei within this QM/MM framework.

\subsection{Harmonic approximation\label{subsec:harmonic}}

Although the formulation of Eq.~\eqref{def_fint_nucl} significantly reduces the dimensionality of the problem, it still requires solving a non factorisable quantum problem involving all nuclear degrees of freedom, which is rapidly cumbersome.  
To address this, and following the treatment proposed by McQuarrie\cite{Quarrie}, we approximate the Gibbs free energy of the nuclei by invoking a local description of the external potential and employing a tractable, integrable model. 
 These two criteria naturally lead to the harmonic oscillator (HO) approximation for the effective potential  $\vtot(\brnucltot)$ of Eq.~\eqref{eq:Vtot}, along with the rigid-rotor harmonic oscillator (RRHO) model to describe internal rotations of the solute.
 Several variants of the HO and RRHO approximations have been developed to estimate the Gibbs free energy\cite{RiMarCraTru-JPCB-11,Grimme-CEJ-12,LiGomMalBelHea-JPCC-15}, particularly aimed at mitigating the divergence of entropic contributions in the low-frequency regime. 
A detailed discussion about the implication of the latter can be found in a recent paper \cite{VelMalGerMed-JCP-25}. 
However, since the primary objective of the present work is to benchmark our solvent model against polarizable continuum models (PCM), we adopt the simplest HO formulation—without low-frequency cutoffs—to remain consistent with widely used implementations in standard electronic structure packages such as Gaussian\cite{Ochterski-Gaussian-00, gaussian-16}, PySCF\cite{sun-pyscf-JCP-2020}, and Orca\cite{Orca-JCP-20}.
Naturally, this choice does not prevent the use of more sophisticated HO or RRHO treatments, which can be incorporated as post-processing steps based on the computed normal modes and their associated vibrational frequencies.

After a series of mathematical  derivation detailed  in appendix \ref{app:HO},  the  grand potential can be approximated as  
\begin{equation}
 \begin{aligned}
 \label{eq:fint_final}
 \Omega  = \vtot(\req) + \sum_k \big(\frac{\omega_k}{2} +  \frac{1}{\beta}\log(1 - e^{-\beta \omega_k })\big),
 \end{aligned}
\end{equation}
 where $ \vtot$ is the effective potential of Eq \eqref{eq:Vtot}, $\req$ is the equilibrium geometry  and $\omega_k$ is the frequency associated to the normal mode $k$.
 
Evaluating $\Omega$ according to  Eq.~\eqref{eq:fint_final} requires the knowledge of the equilibrium geometry 
 $\req$. Moreover, it is necessary to compute the Hessian matrix of $\vtot$ at the equilibrium geometry to compute the normal modes $k$ and their frequencies $\omega_k$. A discussion about the evaluation of the gradient of the total 
 effective potential $\vtot$  with respect to the solute nuclear coordinates can be found in Appendix \ref{app:grad_hessmat}.

\section{Methodology \label{sec:Methology}}

\subsection{Theoretical models \label{subsec:models}}
\indent
As a practical illustration of the proposed framework, we   focus on the computation of redox properties of a series of quinones in aqueous solution.
To evaluate the grand potential according to Eq.~\eqref{eq:fint_final} we must specify how the various terms contributing to the effective potential felt by the nuclei as defined in Eq.~\eqref{eq:MystFunc_small} are modeled. In practice, the electronic Levy-Lieb functional ${\cal F}_\text{el}$ is replaced by an approximated electronic functional within the Kohn-Sham formalism. For the classical universal functional  ${\cal F}_\text{MM}$, we chose a MDFT functional which is well suited for the study of solvated systems. Since detailed descriptions of MDFT can be found in our previous work, we only recall here its essential aspects. As mentioned earlier, the classical solvent is constituted of rigid molecules  and is described by the solvent density $n(\tilde{\bq})=n(\br,\bomega)$. The functional of this solvent density is split into two contributions
\begin{equation}
{\cal F}_\text{MM}[n]={\cal F}_\text{id}[n]+{\cal{F}}_\text{exc}[n] \label{eq:F=Fid+Fexc}.
\end{equation}
The ideal term ${\cal F}_\text{id}[n]$ is know exactly and corresponds to the grand potential of a non interacting fluid of density $n$. While no exact practical expression exists for ${\cal F}_\text{exc}[n]$, it can be approximated by performing a Taylor expansion around a homogeneous reference fluid. This leads to:
\begin{equation}    
 \begin{aligned}
  {\cal F}_\text{exc}[n]
  =-\frac{1}{2\beta}\iint & \Delta n(\bm{r}) c(|\bm{r}-\bm{r^{\prime}}|,\bm{\Omega},\bm{\Omega}^\prime;n_0) \\
  & \Delta n(\bm{r}^\prime) d\bm{r}d\bm{r}^\prime d\bm{\Omega}d\bm{\Omega}^{\prime}+{\cal F}_\text{B}[n]\label{eq:fex=fhnc+fb}.
 \end{aligned}
\end{equation}

In Eq.~\eqref{eq:fex=fhnc+fb} the second order term in density involves the second order direct correlation function of the homogeneous fluid at density $n_0$ and $\Delta n=n-n_0$. All the terms of higher order are gathered into the so-called bridge functional ${\cal F}_\text{B}$. While the second order term can be computed thanks to ingenious computational tricks detailed elsewhere\cite{DinLevBorBel-JCP-17},  the higher-order terms are not tractable in practice and require approximations for ${\cal F}_\text{B}$.
The last parts of  Eq.~\eqref{eq:MystFunc_small} that needs to be specify are the QM/MM coupling terms, $\cpf{\esmall}$ and $\cpf{\nsmall}$. Again there is no unique practical definition for these terms. In this work, we adopt a  mean-field Coulombic interaction between the solvent and both the electrons and nuclei, supplemented by
a Lennard-Jones potential to account for the Pauli repulsion and dispersion forces. 
More specifically, the electron-solvent interaction term of Eq.~\eqref{eq:mf_qmmm} is expressed as 
an electrostatic term 
\begin{equation}
 \label{eq:mf_qmmm_2}
 (\rhomm|\cpf{\esmall}|\rhoel) = -\int {d}\bri{1} d\bri{2}\,\rhommc(\bri{1})
  \frac{1}{|\bri{1}-\bri{2}|} \rhoel(\bri{2}),
\end{equation}
where $\rhommc(\br)$ is the classical charge density associated to the classical density $\rhomm$, 
which is given by 
\begin{equation}
	\rhommc(\bm{r})= \iint \rhomm(\bm{r}', \bm{\Omega}) \gamma(\bm{r} - \bm{r}',\bm{\Omega}) d\textbf{r}'d\bm{\Omega},
	\label{charge_dens_solv}
\end{equation}
where $\gamma(\bm{r},\bm{\Omega})$ is the point charge distribution of a water molecule, taken at the origin with orientation $\bm{\Omega}$. 

The nuclear-solvent interaction term, $\cpf{\nsmall}$,  is here expressed as a sum of an electrostatic 
and Lennard-Jones (LJ) potential contributions, \textit{i.e.} 
\begin{equation}
 \label{eq:nucl_mm_2}
 (\cpf{\nsmall}(\brnucltot)|\rhomm) = (w^\text{MM}_\text{C}(\brnucltot)|\rhommc) + (w^\text{MM}_\text{LJ}(\brnucltot)|\rhomm),
\end{equation}
with 
\begin{equation}
 \label{eq:nucl_mm_3}
 (w^\text{MM}_\text{C}(\brnucltot)|\rhommc) = \int {d}\bri{} \rhommc(\bri{})\sum_{A}
  \frac{Z_A}{|\brnucl{A}-\bri{}|} ,
\end{equation}
\begin{equation}
 \label{eq:nucl_mm_4}
 (w^\text{MM}_\text{LJ}(\brnucltot)|\rhomm)  = \int {d}\bqtotone{} \rhomm(\bqtotone{}) \sum_{A}\cpFbis{A}^\text{LJ}(\brnucl{A},\bqtotone{}),
\end{equation}
where $\cpFbis{A}^\text{LJ}$ is the usual LJ interaction between a nucleus $A$ and a classical water molecule.

Regarding the practical implementation of our QM/MM scheme, it is useful to introduce the following potentials
\begin{equation}
 \label{eq:mf_qmmm_2_qm}
 \begin{aligned}
 v_{\text{ext}}^{\text{QM}}(\bri{}) & = \int d\bri{}'\,\rhommc(\bri{}')
  \frac{1}{|\bri{}'-\bri{}|}, \\
  \end{aligned}
\end{equation}
\begin{equation}
 \label{eq:mf_qmmm_2_mm}
 \begin{aligned}
  v_{\text{ext}}^{\text{MM}}(\bri{}) & = \sum_{A}
  \frac{Z_A}{|\brnucl{A}-\bri{}|} - \int d\bri{}'\,\rhoel(\bri{}')\,\,\,
  \frac{1}{|\bri{}'-\bri{}|}.
  \end{aligned}
\end{equation}
The potential $v_{\text{ext}}^{\text{QM}}$ is the electrostatic potential created by the MM region and felt by the QM region, and on the other hand $v_{\text{ext}}^{\text{MM}}$ is the electrostatic potential created by the solute (\textit{i.e.} electrons and nuclei) and felt by the MM region. 
With these definitions, the electron-solvent interaction of Eq.~\eqref{eq:mf_qmmm_2} can then be interpreted as an external electrostatic potential for the QM system, namely
\begin{equation}
 \begin{aligned}
 \label{eq:mf_qmmm_2_bis}
 (\rhomm|\cpf{\esmall}|\rhoel) & = -\int {d}\bri{} 
  v_{\text{ext}}^{\text{QM}}(\bri{}) \rhoel(\bri{}), \\
   \end{aligned}
\end{equation}
and similarly the total electrostatic interaction between the QM and MM region can be obtained as 
\begin{equation}
 (w^\text{MM}_\text{C}(\brnucltot)|\rhommc) + (\rhomm|\cpf{\esmall}|\rhoel) = \int {d}\bri{} v_{\text{ext}}^{\text{MM}}(\bri{}) \rhommc(\bri{}).
\end{equation}
The potentials $v_{\text{ext}}^{\text{QM}}$ and  $v_{\text{ext}}^{\text{MM}}$ will be used in practice to couple the QM and MM calculations as explained in Sec.~\ref{subsec:ComDet}. 


\subsection{Computational details \label{subsec:ComDet}}

This paper focuses on the aqueous solvation of a selection of quinones. The geometries of the 24 quinones displayed in Fig.~\ref{fig:quinones} were extracted from Huynh et al. \cite{HuyAnsCavStaHam-JACS-16}. 

For a given nuclear geometry $\brnucltot$, the methodology of QM/MDFT calculations  has already been reported in our previous work\cite{JeaLevBor-JCTC-20,LabGinJean-JCP-24}. In brief, the total effective nuclear potential $\vtot(\brnucltot)$ of Eqs.~\eqref{eq:Vtot} and \eqref{eq:MystFunc} is obtained from successive QM and MM functional optimizations of both the classical and quantum densities. Each of the QM and MM calculations are coupled to one-another through an effective potential (either $v_{\text{ext}}^{\text{QM}}$ or $v_{\text{ext}}^{\text{MM}}$ of Eqs.~\eqref{eq:mf_qmmm_2_qm} and \eqref{eq:mf_qmmm_2_mm}), and the cycle is stopped when some energetic convergence criteria (detailed below) is reached. Then, once the optimal QM and MM densities are obtained, the nuclear gradient is computed to optimize the nuclear geometry of the solute. The solvent contribution to the nuclear gradient (see Sec.~\ref{subsubsec:comp_grad}) is then added to the usual electronic gradient contribution, forming the total nuclear gradient which is then used to optimize the geometry. The latter is summarized in Fig.~\ref{fig:fancy_maxime_figure}.

In more details, the QM calculations are performed using the version 2.6.2 of the PySCF software\cite{sun-pyscf-JCP-2020}, employing the B3LYP functional \cite{Beck-JCP-93,LeeYanPar-PRB-88} and the 6-31++G** basis set \cite{HehDitPop-JCP-72, HaraPopl-TCA-73,FraPieHehBinGorDefPop-JCP-82,ClaChaSpiSch-JCC-83,SpiClaSchHeh-JCC-87} to approximate the electronic Levy-Lieb functional. This choice of exchange/correlation functional and basis set is similar to that of Huynh et al. \cite{HuyAnsCavStaHam-JACS-16}, such that we can directly compare with the results reported in that previous work. Nevertheless, the QM calculations are influenced by the classical solvent through an additional external electrostatic potential $v_{\text{ext}}^{\text{QM}}(\bri{})$ coming from the classical charge density (see Eq.~\eqref{eq:mf_qmmm_2_qm}). The latter is obtained by a set of point-charges lying at the nodes of a 48$^3$ a.u$^3$ cubic box, with 100 grid nodes in each direction. Once the ground state electronic density is obtained, the corresponding electrostatic potential $v_{\text{ext}}^{\text{MM}}(\bri{})$ (see Eq.~\eqref{eq:mf_qmmm_2_mm}) generated by the solute is computed on the same cubic grid. This potential is then used as an external potential in the subsequent MDFT calculation.

The solvent density is optimized and computed using our in-house Fortran-based MDFT program. The spatial grid used for the solvent density matches that of the electrostatic potential, and 196 discrete orientations are used to sample the three Euler angles $(\theta, \phi, \psi)$. Water is modeled using the non-polarizable SPC/E model  \cite{BerGrigStr-JCP-87}, which then defines the charge density $\gamma(\bm{r},\bm{\Omega})$ used in Eq.~\eqref{charge_dens_solv}. The $c^{(2)}$ direct correlation function, required to compute the second order term of the excess functional of  Eq.~\eqref{eq:fex=fhnc+fb}, was kindly provided by Luc Belloni. It was computed for a homogeneous water density of $n_0=0.0332891$\AA$^{-3}$ using a combination of Monte Carlo simulation  and integral equation theory \cite{Belloni-JCP-17}. 

The bridge functional was neglected, \textit{i.e.} truncating the excess functional  at second order, corresponding to the so-called HNC excess functional \cite{DinLevBorBel-JCP-17}. 

To model dispersion-repulsion interactions and prevent unphysical overlaps between the classical and quantum densities—which could lead to numerical instabilities— LJ sites were added to the solute nuclei following  Eq.\eqref{eq:nucl_mm_4}. The LJ parameters are extracted from the second-generation Generalized Amber Force Field (GAFF2) \cite{HeManYanLeeTaiWan-JCP-20}. All selected LJ parameters are provided in SI. 

While the gradients of  $w^\text{MM}_\text{C}$ and $\cpf{\nsmall}$  are directly computed in Pyscf, the gradient of the LJ potential $ w^\text{MM}_\text{LJ}$ is computed in the MDFT program and then added to Pyscf gradient, before performing  geometry optimization (see Sec.~\ref{subsubsec:comp_grad} for a detailed derivation). A new geometry is proposed using the geomTRIC solver \cite{WangChen-JCP-16} of PySCF and a new QM/MDFT cycle is performed using this updated structure. This process is iterated until the solute geometry reaches convergence. 
The full QM/MM calculation is considered complete when the convergence criterion for the solute free energy is satisfied. The threshold used here is  $10^{-4}$  Hartree. 

Then, the  thermodynamic corrections, including zero-point energy contributions, thermochemical corrections at T~=~298.15~K and P~=~1~atm are computed and added to the solvation free energy and the solute energy yielding the total free energy of the system, which eventually yields to $\Omega$ of Eq.~\eqref{eq:fint_final}. When the full QM/MM calculation is converged, the QM energy, the QM/MM coupling and the classical energy are added to form $\vtot(\brnucltot)$ of Eq.~\eqref{eq:Vtot}. \\

For the reference quinone, the eDFT/PCM calculations took roughly 5 min while the full eDFT/MDFT calculation, with geometry optimization took approximately 5 hours on the same computer. The numerical cost of our eDFT/MDFT procedure might appear prohibitive but one should keep in mind that it is ”naively ” implemented and not optimized at this stage. Moreover, MDFT provides a full molecular description of the solvent structure which is not accessible with continuum methods. In fact, the description of the solvent is theoretically equivalent to simulation based methods such as Molecular Dynamics, which are even more computationally demanding. 

\subsection{Calculations of the 2e$^-$/2H$^+$ Reduction Potentials \label{sec:RedOxPot}}

We now turn to the computation of the redox properties of a series of quinone derivatives. More specificaly, we focus on two electrons reductions of benzoquinone derivatives (QR) into their hydroquinone (H$_2$QR) form:
\begin{equation}
     \text{QR}+ 2\text{e}^- + 2\text{H}^+ \longrightarrow \text{H}_2\text{QR}.
     \label{eq:Q_red}
\end{equation}

It is not possible to compute directly the reaction free energy, or equivalently the redox potential, associated with a redox half reaction such as Eq.~\eqref{eq:Q_red} because it involves the chemical potentials of both the electron and proton, which are ill-defined in solution. Instead, we introduce a reference redox couple—the unfunctionalized benzoquinone/hydroquinone pair Q/H$_2$Q and define the redox process through the following isodesmic reaction:
\begin{equation}
    \text{QR}  +  \text{H}_2\text{Q}        \longrightarrow \text{H}_2\text{QR}        +  \text{Q}
    \label{eq:ref_eq_sys}
    \end{equation}
This reaction avoids the explicit appearance of electrons and protons, and only involves neutral species, allowing us to compute the reaction free energy based solely on the free energies of the quinones involved. 

It is worth noticing that this approach assumes that the two sub-reactions (QR/H$_2$QR and Q/H$_2$Q) are chemically equivalent in terms of the redox process. In particular, it relies on the assumption that the electrons and protons involved in Eq.~\eqref{eq:Q_red} are effectively identical and can be canceled out between the two half-reactions. This assumption is reasonable when comparing structurally similar molecules, such as benzoquinone derivatives, where redox mechanisms are expected to be analogous. However, if the electronic or protonation environments differ significantly between the systems, this simplification can introduce non-negligible errors in the computed redox properties.

The  redox potential for the half-cell reaction appearing in Eq.~\eqref{eq:Q_red}, $\text{E}^\circ_{\text{QR}/\text{H}_2\text{QR}}$, is related to the reaction free energy, $\Delta_\text{r} \text{G}_{\text{QR}/\text{H}_2\text{QR}}$, through 
\begin{equation}
\Delta_\text{r} \text{G}_{\text{QR}/\text{H}_2\text{QR}}=-2 \text{F} \text{E}^\circ_{\text{QR}/\text{H}_2\text{QR}} \label{eq:DeltaG=-nFE}
\end{equation}
where F~=~96,485.3321~C.mol$^{-1}$ is the Faraday constant. 
The overall reaction free energy of Eq.~\eqref{eq:ref_eq_sys} is the difference of the half-cell reaction free energies of the QR/H$_2$QR and of   the reference couple, Q/H$_2$Q. By using Eq.~\eqref{eq:DeltaG=-nFE}, this allows for expressing the redox potential of the considered couple as
\begin{equation}
    \text{E}^\circ_{\text{QR}/\text{H}_2\text{QR}} = - \dfrac{\Delta_\text{r} \text{G}^\circ_{}}{2\text{F}}+ \text{E}^\circ_{\text{Q}/\text{H}_2\text{Q}},
     \label{eq:quinone_red_pot}
\end{equation}
The reduction potential of Q/H$_2$Q   is set to 0~V, and the redox potentials of functionalized quinones can then be computed from the free energy difference between the reduced, $\text{H}_2\text{QR}$, and oxydized form, Q.

\section{Applications \label{sec:Applications}}    

\subsection{Redox Potential calculations\label{subsec:RedPot}}

\begin{figure}[t]
	\centering
	\includegraphics[width=\columnwidth]{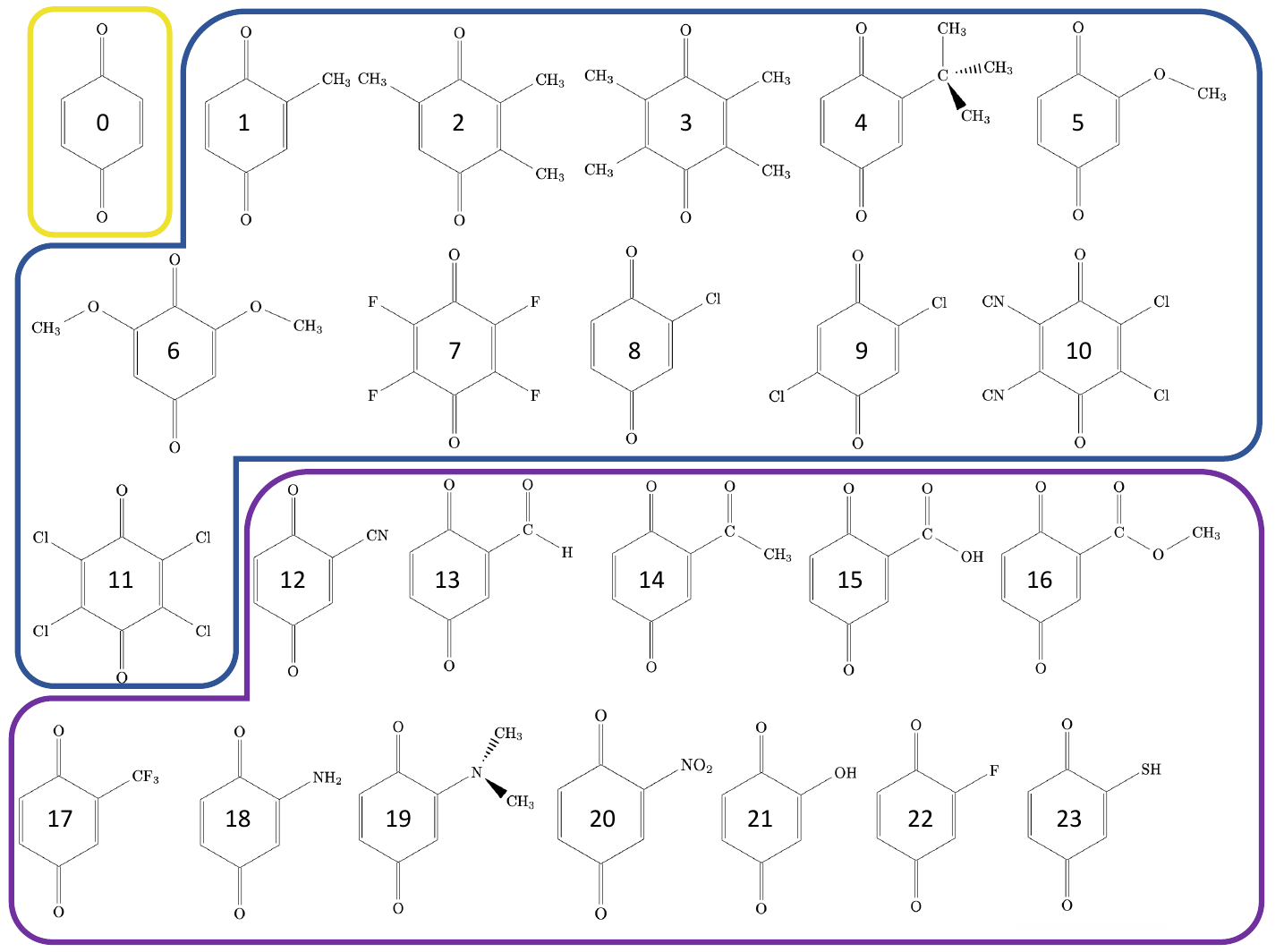}
	\caption{Quinones studied in this paper. The reference Benzoquinone, labeled 0 is circled in yellow. Quinones from 1 to 11, circled in blue, are compared to experimental data and to DFT/C-PCM calculations in figure \ref{fig:red_pot_exp} while quinones from 12 to 23, circled in purple are only compared to DFT/C-PCM predictions in figure \ref{fig:red_pot_comp}, due to the lack of experimental data.}
	\label{fig:quinones}
\end{figure}

\begin{figure}[t]
    \centering
	\includegraphics[width=\columnwidth]{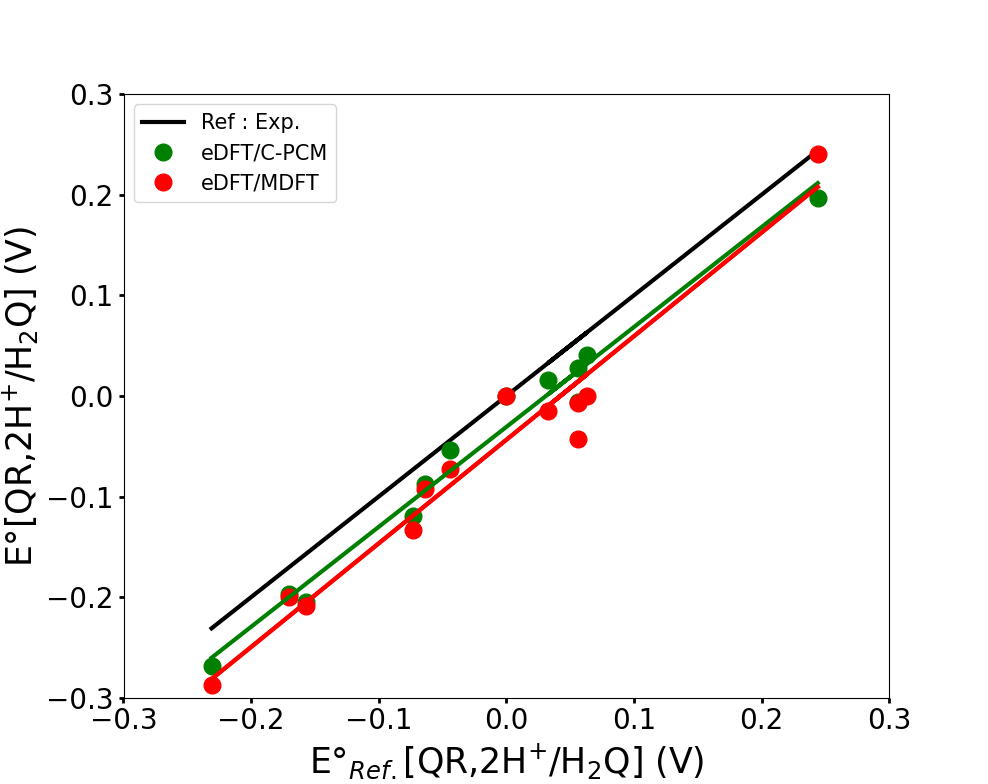}
    \caption{Comparison of 2e$^-$/2H$^+$ reduction potentials (vs.Q,2H$^+$/H$_2$Q) for various quinone couples in water. 
    The results obtained with eDFT/MDFT simulations are reported in red, while the results obtained with eDFT/C-PCM are reported in green and experimental data in black \cite{HuyAnsCavStaHam-JACS-16}. }
    \label{fig:red_pot_exp}
\end{figure}

\begin{figure}[t]
    \centering
    \includegraphics[width=\columnwidth]{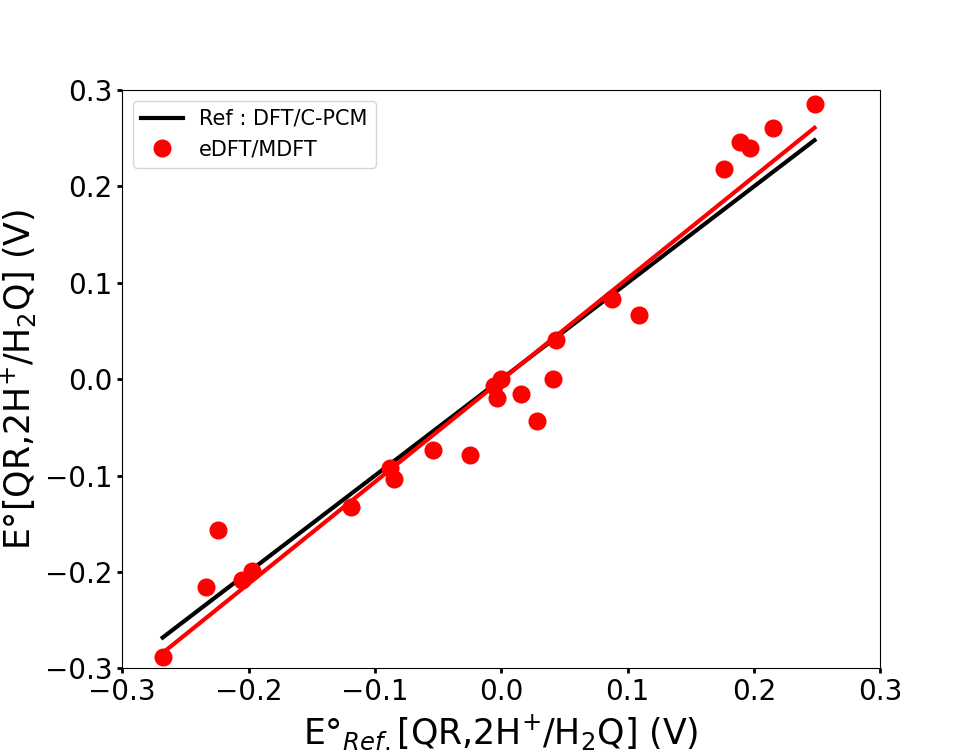}
    \caption{Comparison of 2e$^-$/2H$^+$ reduction potentials (vs.Q,2H$^+$/H$_2$Q) for various quinone couples in water. Our results obtained with eDFT/MDFT simulations in red and the references obtained with eDFT/C-PCM in black\cite{HuyAnsCavStaHam-JACS-16}.} 
    \label{fig:red_pot_comp}
\end{figure}

We computed the redox potential for a set of 23 quinone/hydroquinone couples displayed in Fig.~\ref{fig:quinones} using Eq.~\eqref{eq:quinone_red_pot}. The required reactions free energies were computed using the geometries optimized with the QM/MDFT scheme, including  ZPE and thermochemical corrections. The predicted redox potentials are reported in {Fig.~\ref{fig:red_pot_exp}}  and compared to the  experimental and  theoretical results of Huynh~\textit{et al.} who employed electronic DFT (eDFT) combined with the conductor-like polarizable continuum model (C-PCM)\cite{HuyAnsCavStaHam-JACS-16}.  
As can be seen from Fig.~\ref{fig:red_pot_exp}, the redox potential predicted using the QM/MDFT are consistent with the eDFT/C-PCM predictions. 
With respect to experimental results, both methods underestimate the redox potential by a few tenth of a volt on average. 
To further test our approach,  we extended the study to 12 additional quinone derivatives for which no experimental redox potentials were reported in Huynh’s work.  The redox potentials for these 12 new compounds, together with the 11 original ones, are presented in {Fig.~\ref{fig:red_pot_comp}} where they are compared to Huynh's predictions using eDFT/C-PCM. Again, the agreement is satisfactory, with an average deviation of only 0.007 V between our results and theirs.

Overall, the prediction of the redox potential with the QM/MDFT method  agrees quite well both with the experimental values and the eDFT/C-PCM method. This is rather satisfactory since we employed the simplest mean-field coupling between eDFT and MDFT. Moreover, the dispersion-repulsion interaction has been modeled with a simple Lennard-Jones interaction whose parameters were not optimized. 
Several potential improvements, such as including explicit polarization effects or refining the QM/MM coupling, remain to be explored to further enhance the accuracy.

\begin{figure}[t]
	\centering
	\includegraphics[width=\columnwidth]{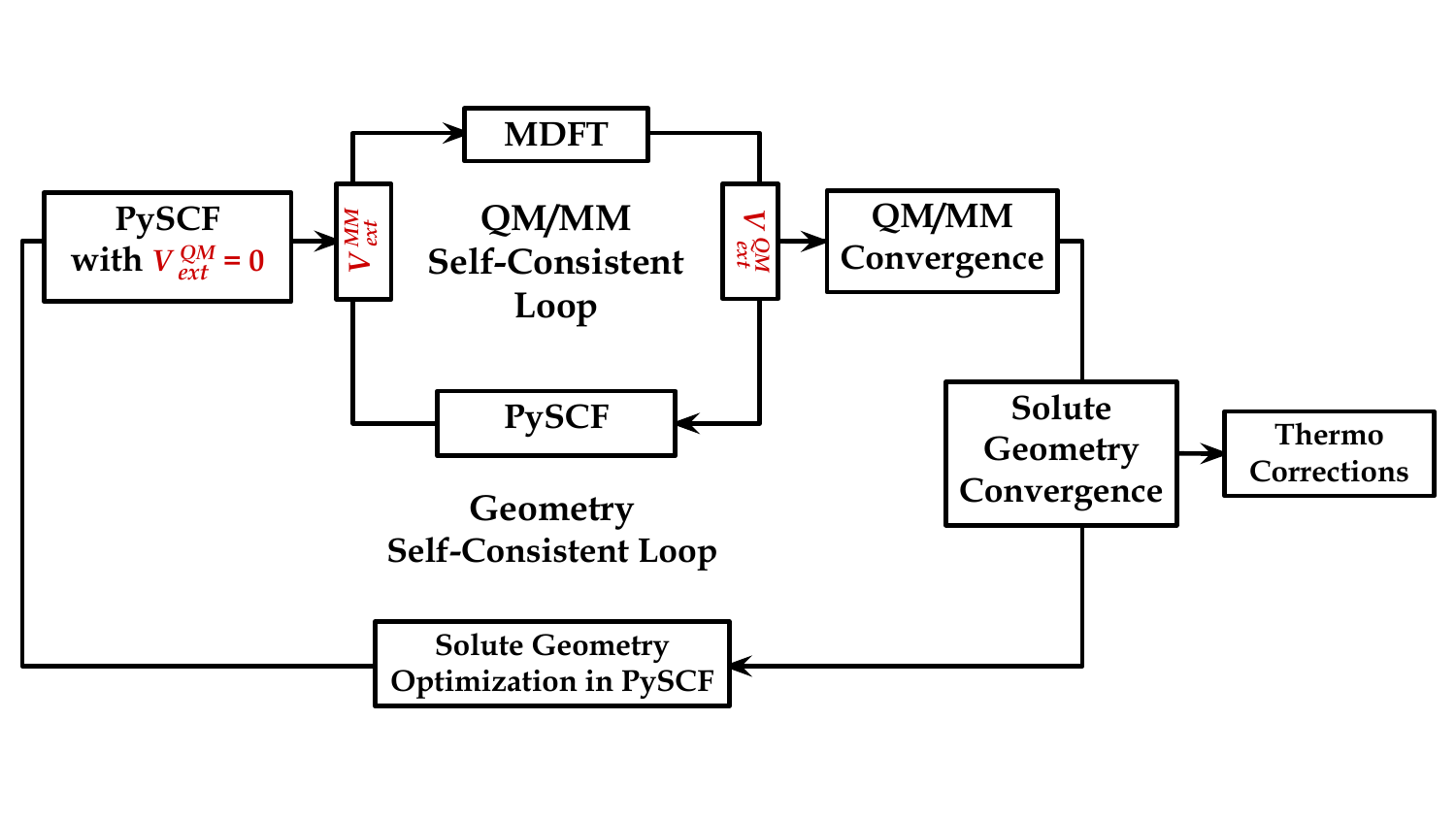}
	\caption{Schematic representation of the overall QM/MM scheme. The cycle start by a eDFT calculation in vacuum, which  produces the electrostatic potential, $v_{\text{ext}}^{\text{MM}}$, of Eq.~\eqref{eq:mf_qmmm_2_mm}. This potential is read by the MDFT program to optimize the MM density. Then, the electrostatic potential generated by the MM region,  $v_{\text{ext}}^{\text{QM}}$, of Eq.~\eqref{eq:mf_qmmm_2_qm} is read by the QM software, and the cycle is iterated until convergence of the QM/MM energy. 
     The nuclear gradient is then computed and a geometry optimization step is performed. Once the optimal nuclear geometry is obtained, the thermodynamical corrections are computed, to obtain the final estimate of the grand potential of the system.}
	\label{fig:fancy_maxime_figure}
\end{figure}


\subsection{Electronic and Solvent Charge Densities \label{subsec:Densities}}

One clear advantage of the QM/MDFT approach over implicit solvation models is its ability to capture the detailed solvent structure around the solute. Specifically, the spatially and orientationally dependent solvent density $n(\bm{r},\bm{\Omega})$ is obtained directly from the self-consistent functional minimization. A more intuitive and visually accessible quantity is the inhomogeneous solvent charge density $n_c(\bm{r})$ (see Eq.~\eqref{charge_dens_solv}), which is derived from the molecular solvent distribution and allows further analysis of the solvation environment.
The charge density directly impacts the electronic structure calculation through the electrostatic external potential $v_{\text{ext}}^{\text{QM}}$ it creates (see Eq.~\eqref{eq:mf_qmmm_2_qm}).

\begin{figure}[!htbp]
    \centering
    \includegraphics[width=\columnwidth]{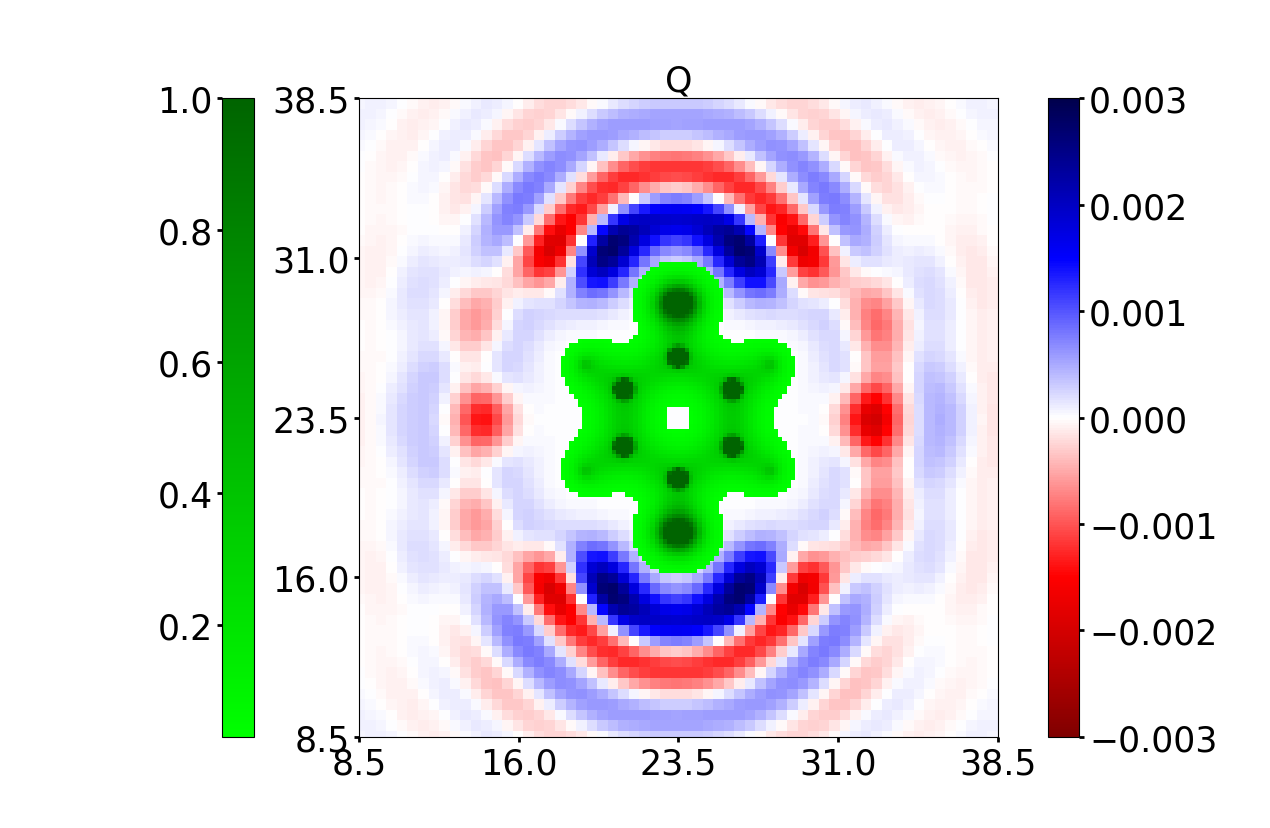}
    \includegraphics[width=\columnwidth]{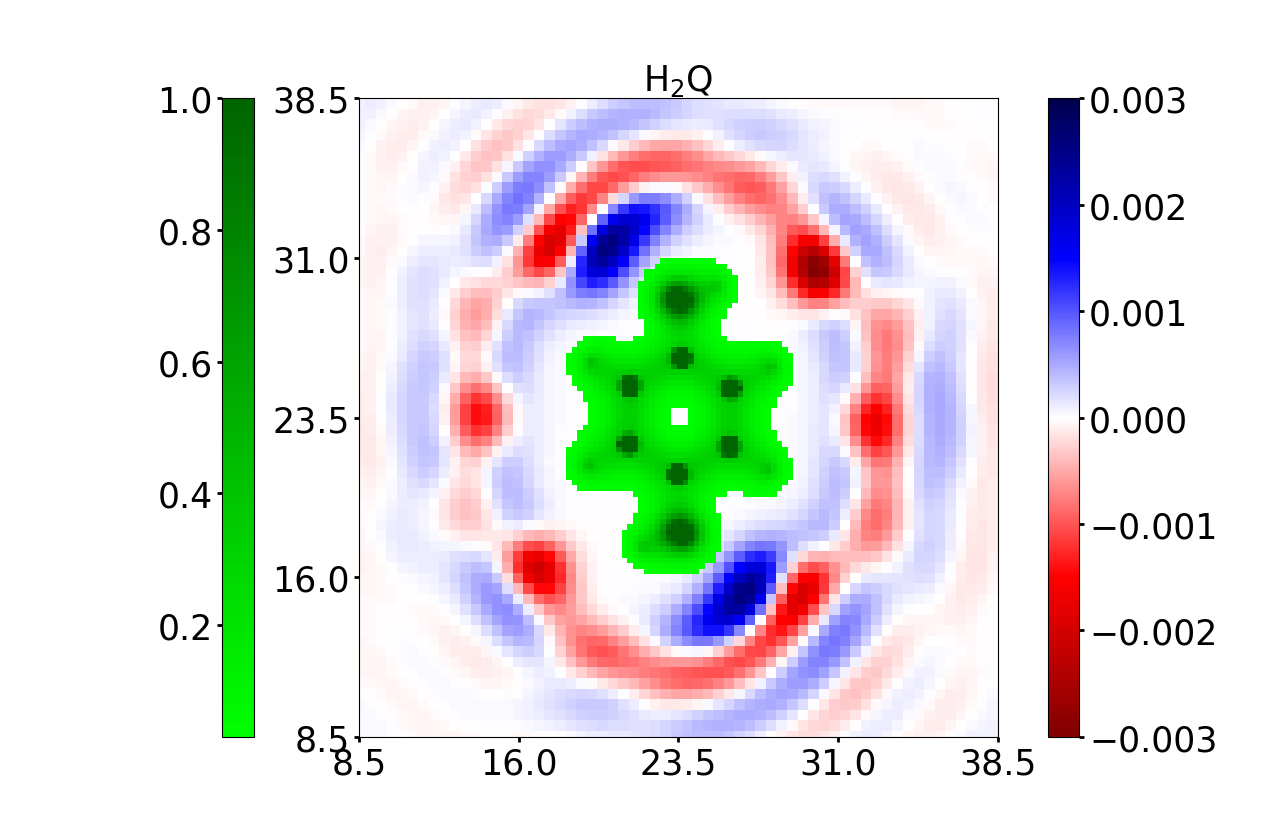}
    \caption{Electronic densities (a.u.){$^{-3}$} computed with eDFT/MDFT  around Q (top) and H$_2$Q (bottom) are depicted in green. The associated solvent charge density (atomic unit) are displayed in blue and red for the positive and negative regions respectively.}
    \label{fig:solvcharge_ref}
\end{figure}

The solvent charges and electronic densities in the plane of the aromatic ring of the Q/H$_2$Q couple are reported in Fig.~\ref{fig:solvcharge_ref} for the optimized solute geometries. Starting with the benzoquinone, the simplest molecule of our benchmark, the  electronic density exhibits the expected pattern: it is largely concentrated around the atomic nuclei.

In contrast, the solvent charge density shows a more complex structure. Although not clearly visible in {Fig.~\ref{fig:solvcharge_ref}}, there is a vanishing region in the vicinity of the solute, due to the exclusion of  solvent molecules. There are two banana shape lobes of positive charge close to the oxygen atoms, with two marked maxima. These features reflect the formation of hydrogen bonds between the hydrogen atoms of nearby water molecules and the lone pairs of the carbonyl oxygens. Each of these positively charged lobes is followed by a negatively charged region, indicating that the water molecules are oriented such that the hydrogen atoms point toward the carbonyl groups while the oxygen atoms pointing outwards.

In the vicinity of the benzoquinone's hydrogen atoms, three negatively charged regions are observed: two aligned with the C–H bonds and a third, more intense one located on the perpendicular bisector of the H–H segment. This pattern indicates preferential orientation of water molecules such that their oxygen atoms face the hydrogen atoms of Q.

A diffuse region of low positive charge is also visible closer to the solute. This  can be an artefact due to  the usage of the SPC/E model, in which only the oxygen atom carries a Lennard-Jones  site.  The excluded volume, \textit{i.e.} the region where the water density vanishes, is essentially governed by the Lennard-Jones interaction between the oxygen of the solvent water molecules and the solute. However, due to molecular rotations, hydrogen atoms of water molecules may enter this excluded zone.  This is confirmed by examining the map of the oxygen density, where the frontier between the first solvation shell and the excluded volume essentially correspond to the beginning of negative charge surfaces. 

\begin{figure}[!htbp]
    \centering
    \includegraphics[width=\columnwidth]{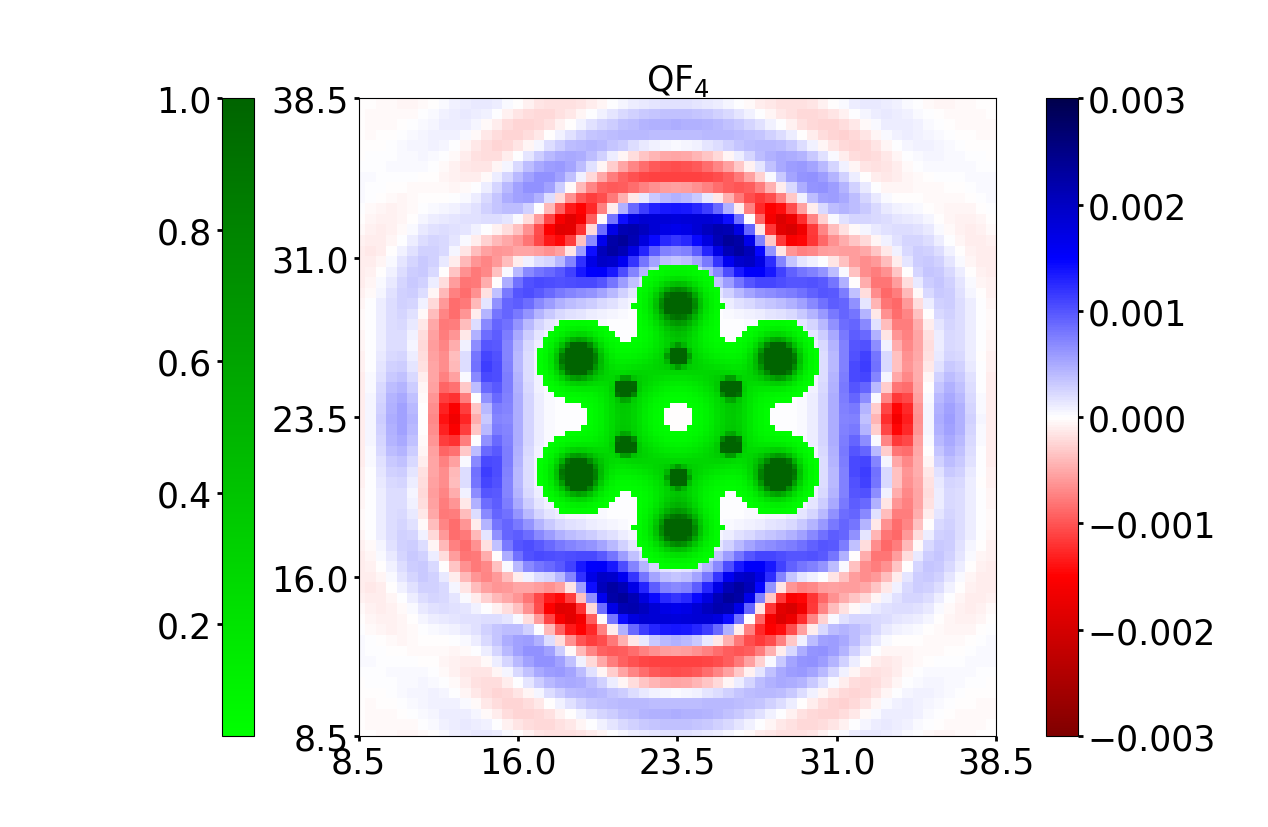}
    \includegraphics[width=\columnwidth]{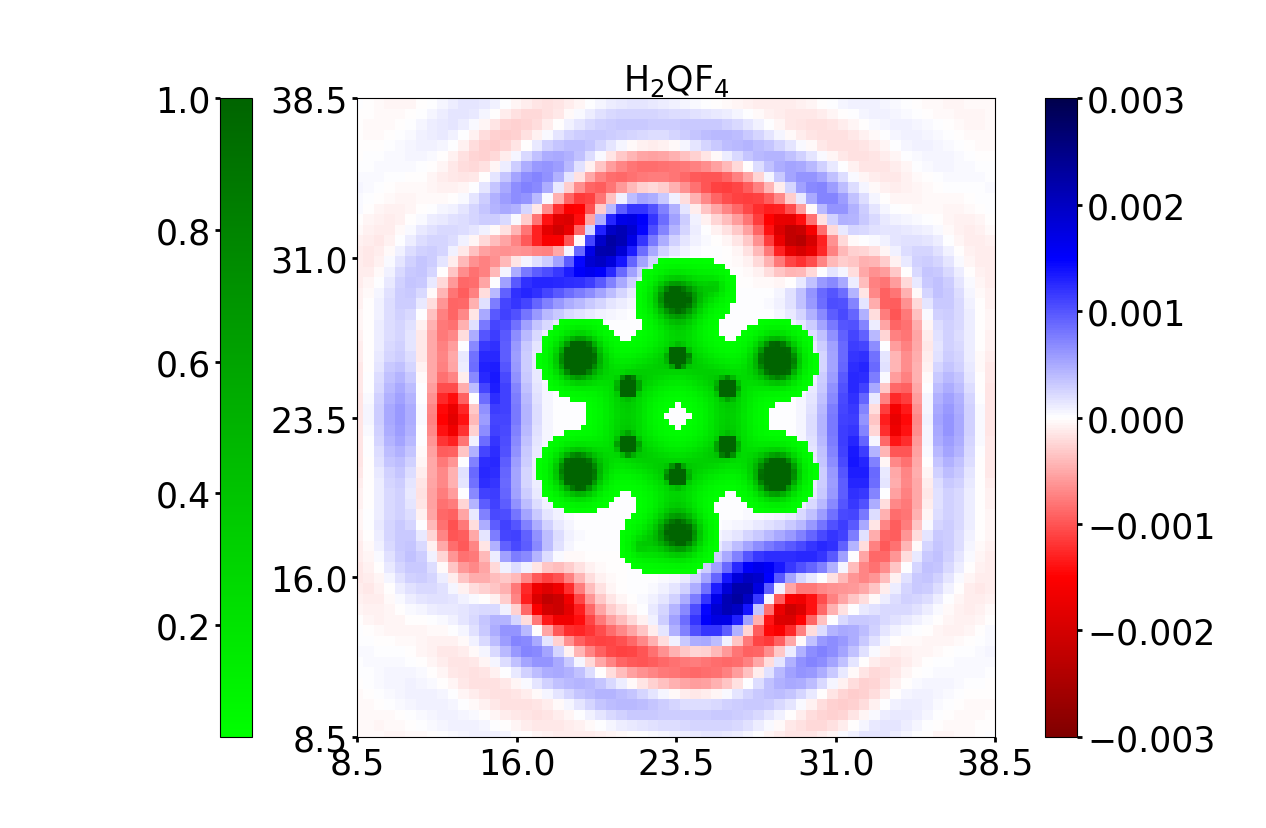}
    \caption{Same as figure \ref{fig:solvcharge_ref} for the QF$_4$/H$_2$QF$_4$.}
    \label{fig:solvcharge_4f}
\end{figure}
 
The second solvation shell is also well captured, as evidenced by the oscillatory pattern of the charge distribution. Finally, as moving away from the solute, the charge density tends toward
zero indicating that all orientations of the solvent become equivalent.

The solvation structure around H$_2$Q differs notably from that around Q as evidenced by the solvent charge density displayed in the bottom of Fig. {\ref{fig:solvcharge_ref}}. The most notable change is the diminution of the positively charged region near the oxygen atom, due to the presence of OH bond. It also modifies the negatively charged region, with the apparition of a marked maximum aligned with the OH bond, and of two lateral "bumps". The second solvation shell is also slightly modified.

The functionalization of  quinone molecules naturally impacts their solvation structure. As an illustration, the equilibrium solvent charge density and electronic ground-state density around the tetrafluorinated  quinone and hydroquinone are shown in {Fig.~\ref{fig:solvcharge_4f}}. As expected, the electronic density exhibits an increase of the density in the region where  fluorine atoms replaced the hydrogen atoms. Regarding the solvent charge, the most notable change is that the QF$_4$ molecule is now surrounded by a marked  region of positive charge, followed by a region of negative charge. In particular, there is a change of sign in the region in the vicinity of fluorine atoms compared to the benzoquinone, indicating a flip in the orientation of water molecules in these regions. It is noteworthy that the solvent charge is more polarized around the O atoms than around the F atoms. 

\begin{figure}[!htbp]
    \centering
    \includegraphics[width=0.8\columnwidth]{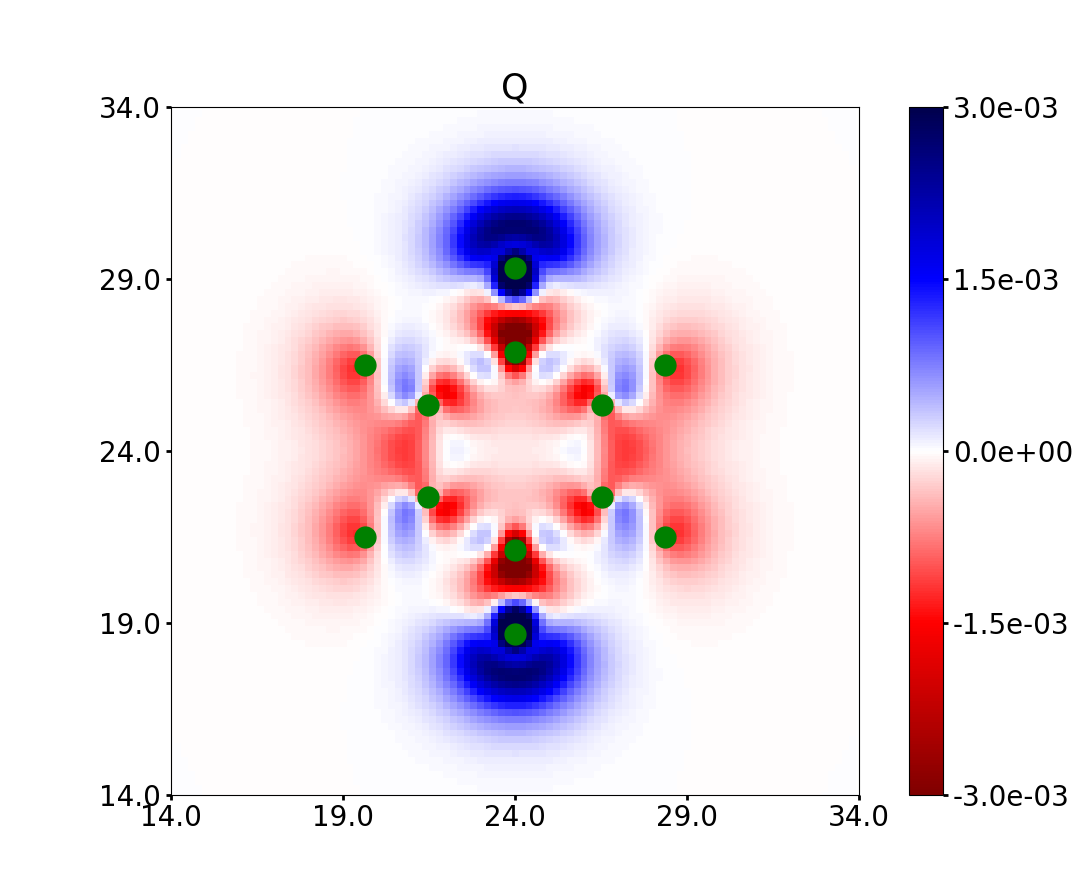}
    \includegraphics[width=0.8\columnwidth]{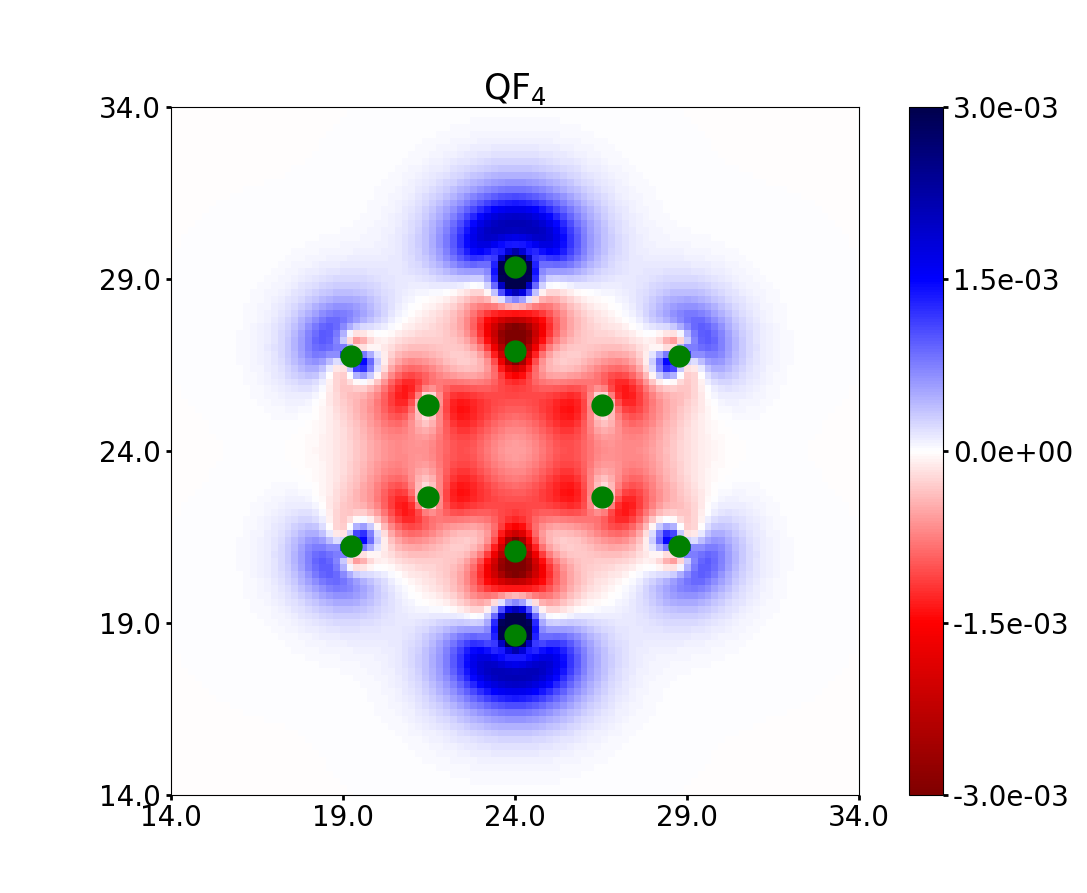}
    \caption{Difference of solute electronic density (a.u.) in blue and red between eDFT/MDFT simulation and in vacuum. The solutes, represented in green, are Q (top) and Q4F (bottom)}
    \label{fig:elecdensdiff_mdft}
\end{figure}

The modification of the charge density in the reduced form are similar to what is observed for the unsubstituted hydroquinone \textit{i.e.} the positive charge is reduced close to the O-H bond while negative charge increases. 

Overall, MDFT effectively captures the non-trivial three-dimensional solvent structure around the quantum solute, as evidenced in 
{Fig.~\ref{fig:solvcharge_ref} and Fig.~\ref{fig:solvcharge_4f}}. In particular, the tetrahedral shape of the charge distribution, resulting from hydrogen bonding, is well rendered. This is a clear advantage of MDFT compared to simpler continuum solvation models, which ignore the molecular nature of the solvent.
\begin{figure}[!htbp]
    \centering
    \includegraphics[width=0.8\columnwidth]{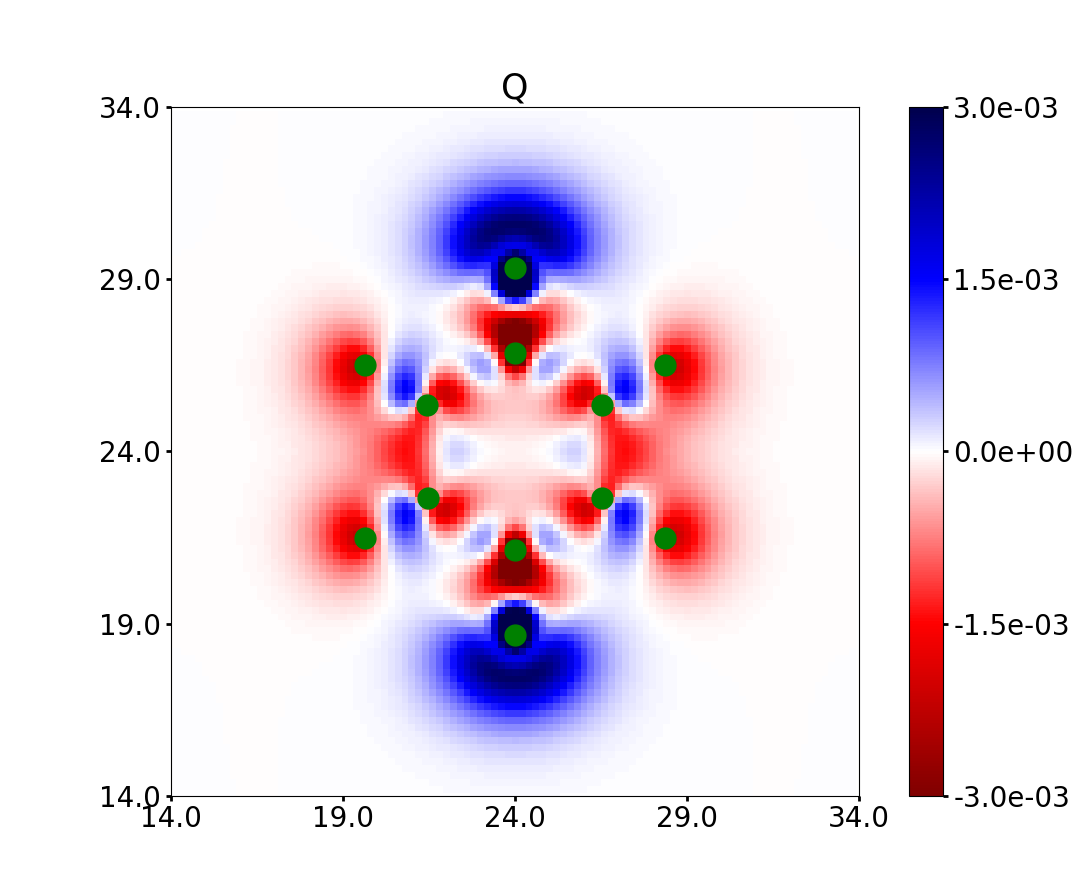}
    \includegraphics[width=0.8\columnwidth]{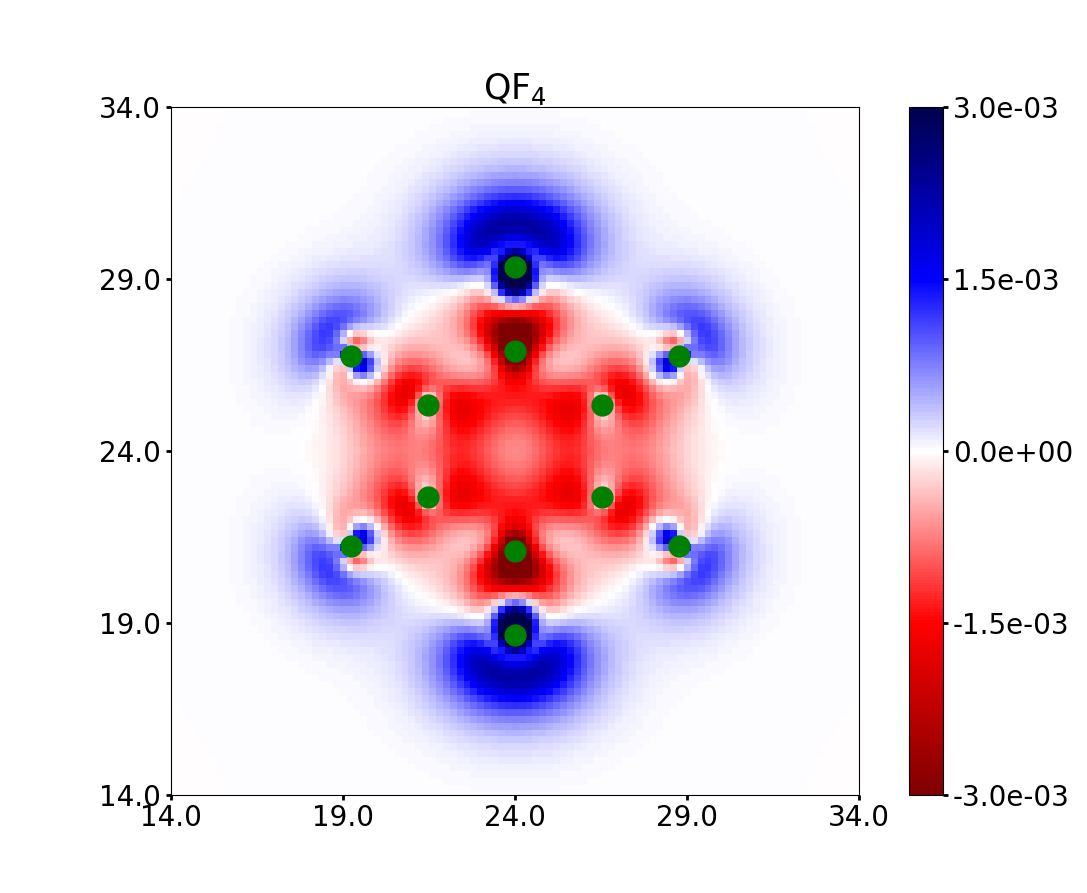}
    \caption{Difference of solute electronic density (a.u.) in blue and red between eDFT/C-PCM simulation and in vacuum. The color code is the same than in Fig.~\ref{fig:elecdensdiff_mdft}}
    \label{fig:elecdensdiff_cpcm}
\end{figure}
\begin{figure}[!htbp]
    \centering
    \includegraphics[width=0.8\columnwidth]{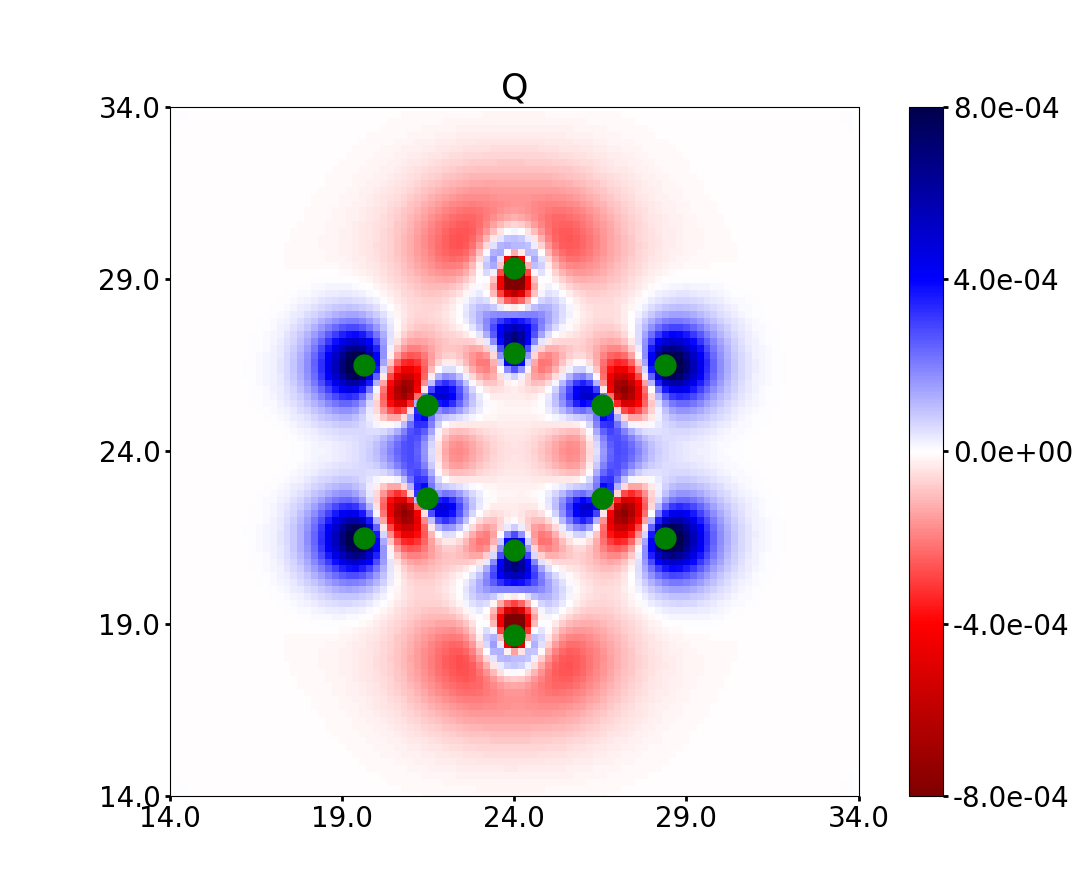}
    \includegraphics[width=0.8\columnwidth]{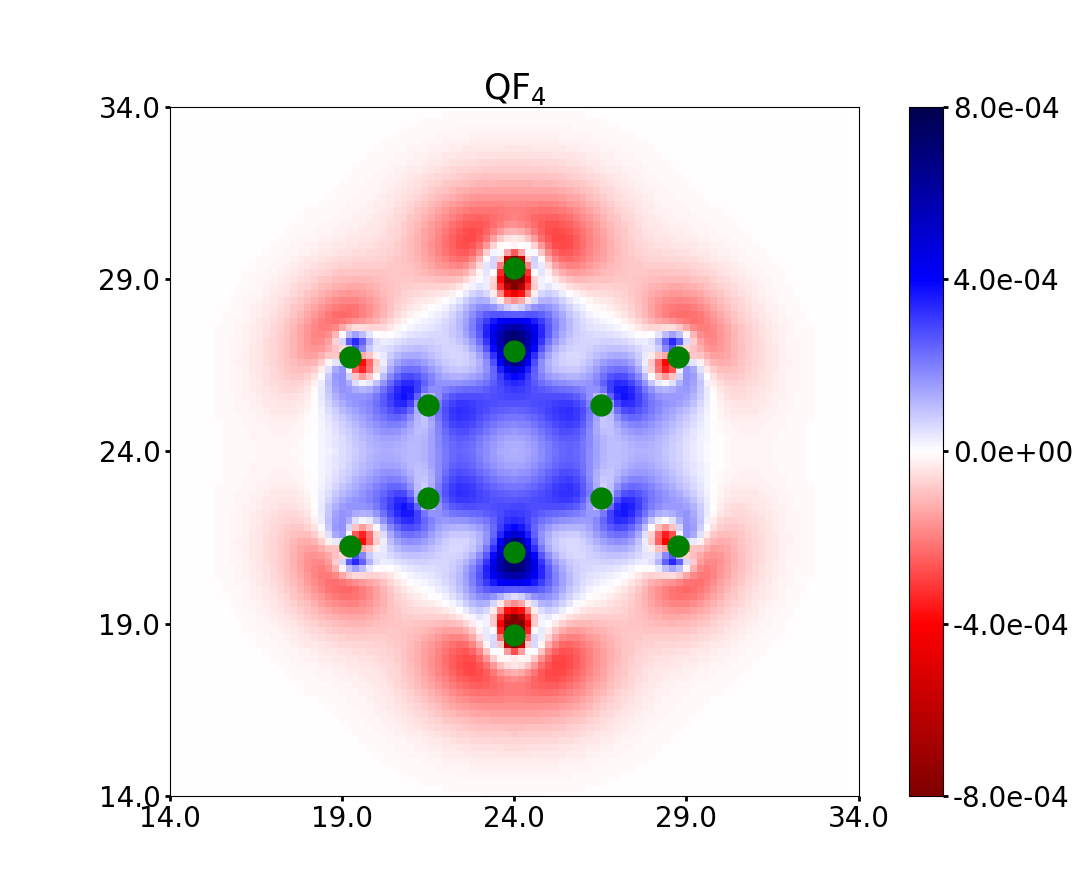}
    \caption{Difference of solute electronic density (a.u.) between eDFT/MDFT and eDFT/C-PCM simulation. The color code is the same than in Fig.~\ref{fig:elecdensdiff_mdft}}
    \label{fig:elecdensdiff_solv}
\end{figure}

To assess the impact of the solvent on the electronic structure of the solute, we also computed the difference between the electronic densities obtained in the presence of solvent and in vacuum. These differences are displayed in {Fig.~\ref{fig:elecdensdiff_mdft}}  for the Q and QF$_4$ couple. Blue regions correspond to an increase in electronic density compared to vacuum, while red regions indicate a depletion.
 Overall, there is a clear connection between the maps of solvent charge density displayed in {Fig.~\ref{fig:solvcharge_ref} and Fig.~\ref{fig:solvcharge_4f}} and the maps of electronic density difference.

In both Q and QF$_4$, electronic density increases primarily around the oxygen atoms, indicating strong polarization effects induced by the positively charged solvent environment in those regions. This coincides with a decrease in electronic density  around the aromatic ring, particularly on the carbon atoms bonded to oxygen.

 However, while the entire cycle undergoes a decrease in electronic density in QF$_4$, the picture is more subtle in Q where some density depletions are observed along the C-H bonds and along the 4 C-C bonds connected to the carbonyl. 
 In Q, the H atoms are surrounded by  negative solvent charges which tend to repel the electrons towards the center of the molecule while the F atoms in  QF$_4$ are surrounded by a positive solvent charge density which tends to increase the polarization of the C-F bound with respect to vacuum. 
The stronger solvent-induced polarization of the oxygen atoms relative to the fluorine atoms in QF$_4$ is consistent with the known differences in their electric dipole polarizabilities: the polarizability of oxygen is approximately 30-45\% higher than that of fluorine \cite{TeachPack-ADNDT-71,AndersSadlej-PRA-92,SchwerNagle-MolPhys-19}.

In {Fig.~\ref{fig:elecdensdiff_cpcm}}, we display the difference between the electronic densities computed using a C-PCM model and  in vacuum for the Q/HQ couple. The maps are almost identical to the one displayed in {Fig.~\ref{fig:elecdensdiff_mdft}}, indicating that modeling the solvent with MDFT yields an electronic density comparable to that obtained with using a C-PCM model. A closer look reveals that the charges are slightly more intense when the C-PCM model is used. 

To further examinate this behaviour we report in Fig.~\ref{fig:elecdensdiff_solv} the  difference between the electronic densities obtained with MDFT and C-PCM. Here, the blue regions indicate an excess of electronic density predicted by the eDFT/MDFT calculation with respect to  eDFT/C-PCM one.  For both molecules, the electronic density predicted  around the O atoms is greater with the C-PCM model than with eDFT/MDFT. In benzoquinone, the electronic density surrounding the hydrogen and carbon atoms is lower with C-PCM than with MDFT.

 In the case of QF$_4$, electrons are more localized around O and F atoms when the C-PCM is used than when MDFT is. The trend is opposite around the aromatic ring, where MDFT predicts an excess of electrons with respect to C-PCM.
The prediction of the electronic structure using both  solvent models are in good agreement overall. 

In summary, modeling the solvent using  MDFT calculations clearly impacts  the electronic density of the solute, which differs markedly with respect to in-vacuum predictions. These variations are well rationalized by examining the solvent charge computed by MDFT. The comparison with eDFT/C-PCM shows an overall good agreement even if some differences have been highlighted.

Importantly, the QM/MDFT calculations rely on only one set of \textit{ad hoc} parameters—those from the Lennard-Jones force field used to model dispersion-repulsion interactions between solute and solvent. These parameters were taken directly from the generic GAFF2 force field and were not fine-tuned. The fact that MDFT-based predictions closely match those of the well-established and heavily parametrized C-PCM method is therefore quite encouraging.

In addition, MDFT allows for the prediction of the detailed  structure of the solvent around the solute which is not possible with C-PCM but requires the use of more expansive techniques, such as Molecular Dynamics.

\section{Conclusions \label{sec:Conclusions}}

In this paper, we propose a variational principle of the grand potential of a QM/MM system as a functional of the one body electronic and classical densities. This framework is constructed through a series of well-defined approximations, eventually leading to a practical mean-field computational scheme. We demonstrate the applicability of this approach by computing the redox properties of a series of quinones.


This work begins by describing in Sec.~\ref{subsec:ExGSEDMat} the generic QM/MM system treated here, 
which consists of a QM solute made of electrons and nuclei, both interacting with classical rigid solvent molecules, whose number is allowed to vary. This QM/MM system is therefore described in a semi-grand canonical statistical ensemble. We introduce the corresponding equilibrium density matrix, partition function and grand potential which are the three key objects characterizing the system. Under the Born-Oppenheimer and zero electronic entropy approximations, we show in Sec.~\ref{sec:bo_zee} that the equilibrium density matrix can be  written as the product of a nuclear and an electronic density matrices. Nevertheless, these QM density matrices are parametrized by the classical solvent coordinates, thereby introducing  QM/MM correlation terms.

Based on this findings, we propose in Sec.~\ref{subsec:VarForm} a variational formulation of the grand potential in terms of both QM density matrices and classical probability distributions. This expression is then simplified using a mean-field approximation which neglects QM/MM correlation effects, allowing us to write an upper bound to the exact grand potential.

By properly integrating over quantum electrons and classical degrees of freedom, 
we reformulate in Sec.~\ref{subsec:NuclEffectPot} the mean-field grand potential only in terms of nuclear density matrices. This naturally defines an effective potential experienced by the nuclei due to the presence of the electrons and classical molecules. This effective potential is then identified as a special case of QM/MM object established by some of us in a previous work \cite{JeanLabGiner-arXiv-24}. This allows us to reformulate the effective potential as a coupled minimization problem over one-body QM and MM densities. 

We then introduce the standard Harmonic approximation in Sec.~\ref{subsec:harmonic} to obtain a tractable expression of the grand potential. Eventually, computing the grand potential requires the knowledge of the equilibrium geometry of the nuclei  and of  the Hessian matrix of the total effective potential experienced by the nuclei at this geometry. Details about the computation of nuclear gradients and Hessian matrix are presented in the Appendix \ref{app:grad_hessmat}.

 Therefore, accessing the grand potential  reduces to a sequence of density optimizations. First, the solvent density and the solute electronic density are optimized for a fixed solute nuclear geometry. The self-consistent QM/MM procedure allowing the relaxation of the solvent density around the solute was presented in a previous work\cite{LabGinJean-JCP-24}. Subsequently, the solute geometry is optimized for a fixed solvent configuration. This process is iterated until the overall convergence is reached. A schematic description of this QM/MM scheme is presented in Fig.~\ref{fig:fancy_maxime_figure}.
 
To validate this methodology, we applied the proposed QM/MM scheme to the calculation of the redox properties of a series of functionalized quinones in aqueous solution. A selection of 23 quinone couples (see Fig.~\ref{fig:quinones}) was studied and predictions were compared to experimental  data and to computations using electronic DFT in a C-PCM model extracted  from Huynh~\textit{et al.}\cite{HuyAnsCavStaHam-JACS-16}. The quantitative results are presented in Sec.~\ref{subsec:RedPot}.

 The redox potentials predicted using QM/MDFT are in good agreement with the eDFT/C-PCM predictions, but both methods underestimate the experimental values. It is worth emphasizing that the MDFT framework offers a more detailed representation of the solvent structure surrounding the solute than conventional PCM or other continuum solvation models. This level of detail allowed us to thoroughly discuss the interplay between the electronic density and the surrounding solvent, as presented in Sec.~\ref{subsec:Densities}, where we plotted the solvent charge density around the solute for two quinone couples. Overall, we found that MDFT accurately captures the complex, three-dimensional solvent structure around the quantum solute.

To further assess the interplay between the solvent environment and the electronic density, we computed the difference between the electronic densities computed in the presence of the solvent moldeled by MDFT and in vacuum. As expected,  the electronic density is impacted by the presence of a non trivial solvent charge density. These electronic density variation can be understood by simple electrostatic consideration. We also compared the electronic density variations from eDFT/C-PCM calculations with respect to the vacuum. 

Eventually, the proposed eDFT/MDFT scheme is able to obtain comparable results to the well-known C-PCM, whose parameters have been subtly optimized over the years \cite{BaroCossi-JPCA-98, CosRegQcaBar-JCC-03}, 
even if the only ad hoc parameters introduced in our model, namely the Lennard-Jones parameters, were not tuned at all. 
In addition,  MDFT provides a detailed picture of the solvent structure around the solute, a feature that is out of reach for continuum models.

Overall, these results are rather encouraging, particularly considering  there are still many routes to explore which might improve the proposed QM/MDFT scheme. Among those, the QM/MM coupling scheme can certainly be improved, using for instance some overlap models for the Pauli repulsion instead of LJ potentials. Another obvious limitation is the lack of polarizability of the classical solvent molecules in the current MDFT functional. We believe these are the two main sources of errors, which should be addressed in priority to improve the precision of the QM/MDFT approach.

\section*{Acknowledgements}
The authors want to thank Luc Belloni for providing the SPC/E direct correlation functions. M.L. acknowledges the Institute of Materials Science (iMAT) of the Alliance Sorbonne Université for financial support.This work was supported by
the French National Research Agency under the France
2030 program (Grant ANR-22-PEBA-0002).

\section*{Author Declarations}
\subsection*{Conflict of Interest}
The authors have no conflicts to disclose.

\section*{Author Contribution}
\textbf{Maxime Labat}: Conceptualization (equal); Data curation (equal); Investigation (equal); Methodology (equal); Project administration (equal); Software (equal); Supervision (equal); Validation (equal); Visualization (equal); Writing – original draft (equal); Writing – review \& editing (equal). \textbf{Emmanuel Giner}: Conceptualization
(equal); Investigation (equal); Methodology (equal); Project administration (equal); Software (equal); Supervision (equal); Validation (equal); Writing – original draft (equal); Writing – review \& editing (equal). \textbf{Guillaume Jeanmairet}: Conceptualization (equal); Data curation (equal); Investigation (equal); Methodology (equal); Project administration (equal); Software (equal); Supervision (equal); Validation (equal); Visualization (equal); Writing – original draft (equal); Writing – review \& editing (equal).

\section*{DATA AVAILABILITY}
The data that support the findings of this study are available
from the corresponding author upon reasonable request.
\appendix
\subsection{Appendix: Derivation of $\omegamf$ \label{app:omegamf}}
{
We aim at deriving the variational principle for the grand potential involving only the nuclear density matrix.  We start by inserting the mean-field approximation of the electron-nuclei-solvent density matrix of Eq.~\eqref{def_gen_prho_mf} in the grand potential functional for the
QM/MM system of Eq.~\eqref{eq:omega}
\begin{equation}  
\begin{split}
 \omegamf &= \min_{\Prhob{MF}{GC}} \big( \Uint[\Prhob{MF}{GC}] - \frac{1}{\beta} \Sint[\Prhob{MF}{GC}] \big) \\
   &= \min_{p^{\text{GC}} \Orho{\nsmall}\, \Orho{\elsmall}} \big( \Uint[p^{\text{GC}} \Orho{\nsmall}\, \Orho{\elsmall}] - \frac{1}{\beta} \Sint[p^{\text{GC}} \Orho{\nsmall} ] \big).
\end{split}
\end{equation}
As a reminder, $\Orho{\elsmall}$ is not included in $\Sint$ due to the ZEE approximation.}

{Let us start with the entropic term $\Sint$ of Eq.~\eqref{eq:Sint1}. Thanks to the mean field approximation, it can be split into a purely nuclear quantum entropy, $\Sqm$, of Eq.~\eqref{eq:Sqm} and a purely classical term, $\Smm$, as follow
\begin{align}
   \Sint[\Prhob{MF}{GC}] &=-\Trac_\text{GC}\left\{ p^{\text{GC}} \Orho{\nsmall}\,\,\text{log} ( p^{\text{GC}} \Orho{\nsmall}\ ) \right\} \nonumber \\
   &=-\Trac_\text{GC}\left\{p^{\text{GC}} \Orho{\nsmall}\,\,\left[\text{log}  (p^{\text{GC}})+\text{log}( \Orho{\nsmall})\right] \right\} \nonumber \\
      &=-\text{Tr}\left\{ \Orho{\nsmall} \,\, \text{log} ( \Orho{\nsmall}) \right\} \nonumber \\
   &\phantom{=} - \sum_{\NMM=0}^{\infty} \left [\iint  d\bqtot d\textbf{P}^{\NMM} \, \, \pgc \log(\pgc) \right ]\nonumber \\
   &=\Smm[\pgc]+\Sqm[\Orho{\nsmall}].
\label{eq:Sint_appendix}
\end{align}
}

{We now move to the energetic term,  $\Uint$, \textit{i.e.}
\begin{align}
   \Uint[\Prhob{MF}{GC}] &= \Trac_\text{GC} \{ 
  p^{\text{GC}} \Orho{\nsmall}\, \Orho{\elsmall} \big[  \hkinn + \cpF{\nsmall}   \nonumber \\ &+\hkine + \hwee + \hvne (\brnucltot) +\vnn(\brnucltot)  \nonumber \\
    &+ \cpF{\esmall}  + W   + T -\mu N\big]  \} \nonumber,
\end{align}
 where we explicited the whole Hamiltonian. Isolating the terms which depends only on $\brnucltot$ leads to
\begin{align}
   &\Uint[\Prhob{MF}{GC}] = \text{Tr} \{ \Orho{\nsmall} [\hkinn +\vnn(\brnucltot) ] \}+ \Trac_\text{GC} \{ p^{\text{GC}} \Orho{\nsmall}\, \Orho{\elsmall} \nonumber \\ \big[  
   &   \cpF{\nsmall}   \nonumber +\hkine + \hwee + \hvne (\brnucltot) +\cpF{\esmall}  \nonumber \\ 
   & + W  + T -\mu N \big ]  \} .\nonumber
\end{align}
Inserting the electron and classical particles embedding potentials defined in Eqs.~\eqref{eq:ven}-\eqref{eq:vmmn}
\begin{align}
   &\Uint[\Prhob{MF}{GC}] =  \nonumber\\ & \text{Tr}\{ \Orho{\nsmall} [\hkinn +\vnn(\brnucltot) +V_{\elsmall}^{\nsmall}[\Orho{\elsmall};\brnucltot] +V_{\mm}^{\nsmall}[\pgc;\brnucltot] ] \}+ \nonumber \\ 
   &\Trac_\text{GC} \{ p^{\text{GC}} \Orho{\nsmall}\, \Orho{\elsmall}\big[  \hkine + \hwee + \cpF{\esmall}  + W   + T -\mu N \big ]  \}. \nonumber
\end{align}
Gathering the electronic terms, we recognize the expression of the intrinsic electronic functional, $f_\elsmall$,  given in   Eq.~\eqref{eq:fel}. If the entropic contribution due to the classical solvent, $\Smm$, is added we can similarly identify the intrinsic electronic functional, $f_{\mm}^{\gc}$, in Eq.~\eqref{eq:fmm}
\begin{align}
    &\Uint[\Prhob{MF}{GC}]-\frac{1}{\beta}\Smm[ p^{\text{GC}}] = \text{Tr}\{ \Orho{\nsmall} [\hkinn +\vnn(\brnucltot)  ] \}+ \nonumber  \\ 
    &\text{Tr}\{ \Orho{\nsmall}  \big[ V_{\elsmall}^{\nsmall}[\Orho{\elsmall};\brnucltot] +V_{\mm}^{\nsmall}[\pgc;\brnucltot]+ f_{\elsmall}[\Orho{\elsmall}] 
         + f_{\mm}^{\gc}[\pgc] \nonumber\\
    &+ E_{\qm}^{\mm}[\Orho{\elsmall},\pgc ] \big ]  \} .\nonumber
\end{align}
By adding $\Sqm[\Orho{\nsmall}]$, we can now go back to expression of the grand potential, $\omegamf$, wich requires minimizations over $p^{\text{GC}} \Orho{\nsmall}\, \Orho{\elsmall}$
\begin{align}
    \omegamf = & \min_{p^{\text{GC}} \Orho{\nsmall}\, \Orho{\elsmall}}  \big( \text{Tr}\{ \Orho{\nsmall} [\hkinn +\vnn(\brnucltot) + \beta^{-1} \ln (\Orho{\nsmall})  ]  \nonumber\\
& + \Orho{\nsmall}  \big[ V_{\elsmall}^{\nsmall}[\Orho{\elsmall};\brnucltot] +V_{\mm}^{\nsmall}[\pgc;\brnucltot]+ f_{\elsmall}[\Orho{\elsmall}]  \nonumber\\
   & + f_{\mm}^{\gc}[\pgc]+ f_{\qm}^{\mm}[\Orho{\elsmall},\pgc ] \big ]  \} \big) \nonumber. 
\end{align}
}

{Here, if one perform a first optimization over the electronic and classical density matrix, \textit{i.e.}
\begin{equation}
\min_{p^{\text{GC}} \Orho{\nsmall}\, \Orho{\elsmall}}=\min_{\Orho{\nsmall}} \{ \min_{p^{\text{GC}}  \Orho{\elsmall}}\}
\end{equation}
one recover the expression of the total effective potential felt by the nuclei, $\MystFunc(\brnucltot)$ and finally the expression of the grand potential, $\omegamf$ of Eq.\eqref{def_fint_nucl}.}

\subsection{Appendix: The Harmonic Approximation\label{app:HO}}
This Appendix briefly summarizes the basic equations and concept of harmonic approximation allowing us to obtain Eq.~\eqref{eq:fint_final} as an estimate of the Grand potential. 
We therefore recapitulate here the main steps proposed originally by McQuarrie\cite{Quarrie}, which were further explicited by Ochterski in the documentation of the thermodynamic corrections as implemented in the Gaussian software\cite{Ochterski-Gaussian-00,Ochterski-Gaussian-01}. 

We consider a system of quantum nuclei evolving in a generic potential, which in our case is 
$\vtot(\brnucltot)$. 
The first step of the 
harmonic approximation is to approximate up to the second-order the potential $\vtot(\brnucltot)$ around the equilibrium geometry, \textit{i.e.}
\begin{equation}
	\vtot(\brnucltot) \approx \vtot(\brnucl{\text{eq}})   
	+ \frac{1}{2} \delta^\dagger_{\brnucltot} \textbf{h} \delta_{\brnucltot},
\end{equation}
where $\brnucl{\text{eq}}$ is the equilibrium geometry verifying 
\begin{equation}
	\label{eq:def_req}
	\nabla_{\brnucltot} \vtot(\brnucltot)\big|_{\brnucltot = \brnucl{\text{eq}}} = 0,
\end{equation}
$\delta_{\brnucltot} = \brnucltot - \brnucl{\text{eq}}$ is the displacement vector with respect to $\brnucl{\text{eq}}$, 
and $\textbf{h}$ is the Hessian matrix defined as 
\begin{equation}
	\label{eq:def_hess}
	\textbf{h}_{{i}{j}} = \frac{\partial^2 \vtot(\brnucltot)}{\partial \brnucl{i} \partial\brnucl{j}}\big|_{\brnucltot = \brnucl{\text{eq}}}.
\end{equation}
Written in the basis of the $N_\nu$ normal modes (\textit{i.e.} the eigenvectors of the Hessian matrix), 
the total nuclear Hamiltonian become 
a sum of uncoupled harmonic oscillators, such that $\Unucl$ in Eq.~\eqref{eq:Uint2} can be written as 
\begin{equation}
	\label{unucl_ho}
	\begin{aligned}
		&\Unucl [\Orho{\nsmall} ] =   \vtot(\brnucl{\text{eq}})  +\sum_{k=1}^{N_\nu} \text{Tr}_\text{QM}\{ \Orho{\nsmall} \hat{h}_{\omega_k}\}, \\
		&\quad \hat{h}_{\omega_k} = 
		-\frac{1}{2\mu_k} \frac{\partial}{\partial q_k^2} + \frac{1}{2} \mu_k (\omega_k)^2 q_k^2, 
	\end{aligned}
\end{equation}
where $q_k$, $\mu_k$ are the displacement and effective mass associated to the normal mode $k$, respectively,  
and $\omega_k$ is the associated frequency defined as 
\begin{equation}
	\omega_k = \sqrt{\frac{\lambda_k}{\mu_k} },
\end{equation}
where $\lambda_k$ is the corresponding eigenvalue of the Hessian matrix. 
Provided that $\brnucl{\text{eq}}$ is not a saddle point, the eigenvalues are all non negative, 
but nevertheless there are small eigenvalues corresponding to rotational and translational degrees of freedom. 
As such motions cannot be described by the harmonic oscillators, one simply neglects the lowest eigenvalues to retain only vibrational degrees of freedom. 
Under these conditions, minimizing the functional as in Eq.~\eqref{def_fint_nucl} with the approximated potential of Eq.~\eqref{unucl_ho}
is equivalent to find the partition function of a sum of uncoupled oscillators. 
As a consequence, the total partition function is simply the product of the individual partition functions for each vibration mode
\begin{equation}
	\PartF_{\bqtot} = e^{-\beta \vtot(\brnucl{\text{eq}})  }\prod_{k=1}^{N_\nu}  \PartF_{\text{vib}}^k,
\end{equation}
where $\PartF_{\text{vib}}^k$ is the partition function of the $k$-th vibration mode, 
\begin{equation}
	\PartF_{\text{vib}}^k = \frac{e^{-\frac{\beta \omega_k}{2} }}{1 - e^{-\beta \omega_k }}.
\end{equation}
When this partition function is inserted into the definition of the grand potential
\begin{equation}
	\Omega = -\frac{1}{\beta}\log( \PartF_{\bqtot}),
\end{equation}
it yields the approximate expression of the grand potential of Eq. \eqref{eq:fint_final}.

\subsection{Appendix: Evaluating the nuclear gradient and Hessian\label{app:grad_hessmat}}
As shown in Sec.~\ref{app:HO}, the computation of $\Omega$ requires the gradient and Hessian of the total 
effective potential $\vtot(\brnucl{})^{\NN}$ with respect to the solute nuclear coordinates, 
and the non trivial part is that coming from $ \MystFunc (\brnucltot)$. 
We provide here a summary of the main equations involved to compute these quantities. 

\subsubsection{Computation of the total nuclear gradient of $ \MystFunc (\brnucltot)$ \label{subsubsec:comp_grad}}
Let us first define the optimal classical and quantum one-body densities for a given nuclear geometry 
\begin{equation}
	\begin{aligned}
		\big(\rhoelopt(\brnucltot),\rhommopt(\brnucltot) \big) = \underset{\rhoel,\rhomm}{\text{argmin}}\,\,
		\mystFunc[\rhoel,\rhomm,\brnucltot].
	\end{aligned}
\end{equation}
Then, the gradient of $ \MystFunc (\brnucltot)$ with respect to the nuclear geometry is defined as
\begin{equation}
	\begin{aligned}
		\nabla_{\brnucltot} \MystFunc (\brnucltot) & = \nabla_{\brnucltot} \minu{\rhoel,\rhomm} \,\,\mystFunc[\rhoel,\rhomm,\brnucltot] \\
		& = \nabla_{\brnucltot} \mystFunc[\rhoelopt(\brnucltot),\rhommopt(\brnucltot),\brnucltot].
	\end{aligned}
\end{equation}
The gradient contains therefore two terms: one coming from the explicit dependency of the functional $\mystFunc$ with respect 
to the nuclear geometry, and another one coming from the implicit dependence of $\big(\rhoelopt(\brnucltot),\rhommopt(\brnucltot) \big)$ with respect to the nuclear geometry. 
Because $\big(\rhoelopt(\brnucltot),\rhommopt(\brnucltot) \big)$ minimizes the functional $\mystFunc$, 
the equivalent of the Hellman-Feynman theorem can be used such that only the contribution from coming from the explicit dependency of the functional is non zero, 
which yields to 
\begin{equation}\label{eq:grad_final}
	\begin{aligned}
		\nabla_{\brnucltot} \MystFunc (\brnucltot) =& (\nabla_{\brnucltot} \vne(\brnucltot)|\rhoel_{\opt}(\brnucltot)) \\
		+&(\nabla_{\brnucltot} \cpf{\nsmall}(\brnucltot)|\rhomm_{\opt}(\brnucltot)).
	\end{aligned}
\end{equation}
In Eq.~\eqref{eq:grad_final}, one recognizes the usual electronic gradient term $(\nabla_{\brnucltot} \vne(\brnucltot)|\rhoel_{\opt}(\brnucltot))$, and the other one is the contribution to the classical density, which is simply computed as an integral 
\begin{equation}
  \label{eq:grad_nucl}
	\begin{aligned}
		(\nabla_{\brnucltot} &\cpf{\nsmall}(\brnucltot)|\rhomm_{\opt}(\brnucltot)) = \\
		& \int {d}\bqtotone{} \rhomm_{\opt}(\bqtotone{}) \nabla_{\brnucltot} \cpf{\nsmall}(\bqtotone{},\brnucltot).
	\end{aligned}
\end{equation}
Therefore, for each component of the nuclear coordinates, one just needs to compute the integral of the nuclear gradient 
of the nuclear-solvent interaction potential (\textit{i.e.} $\nabla_{\brnucltot} \cpf{\nsmall}(\bqtotone{},\brnucltot)$ in Eq.~\eqref{eq:grad_nucl}) multiplied by the classical solvent density (\textit{i.e.} $\rhomm_{\opt}(\bqtotone{})$). 
Each of these integrals is typically approximated by a numerical integral as usually done in MDFT, and 
can then be added to the usual electronic gradient to obtain the total nuclear gradient for the QM/MM system. 

\subsubsection{Computation of the exact and approximated Hessian matrix of $ \MystFunc (\brnucltot)$ \label{subsubsec:comp_hessmat}}

Regarding now the second-order derivative of $\MystFunc (\brnucltot)$, applying the linear response formalism to the QM/MM functional of Eqs.~\eqref{eq:MystFunc} and \eqref{eq:MystFunc_small} leads to the following expression 
\begin{equation}
	\begin{aligned}
		&\frac{\partial^2 \MystFunc (\brnucltot)}{\partial \brnucl{i}\partial \brnucl{j}} \!= \!
		\dertwoel(\brnucl{i},\brnucl{j}) \!+ \!\dertwomm (\brnucl{i},\brnucl{j})\!+ \!\dertwommel(\brnucl{i},\brnucl{j}) ,
	\end{aligned}
\end{equation}
with 
\begin{equation}
	\begin{aligned}
		\dertwoel(\brnucl{i},\brnucl{j})\!=\! \big(\frac{\partial^2 \vne(\brnucltot)}{\partial \brnucl{i}\partial \brnucl{j}} | \rhoelopt\big) 
		\!+\!\big(\nabla_{\brnucl{i}}\vne(\brnucltot)|\nabla_{\brnucl{j}}\rhoelopt\big), 
	\end{aligned}
\end{equation}
\begin{equation}
	\dertwomm(\brnucl{i},\brnucl{j})\!=\! 
	\big(\frac{\partial^2 {\cpf{\nsmall}}(\brnucltot)}{\partial \brnucl{i}\partial \brnucl{j}} | \rhommopt\big) 
	\!+\!\big(\nabla_{\brnucl{i}}\cpf{\nsmall}(\brnucltot)|\nabla_{\brnucl{j}}\rhommopt),
\end{equation}
\begin{equation}
	\dertwommel(\brnucl{i},\brnucl{j}) = \big(\nabla_{\brnucl{i}}\rhommopt | \cpf{\esmall} | \nabla_{\brnucl{j}}\rhoelopt).
\end{equation}
Here, the response vectors $\nabla_{\brnucltot}\rhoelopt$ and $\nabla_{\brnucltot}\rhommopt$ can be obtained 
by solving the coupled linear response equations which, after some derivations, yield  
\begin{equation}
	\begin{aligned}
		\nabla_{\brnucltot}\frac{\delta }{\delta \rhoel(\br{})}\mystFunc[\rhoelopt,\rhommopt,\brnucltot] = 
		&\langle \hat{\chi}^{\elsmall} \cdot \nabla_{\brnucltot}\rhoelopt \rangle_{\br{}} \\
		+&\langle \hat{\chi}^{\mm}_{\elsmall} \cdot \nabla_{\brnucltot}\rhommopt \rangle_{\br{}} ,
	\end{aligned}
\end{equation}
\begin{equation}
	\begin{aligned}
		\nabla_{\brnucltot}\frac{\delta }{\delta \rhomm(\bqtotone)}\mystFunc[\rhoelopt,\rhommopt,\brnucltot] = 
		&\langle \hat{\chi}^{\mm} \cdot \nabla_{\brnucltot}\rhommopt \rangle_{\br{}} \\
		+&\langle \hat{\chi}_{\mm}^{\elsmall} \cdot \nabla_{\brnucltot}\rhoelopt \rangle_{\br{}} ,
	\end{aligned}
\end{equation}
where the notation $\langle \cdot \rangle_{\br{}}$ denotes the convolution product 
\begin{equation}
	\langle \hat{f} \cdot g \rangle_{\br{}} = \int d \br{}' f(\br{},\br{}') g(\br{}'),
\end{equation}
and where the response functions are defined as 
\begin{equation}
	\chi^{\elsmall}(\br{},\br{}')^{-1} = -\frac{\delta^2 }{\delta \rhoel(\br{})\delta \rhoel(\br{}')}\mystFunc[\rhoel,\rhomm,\brnucltot],  
\end{equation}
\begin{equation}
	\chi^{\mm}(\bqtotone,\bqtotone')^{-1} = -\frac{\delta^2 }{\delta \rhomm(\bqtotone)\delta \rhomm(\bqtotone')}\mystFunc[\rhoel,\rhomm,\brnucltot],  
\end{equation}
\begin{equation}
	\chi^{\mm}_{\elsmall}(\bqtotone,\br{})^{-1} 
	= -\frac{\delta^2 }{\delta \rhomm(\bqtotone)\delta \rhoel(\br{})}\mystFunc[\rhoel,\rhomm,\brnucltot],  
\end{equation}
\begin{equation}
	\chi^{\elsmall}_{\mm}(\br{},\bqtotone)^{-1} = -\frac{\delta^2 }{\delta \rhoel(\br{})\delta \rhomm(\bqtotone)}\mystFunc[\rhoel,\rhomm,\brnucltot].  
\end{equation}
One sees therefore that the total Hessian matrix contains both electron-electron (\textit{i.e.} $\chi^{\elsmall}$), electron-solvent (\textit{i.e.} $\chi^{\mm}_{\elsmall}$ and $\chi^{\elsmall}_{\mm}$) and solvent-solvent ($\chi^{\mm}$) response functions. 

In the computations presented here, we neglected the response functions $\chi^{\elsmall}_{\mm}$, $\chi^{\mm}_{\elsmall}$ and $\chi^{\mm}$. 
This leads then to the standard electronic linear response equations, which are 
nevertheless evaluated at the optimal electronic 
density $\rhoelopt$ which takes into account the influence of the solvent, \textit{i.e.}  
\begin{equation}
	\begin{aligned}
		&\frac{\partial^2 \MystFunc (\brnucltot)}{\partial \brnucl{i}\partial \brnucl{j}} \approx 
		\dertwoel(\brnucl{i},\brnucl{j}).  
	\end{aligned}
\end{equation}
Therefore, under this approximation, no additional computation is needed to obtain the Hessian matrix. 

\section*{Supporting information}
The supporting information file contains i) the redox potentials computed within our eDFT/MDFT scheme together with the experimental and eDFT/C-PCM data of Ref. \onlinecite{HuyAnsCavStaHam-JACS-16}, ii) the Lennard-Jones parameters used in our simulations, and iii) the Cartesian coordinates obtained with our eDFT/MDFT geometry optimization scheme.  

\bibliography{reference_quinone}
\begin{figure}[!htbp]
	\centering
	\includegraphics[width=\columnwidth]{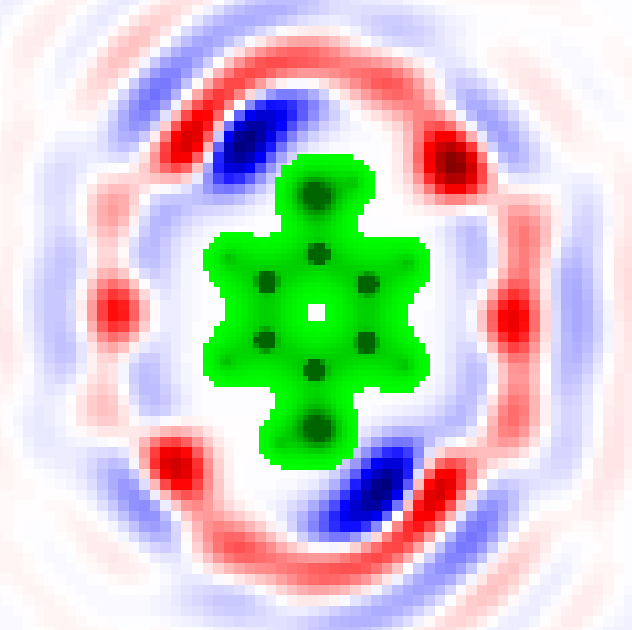}
	\caption{For Table of Contents Only}
\end{figure}

\end{document}


\maketitle
\tableofcontents
\thispagestyle{empty}
\newpage
\section*{Computed and Experimental $E$°[BQ,2H$^+$/HQ] in water : \label{sec:Red_Pot_Quinones}}

\begin{table}[H]
    \centering
\begin{tabular}{lccccccc}
\hline
\textbf{Atom} & \hspace{1cm} & \textbf{Exp.}\cite{HuyAnsCavStaHam-JACS-16} & \hspace{1cm} & \textbf{eDFT/C-PCM}\cite{HuyAnsCavStaHam-JACS-16} & \hspace{1cm} & \textbf{eDFT/MDFT} \\ \hline                           
H              & \hspace{1cm} &   0.000  & \hspace{1cm} &   0.000  & \hspace{1cm} &   0.000 \\
CH$_3$         & \hspace{1cm} &  -0.044  & \hspace{1cm} &  -0.054  & \hspace{1cm} &  -0.073 \\
3CH$_3$        & \hspace{1cm} &  -0.157  & \hspace{1cm} &  -0.205  & \hspace{1cm} &  -0.208 \\
4CH$_3$        & \hspace{1cm} &  -0.231  & \hspace{1cm} &  -0.268  & \hspace{1cm} &  -0.288 \\
C(CH$_3$)$_2$  & \hspace{1cm} &  -0.064  & \hspace{1cm} &  -0.088  & \hspace{1cm} &  -0.092 \\
CN             & \hspace{1cm} &  -       & \hspace{1cm} &   0.088  & \hspace{1cm} &   0.083 \\
CHO            & \hspace{1cm} &  -       & \hspace{1cm} &   0.215  & \hspace{1cm} &   0.261 \\
COCH$_3$       & \hspace{1cm} &  -       & \hspace{1cm} &   0.189  & \hspace{1cm} &   0.246 \\
COOH           & \hspace{1cm} &  -       & \hspace{1cm} &   0.109  & \hspace{1cm} &   0.067 \\
COOCH$_3$      & \hspace{1cm} &  -       & \hspace{1cm} &   0.176  & \hspace{1cm} &   0.218 \\
CF$_3$         & \hspace{1cm} &  -       & \hspace{1cm} &   0.043  & \hspace{1cm} &   0.041 \\
NH$_2$         & \hspace{1cm} &  -       & \hspace{1cm} &  -0.234  & \hspace{1cm} &  -0.215 \\
N(CH$_3$)$_2$  & \hspace{1cm} &  -       & \hspace{1cm} &  -0.224  & \hspace{1cm} &  -0.156 \\
OH             & \hspace{1cm} &  -       & \hspace{1cm} &  -0.085  & \hspace{1cm} &  -0.104 \\
OCH$_3$        & \hspace{1cm} &  -0.073  & \hspace{1cm} &  -0.119  & \hspace{1cm} &  -0.133 \\
2OCH$_3$       & \hspace{1cm} &  -0.170  & \hspace{1cm} &  -0.197  & \hspace{1cm} &  -0.199 \\
NO$_2$         & \hspace{1cm} &  -       & \hspace{1cm} &   0.248  & \hspace{1cm} &   0.286 \\
F              & \hspace{1cm} &  -       & \hspace{1cm} &  -0.003  & \hspace{1cm} &  -0.019 \\
4F             & \hspace{1cm} &   0.063  & \hspace{1cm} &   0.041  & \hspace{1cm} &   0.000 \\
SH             & \hspace{1cm} &  -       & \hspace{1cm} &  -0.025  & \hspace{1cm} &  -0.079 \\
Cl             & \hspace{1cm} &   0.033  & \hspace{1cm} &   0.016  & \hspace{1cm} &  -0.015 \\
2Cl            & \hspace{1cm} &   0.056  & \hspace{1cm} &   0.028  & \hspace{1cm} &  -0.043 \\
4Cl            & \hspace{1cm} &   0.056  & \hspace{1cm} &  -0.006  & \hspace{1cm} &  -0.007 \\
2Cl2CN         & \hspace{1cm} &   0.244  & \hspace{1cm} &   0.197  & \hspace{1cm} &   0.241 \\ \hline
\end{tabular}

    \caption{All $E$°[BQ,2H$^+$/HQ] are in V and Vs Ref-BQ, 2H$^{+}$/Ref-HQ. Both Exp }
    \label{tab:red_pot_quinone}
\end{table}

\newpage
\section*{Lennard-Jones Parameters: \label{sec:LJ_param}}
\begin{table}[H]
    \centering
\begin{tabular}{ccccccl}                               
\hline
\textbf{Atom} & & $\bm{\sigma}$ \textbf{(\AA)} & & $\bm{\epsilon}$ \textbf{(kJ/mol)} & & \textbf{Description}\\ \hline 
\textbf{C2}    & & 3.31521        & & 0.41338 & & sp2 C, including pure aromatic systems and carbonyl groups.  \\
\textbf{C3}    & & 3.39771        & & 0.45101 & & sp3 C.  \\
\textbf{H }    & & 2.62548        & & 0.06736 & & All H.  \\
\textbf{Cl}    & & 3.46600        & & 1.10370 & & All Cl. \\
\textbf{F }    & & 3.03422        & & 0.34811 & & All F  \\
\textbf{N }    & & 3.36510        & & 0.35899 & & Sp3 N with three connected atoms or nitro N. \\
\textbf{Nh}    & & 3.18995        & & 0.89956 & & Amine N connected one or more aromatic rings.  \\
\textbf{O }    & & 3.04812        & & 0.61212 & & O with one connected atom.   \\
\textbf{Oh}    & & 3.24287        & & 0.38911 & & O in hydroxyl group. \\
\textbf{S }    & & 3.53241        & & 1.18156 & & Sp3 S connected with hydrogen.  \\ \hline
\end{tabular}
\end{table}

\newpage
\section*{Cartesian Coordinates (\AA) for Quinone Couples: \label{sec:Quinones_xyz}}
\subsection*{For R-BQ (left) and R-HQ (right) with R = H}
\begin{table}[H]
\centering
\begin{tabular}{ccccccc}
\hline
\textbf{Atom} & & \textbf{x} & & \textbf{y} & & \textbf{z} \\ 
\hline
C & & -1.069 & & 0.026 & &  -0.839 \\ 
C & & 0.200 & & 0.018 & &  -1.602 \\ 
C & & 1.469 & & 0.013 & &  -0.839 \\ 
C & & 1.469 & & 0.014 & &  0.504 \\ 
C & & 0.200 & & 0.020 & &  1.266 \\ 
C & & -1.069 & & 0.027 & &  0.503 \\ 
H & & -1.982 & & 0.031 & &  -1.422 \\ 
H & & 2.382 & & 0.008 & &  -1.421 \\ 
H & & 2.382 & & 0.010 & &  1.087 \\ 
H & & -1.983 & & 0.033 & &  1.086 \\ 
O & & 0.200 & & 0.020 & &  2.497 \\ 
O & & 0.200 & & 0.015 & &  -2.833 \\ 
\hline
\end{tabular}
\hspace{2.5cm}
\begin{tabular}{ccccccc}
\hline
\textbf{Atom} & & \textbf{x} & & \textbf{y} & & \textbf{z} \\ 
\hline
C & & -1.223 & & 0.000 & &  -0.670 \\ 
C & & -0.030 & & 0.000 & &  -1.396 \\ 
C & & 1.192 & & -0.000 & &  -0.724 \\ 
C & & 1.223 & & -0.000 & &  0.670 \\ 
C & & 0.030 & & -0.000 & &  1.396 \\ 
C & & -1.192 & & 0.000 & &  0.724 \\ 
H & & -2.176 & & 0.001 & &  -1.187 \\ 
H & & 2.114 & & -0.001 & &  -1.292 \\ 
H & & 2.176 & & -0.001 & &  1.187 \\ 
H & & -2.114 & & 0.001 & &  1.292 \\ 
O & & 0.000 & & -0.000 & &  2.770 \\ 
O & & -0.000 & & 0.000 & &  -2.770 \\ 
H & & 0.901 & & -0.001 & &  3.120 \\ 
H & & -0.901 & & 0.000 & &  -3.120 \\ 
\hline
\end{tabular}
\end{table}
 \newpage
\subsection*{For R-BQ (left) and R-HQ (right) with R = CH$_3$}
\begin{table}[H]
\centering
\begin{tabular}{ccccccc}
\hline
\textbf{Atom} & & \textbf{x} & & \textbf{y} & & \textbf{z} \\ 
\hline
C & & -1.255 & & 0.015 & &  -0.678 \\ 
C & & 0.010 & & -0.000 & &  -1.430 \\ 
C & & 1.279 & & -0.016 & &  -0.667 \\ 
C & & 1.267 & & -0.015 & &  0.674 \\ 
C & & -0.006 & & -0.000 & &  1.429 \\ 
C & & -1.295 & & 0.015 & &  0.671 \\ 
H & & -2.162 & & 0.026 & &  -1.272 \\ 
H & & 2.196 & & -0.027 & &  -1.245 \\ 
H & & 2.174 & & -0.026 & &  1.267 \\ 
O & & 0.000 & & 0.000 & &  2.659 \\ 
O & & 0.021 & & -0.000 & &  -2.663 \\ 
C & & -2.559 & & 0.029 & &  1.467 \\ 
H & & -2.585 & & 0.897 & &  2.130 \\ 
H & & -2.617 & & -0.853 & &  2.109 \\ 
H & & -3.430 & & 0.052 & &  0.813 \\ 
\hline
\end{tabular}
\hspace{2.5cm}
\begin{tabular}{ccccccc}
\hline
\textbf{Atom} & & \textbf{x} & & \textbf{y} & & \textbf{z} \\ 
\hline
C & & -1.214 & & 0.000 & &  -0.667 \\ 
C & & -0.021 & & 0.000 & &  -1.394 \\ 
C & & 1.201 & & -0.000 & &  -0.726 \\ 
C & & 1.219 & & -0.000 & &  0.668 \\ 
C & & 0.026 & & -0.000 & &  1.389 \\ 
C & & -1.212 & & 0.000 & &  0.730 \\ 
H & & -2.165 & & 0.000 & &  -1.191 \\ 
H & & 2.125 & & -0.000 & &  -1.291 \\ 
H & & 2.169 & & -0.000 & &  1.193 \\ 
O & & 0.000 & & 0.000 & &  2.765 \\ 
O & & 0.002 & & 0.000 & &  -2.770 \\ 
C & & -2.493 & & 0.000 & &  1.516 \\ 
H & & -2.558 & & 0.877 & &  2.166 \\ 
H & & -2.558 & & -0.877 & &  2.166 \\ 
H & & -3.358 & & -0.000 & &  0.852 \\ 
H & & 0.903 & & -0.001 & &  3.109 \\ 
H & & -0.902 & & 0.001 & &  -3.112 \\ 
\hline
\end{tabular}
\end{table}
 \newpage
\subsection*{For R-BQ (left) and R-HQ (right) with R = 3CH$_3$}
\begin{table}[H]
\centering
\begin{tabular}{ccccccc}
\hline
\textbf{Atom} & & \textbf{x} & & \textbf{y} & & \textbf{z} \\ 
\hline
C & & -1.291 & & -0.000 & &  -0.688 \\ 
C & & 0.011 & & -0.000 & &  -1.416 \\ 
C & & 1.274 & & 0.000 & &  -0.662 \\ 
C & & 1.300 & & 0.000 & &  0.682 \\ 
C & & -0.000 & & -0.000 & &  1.416 \\ 
C & & -1.293 & & -0.000 & &  0.669 \\ 
H & & 2.182 & & 0.001 & &  -1.255 \\ 
O & & -0.000 & & -0.000 & &  2.647 \\ 
O & & 0.034 & & -0.000 & &  -2.651 \\ 
C & & -2.514 & & -0.001 & &  1.539 \\ 
H & & -2.510 & & 0.872 & &  2.197 \\ 
H & & -2.511 & & -0.876 & &  2.194 \\ 
H & & -3.436 & & 0.001 & &  0.963 \\ 
C & & 2.553 & & 0.001 & &  1.495 \\ 
H & & 2.588 & & 0.876 & &  2.149 \\ 
H & & 2.588 & & -0.875 & &  2.148 \\ 
H & & 3.432 & & 0.001 & &  0.851 \\ 
C & & -2.507 & & -0.001 & &  -1.565 \\ 
H & & -2.498 & & 0.873 & &  -2.222 \\ 
H & & -2.499 & & -0.876 & &  -2.220 \\ 
H & & -3.432 & & 0.000 & &  -0.995 \\ 
\hline
\end{tabular}
\hspace{2.5cm}
\begin{tabular}{ccccccc}
\hline
\textbf{Atom} & & \textbf{x} & & \textbf{y} & & \textbf{z} \\ 
\hline
C & & -1.264 & & 0.001 & &  -0.731 \\ 
C & & -0.042 & & -0.000 & &  -1.426 \\ 
C & & 1.179 & & -0.002 & &  -0.762 \\ 
C & & 1.229 & & -0.002 & &  0.634 \\ 
C & & 0.014 & & 0.000 & &  1.329 \\ 
C & & -1.231 & & 0.002 & &  0.678 \\ 
H & & 2.096 & & -0.003 & &  -1.340 \\ 
O & & -0.019 & & -0.000 & &  2.710 \\ 
O & & 0.005 & & -0.001 & &  -2.807 \\ 
C & & -2.462 & & 0.004 & &  1.548 \\ 
H & & -2.463 & & 0.873 & &  2.213 \\ 
H & & -2.481 & & -0.877 & &  2.194 \\ 
H & & -3.385 & & 0.019 & &  0.974 \\ 
C & & 2.543 & & -0.004 & &  1.371 \\ 
H & & 2.657 & & 0.879 & &  2.009 \\ 
H & & 2.655 & & -0.888 & &  2.009 \\ 
H & & 3.377 & & -0.005 & &  0.668 \\ 
C & & -2.539 & & 0.002 & &  -1.540 \\ 
H & & -2.606 & & 0.887 & &  -2.184 \\ 
H & & -2.605 & & -0.881 & &  -2.187 \\ 
H & & -3.430 & & 0.000 & &  -0.918 \\ 
H & & 0.881 & & -0.000 & &  3.059 \\ 
H & & -0.890 & & 0.000 & &  -3.168 \\ 
\hline
\end{tabular}
\end{table}
 \newpage
\subsection*{For R-BQ (left) and R-HQ (right) with R = 4CH$_3$}
\begin{table}[H]
\centering
\begin{tabular}{ccccccc}
\hline
\textbf{Atom} & & \textbf{x} & & \textbf{y} & & \textbf{z} \\ 
\hline
C & & -1.300 & & 0.002 & &  -0.677 \\ 
C & & -0.000 & & 0.003 & &  -1.407 \\ 
C & & 1.300 & & -0.002 & &  -0.677 \\ 
C & & 1.300 & & -0.003 & &  0.677 \\ 
C & & 0.000 & & -0.001 & &  1.407 \\ 
C & & -1.300 & & 0.000 & &  0.677 \\ 
O & & -0.000 & & -0.000 & &  2.640 \\ 
O & & -0.000 & & 0.009 & &  -2.640 \\ 
C & & -2.516 & & -0.001 & &  1.553 \\ 
H & & -2.508 & & 0.870 & &  2.213 \\ 
H & & -2.513 & & -0.878 & &  2.205 \\ 
H & & -3.440 & & 0.004 & &  0.980 \\ 
C & & 2.516 & & -0.005 & &  1.553 \\ 
H & & 2.512 & & 0.868 & &  2.210 \\ 
H & & 2.510 & & -0.880 & &  2.208 \\ 
H & & 3.440 & & -0.006 & &  0.980 \\ 
C & & -2.516 & & 0.004 & &  -1.553 \\ 
H & & -2.511 & & 0.879 & &  -2.207 \\ 
H & & -2.510 & & -0.869 & &  -2.211 \\ 
H & & -3.440 & & 0.002 & &  -0.980 \\ 
C & & 2.516 & & -0.005 & &  -1.553 \\ 
H & & 2.512 & & 0.868 & &  -2.211 \\ 
H & & 2.509 & & -0.880 & &  -2.208 \\ 
H & & 3.440 & & -0.005 & &  -0.980 \\ 
\hline
\end{tabular}
\hspace{2.5cm}
\begin{tabular}{ccccccc}
\hline
\textbf{Atom} & & \textbf{x} & & \textbf{y} & & \textbf{z} \\ 
\hline
C & & -1.240 & & -0.073 & &  -0.680 \\ 
C & & -0.024 & & 0.007 & &  -1.370 \\ 
C & & 1.220 & & 0.073 & &  -0.727 \\ 
C & & 1.242 & & 0.077 & &  0.679 \\ 
C & & 0.021 & & -0.001 & &  1.371 \\ 
C & & -1.219 & & -0.082 & &  0.726 \\ 
O & & 0.000 & & -0.000 & &  2.755 \\ 
O & & 0.011 & & 0.020 & &  -2.751 \\ 
C & & -2.483 & & -0.151 & &  1.547 \\ 
H & & -2.977 & & 0.825 & &  1.604 \\ 
H & & -2.264 & & -0.466 & &  2.565 \\ 
H & & -3.204 & & -0.851 & &  1.122 \\ 
C & & 2.515 & & 0.162 & &  1.487 \\ 
H & & 2.461 & & 0.959 & &  2.237 \\ 
H & & 2.726 & & -0.775 & &  2.015 \\ 
H & & 3.385 & & 0.379 & &  0.872 \\ 
C & & -2.523 & & -0.132 & &  -1.473 \\ 
H & & -2.720 & & 0.813 & &  -1.992 \\ 
H & & -2.499 & & -0.925 & &  -2.229 \\ 
H & & -3.387 & & -0.329 & &  -0.841 \\ 
C & & 2.452 & & 0.121 & &  -1.593 \\ 
H & & 2.389 & & 0.941 & &  -2.313 \\ 
H & & 2.550 & & -0.797 & &  -2.180 \\ 
H & & 3.368 & & 0.245 & &  -1.019 \\ 
H & & 0.903 & & -0.023 & &  3.093 \\ 
H & & -0.889 & & 0.008 & &  -3.101 \\ 
\hline
\end{tabular}
\end{table}
 \newpage
\subsection*{For R-BQ (left) and R-HQ (right) with R = C(CH$_3$)$_3$}
\begin{table}[H]
\centering
\begin{tabular}{ccccccc}
\hline
\textbf{Atom} & & \textbf{x} & & \textbf{y} & & \textbf{z} \\ 
\hline
C & & -1.250 & & -0.000 & &  0.664 \\ 
C & & 0.010 & & 0.000 & &  1.428 \\ 
C & & 1.283 & & 0.001 & &  0.682 \\ 
C & & 1.267 & & 0.000 & &  -0.657 \\ 
C & & -0.001 & & -0.000 & &  -1.430 \\ 
C & & -1.309 & & -0.000 & &  -0.687 \\ 
H & & -2.144 & & -0.001 & &  1.273 \\ 
H & & 2.198 & & 0.001 & &  1.262 \\ 
H & & 2.172 & & 0.001 & &  -1.253 \\ 
O & & 0.059 & & -0.000 & &  -2.657 \\ 
O & & 0.000 & & 0.000 & &  2.663 \\ 
C & & -2.620 & & -0.001 & &  -1.468 \\ 
C & & -2.702 & & 1.264 & &  -2.354 \\ 
H & & -1.894 & & 1.297 & &  -3.084 \\ 
H & & -3.654 & & 1.261 & &  -2.890 \\ 
H & & -2.663 & & 2.169 & &  -1.742 \\ 
C & & -2.701 & & -1.266 & &  -2.354 \\ 
H & & -3.654 & & -1.264 & &  -2.890 \\ 
H & & -1.894 & & -1.298 & &  -3.084 \\ 
H & & -2.662 & & -2.171 & &  -1.742 \\ 
C & & -3.833 & & -0.001 & &  -0.525 \\ 
H & & -3.853 & & -0.889 & &  0.113 \\ 
H & & -3.854 & & 0.887 & &  0.113 \\ 
H & & -4.746 & & -0.002 & &  -1.124 \\ 
\hline
\end{tabular}
\hspace{2.5cm}
\begin{tabular}{ccccccc}
\hline
\textbf{Atom} & & \textbf{x} & & \textbf{y} & & \textbf{z} \\ 
\hline
C & & -1.169 & & 0.001 & &  0.647 \\ 
C & & 0.022 & & 0.000 & &  1.376 \\ 
C & & 1.245 & & -0.001 & &  0.713 \\ 
C & & 1.250 & & -0.001 & &  -0.679 \\ 
C & & 0.059 & & 0.000 & &  -1.408 \\ 
C & & -1.193 & & 0.001 & &  -0.755 \\ 
H & & -2.103 & & 0.002 & &  1.194 \\ 
H & & 2.171 & & -0.002 & &  1.274 \\ 
H & & 2.196 & & -0.002 & &  -1.211 \\ 
O & & 0.096 & & 0.000 & &  -2.786 \\ 
O & & 0.036 & & 0.000 & &  2.754 \\ 
C & & -2.520 & & 0.003 & &  -1.530 \\ 
C & & -2.614 & & 1.267 & &  -2.412 \\ 
H & & -1.793 & & 1.319 & &  -3.126 \\ 
H & & -3.557 & & 1.256 & &  -2.968 \\ 
H & & -2.598 & & 2.168 & &  -1.794 \\ 
C & & -2.617 & & -1.261 & &  -2.413 \\ 
H & & -3.559 & & -1.249 & &  -2.968 \\ 
H & & -1.796 & & -1.315 & &  -3.126 \\ 
H & & -2.602 & & -2.163 & &  -1.794 \\ 
C & & -3.738 & & 0.004 & &  -0.591 \\ 
H & & -3.765 & & -0.883 & &  0.047 \\ 
H & & -3.763 & & 0.892 & &  0.047 \\ 
H & & -4.649 & & 0.005 & &  -1.194 \\ 
H & & 1.016 & & -0.001 & &  -3.079 \\ 
H & & -0.871 & & 0.001 & &  3.089 \\ 
\hline
\end{tabular}
\end{table}
 \newpage
 \subsection*{For R-BQ (left) and R-HQ (right) with R = CN}
\begin{table}[H]
\centering
\begin{tabular}{ccccccc}
\hline
\textbf{Atom} & & \textbf{x} & & \textbf{y} & & \textbf{z} \\ 
\hline
C & & -1.266 & & -0.012 & &  -0.675 \\ 
C & & -0.001 & & -0.001 & &  -1.437 \\ 
C & & 1.275 & & 0.011 & &  -0.676 \\ 
C & & 1.272 & & 0.012 & &  0.673 \\ 
C & & -0.014 & & 0.000 & &  1.445 \\ 
C & & -1.267 & & -0.011 & &  0.669 \\ 
H & & -2.181 & & -0.020 & &  -1.255 \\ 
H & & 2.188 & & 0.019 & &  -1.259 \\ 
H & & -2.182 & & -0.018 & &  1.248 \\ 
O & & 0.000 & & -0.000 & &  2.671 \\ 
O & & 0.023 & & -0.002 & &  -2.662 \\ 
C & & 2.487 & & 0.021 & &  1.428 \\ 
N & & 3.473 & & 0.029 & &  2.043 \\ 
\hline
\end{tabular}
\hspace{2.5cm}
\begin{tabular}{ccccccc}
\hline
\textbf{Atom} & & \textbf{x} & & \textbf{y} & & \textbf{z} \\ 
\hline
C & & -1.179 & & 0.003 & &  -0.678 \\ 
C & & 0.011 & & -0.002 & &  -1.421 \\ 
C & & 1.232 & & -0.008 & &  -0.762 \\ 
C & & 1.264 & & -0.009 & &  0.646 \\ 
C & & 0.068 & & -0.003 & &  1.391 \\ 
C & & -1.150 & & 0.002 & &  0.710 \\ 
H & & -2.135 & & 0.007 & &  -1.189 \\ 
H & & 2.154 & & -0.013 & &  -1.327 \\ 
H & & -2.071 & & 0.007 & &  1.279 \\ 
O & & 0.041 & & -0.003 & &  2.753 \\ 
O & & 0.034 & & -0.002 & &  -2.784 \\ 
C & & 2.505 & & -0.015 & &  1.341 \\ 
N & & 3.491 & & -0.020 & &  1.963 \\ 
H & & 0.937 & & -0.006 & &  3.118 \\ 
H & & -0.865 & & 0.005 & &  -3.142 \\ 
\hline
\end{tabular}
\end{table}
 \newpage
\subsection*{For R-BQ (left) and R-HQ (right) with R = CHO}
\begin{table}[H]
\centering
\begin{tabular}{ccccccc}
\hline
\textbf{Atom} & & \textbf{x} & & \textbf{y} & & \textbf{z} \\ 
\hline
C & & -1.322 & & -0.075 & &  -0.651 \\ 
C & & -0.070 & & -0.043 & &  -1.430 \\ 
C & & 1.208 & & -0.008 & &  -0.674 \\ 
C & & 1.253 & & -0.007 & &  0.675 \\ 
C & & -0.022 & & -0.041 & &  1.457 \\ 
C & & -1.290 & & -0.075 & &  0.692 \\ 
H & & -2.247 & & -0.100 & &  -1.213 \\ 
H & & 2.114 & & 0.016 & &  -1.272 \\ 
H & & -2.193 & & -0.099 & &  1.289 \\ 
O & & -0.040 & & -0.041 & &  2.684 \\ 
O & & -0.061 & & -0.045 & &  -2.656 \\ 
C & & 2.597 & & 0.029 & &  1.302 \\ 
H & & 3.418 & & 0.050 & &  0.562 \\ 
O & & 2.839 & & 0.035 & &  2.496 \\ 
\hline
\end{tabular}
\hspace{2.5cm}
\begin{tabular}{ccccccc}
\hline
\textbf{Atom} & & \textbf{x} & & \textbf{y} & & \textbf{z} \\ 
\hline
C & & -1.158 & & -0.032 & &  -0.709 \\ 
C & & 0.043 & & -0.037 & &  -1.440 \\ 
C & & 1.250 & & -0.044 & &  -0.761 \\ 
C & & 1.273 & & -0.045 & &  0.648 \\ 
C & & 0.058 & & -0.039 & &  1.377 \\ 
C & & -1.153 & & -0.033 & &  0.682 \\ 
H & & -2.106 & & -0.028 & &  -1.236 \\ 
H & & 2.182 & & -0.048 & &  -1.316 \\ 
H & & -2.084 & & -0.029 & &  1.235 \\ 
O & & 0.052 & & -0.038 & &  2.731 \\ 
O & & 0.076 & & -0.036 & &  -2.805 \\ 
C & & 2.541 & & -0.052 & &  1.339 \\ 
H & & 3.444 & & -0.057 & &  0.710 \\ 
O & & 2.660 & & -0.053 & &  2.578 \\ 
H & & 0.998 & & -0.044 & &  3.018 \\ 
H & & -0.819 & & -0.027 & &  -3.172 \\ 
\hline
\end{tabular}
\end{table}
 \newpage
\subsection*{For R-BQ (left) and R-HQ (right) with R = COCH$_3$}
\begin{table}[H]
\centering
\begin{tabular}{ccccccc}
\hline
\textbf{Atom} & & \textbf{x} & & \textbf{y} & & \textbf{z} \\ 
\hline
C & & -1.260 & & 0.048 & &  -0.674 \\ 
C & & 0.013 & & 0.001 & &  -1.435 \\ 
C & & 1.277 & & -0.050 & &  -0.674 \\ 
C & & 1.262 & & -0.042 & &  0.669 \\ 
C & & -0.002 & & 0.011 & &  1.441 \\ 
C & & -1.290 & & 0.031 & &  0.673 \\ 
H & & -2.164 & & 0.067 & &  -1.273 \\ 
H & & 2.197 & & -0.083 & &  -1.246 \\ 
H & & 2.172 & & -0.060 & &  1.259 \\ 
O & & 0.025 & & 0.069 & &  2.668 \\ 
O & & 0.000 & & 0.000 & &  -2.667 \\ 
C & & -2.592 & & -0.031 & &  1.428 \\ 
O & & -2.708 & & -0.828 & &  2.351 \\ 
C & & -3.699 & & 0.894 & &  1.011 \\ 
H & & -3.903 & & 0.803 & &  -0.059 \\ 
H & & -3.384 & & 1.928 & &  1.186 \\ 
H & & -4.599 & & 0.678 & &  1.584 \\ 
\hline
\end{tabular}
\hspace{2.5cm}
\begin{tabular}{ccccccc}
\hline
\textbf{Atom} & & \textbf{x} & & \textbf{y} & & \textbf{z} \\ 
\hline
C & & -1.208 & & 0.258 & &  -0.697 \\ 
C & & -0.031 & & -0.034 & &  -1.365 \\ 
C & & 1.132 & & -0.320 & &  -0.634 \\ 
C & & 1.109 & & -0.311 & &  0.752 \\ 
C & & -0.074 & & -0.023 & &  1.439 \\ 
C & & -1.258 & & 0.264 & &  0.711 \\ 
H & & -2.090 & & 0.482 & &  -1.280 \\ 
H & & 2.055 & & -0.552 & &  -1.153 \\ 
H & & 2.004 & & -0.532 & &  1.318 \\ 
O & & -0.055 & & -0.028 & &  2.790 \\ 
O & & -0.062 & & -0.034 & &  -2.729 \\ 
C & & -2.502 & & 0.545 & &  1.431 \\ 
O & & -2.524 & & 0.527 & &  2.681 \\ 
C & & -3.767 & & 0.896 & &  0.698 \\ 
H & & -3.879 & & 0.372 & &  -0.248 \\ 
H & & -3.756 & & 1.970 & &  0.484 \\ 
H & & -4.619 & & 0.679 & &  1.340 \\ 
H & & -0.989 & & 0.181 & &  3.067 \\ 
H & & 0.806 & & -0.252 & &  -3.093 \\ 
\hline
\end{tabular}
\end{table}
 \newpage
\subsection*{For R-BQ (left) and R-HQ (right) with R = COOH}
\begin{table}[H]
\centering
\begin{tabular}{ccccccc}
\hline
\textbf{Atom} & & \textbf{x} & & \textbf{y} & & \textbf{z} \\ 
\hline
C & & -1.276 & & -0.001 & &  -0.669 \\ 
C & & -0.012 & & 0.000 & &  -1.433 \\ 
C & & 1.265 & & 0.002 & &  -0.672 \\ 
C & & 1.286 & & 0.001 & &  0.673 \\ 
C & & 0.000 & & -0.000 & &  1.427 \\ 
C & & -1.264 & & -0.002 & &  0.675 \\ 
H & & -2.195 & & -0.002 & &  -1.241 \\ 
H & & 2.182 & & 0.003 & &  -1.247 \\ 
H & & -2.171 & & -0.003 & &  1.266 \\ 
O & & 0.000 & & 0.000 & &  2.665 \\ 
O & & 0.012 & & -0.000 & &  -2.660 \\ 
C & & 2.607 & & 0.002 & &  1.394 \\ 
O & & 3.673 & & 0.001 & &  0.798 \\ 
O & & 2.558 & & 0.002 & &  2.722 \\ 
H & & 1.598 & & 0.002 & &  3.007 \\ 
\hline
\end{tabular}
\hspace{2.5cm}
\begin{tabular}{ccccccc}
\hline
\textbf{Atom} & & \textbf{x} & & \textbf{y} & & \textbf{z} \\ 
\hline
C & & -1.219 & & 0.056 & &  -0.665 \\ 
C & & -0.025 & & 0.000 & &  -1.402 \\ 
C & & 1.189 & & -0.055 & &  -0.736 \\ 
C & & 1.227 & & -0.056 & &  0.671 \\ 
C & & 0.023 & & -0.001 & &  1.410 \\ 
C & & -1.194 & & 0.056 & &  0.723 \\ 
H & & -2.171 & & 0.101 & &  -1.184 \\ 
H & & 2.111 & & -0.098 & &  -1.300 \\ 
H & & -2.115 & & 0.099 & &  1.290 \\ 
O & & -0.000 & & -0.000 & &  2.765 \\ 
O & & -0.004 & & -0.002 & &  -2.770 \\ 
C & & 2.497 & & -0.114 & &  1.391 \\ 
O & & 2.593 & & -0.123 & &  2.626 \\ 
O & & 3.589 & & -0.158 & &  0.616 \\ 
H & & 4.368 & & -0.195 & &  1.196 \\ 
H & & 0.938 & & -0.044 & &  3.067 \\ 
H & & -0.902 & & 0.032 & &  -3.124 \\ 
\hline
\end{tabular}
\end{table}
 \newpage
\subsection*{For R-BQ (left) and R-HQ (right) with R = COOCH3}
\begin{table}[H]
\centering
\begin{tabular}{ccccccc}
\hline
\textbf{Atom} & & \textbf{x} & & \textbf{y} & & \textbf{z} \\ 
\hline
C & & -1.281 & & 0.046 & &  -0.660 \\ 
C & & -0.029 & & -0.068 & &  -1.433 \\ 
C & & 1.250 & & -0.102 & &  -0.685 \\ 
C & & 1.291 & & -0.019 & &  0.658 \\ 
C & & 0.019 & & 0.040 & &  1.440 \\ 
C & & -1.250 & & 0.103 & &  0.679 \\ 
H & & -2.204 & & 0.073 & &  -1.225 \\ 
H & & 2.153 & & -0.174 & &  -1.277 \\ 
H & & -2.150 & & 0.172 & &  1.278 \\ 
O & & 0.000 & & -0.000 & &  2.665 \\ 
O & & -0.033 & & -0.121 & &  -2.659 \\ 
C & & 2.600 & & 0.065 & &  1.379 \\ 
O & & 2.783 & & 0.730 & &  2.381 \\ 
O & & 3.551 & & -0.636 & &  0.757 \\ 
C & & 4.879 & & -0.563 & &  1.323 \\ 
H & & 5.174 & & 0.477 & &  1.452 \\ 
H & & 5.524 & & -1.064 & &  0.608 \\ 
H & & 4.896 & & -1.076 & &  2.284 \\ 
\hline
\end{tabular}
\hspace{2.5cm}
\begin{tabular}{ccccccc}
\hline
\textbf{Atom} & & \textbf{x} & & \textbf{y} & & \textbf{z} \\ 
\hline
C & & -1.168 & & 0.046 & &  -0.666 \\ 
C & & 0.021 & & -0.081 & &  -1.399 \\ 
C & & 1.231 & & -0.197 & &  -0.732 \\ 
C & & 1.270 & & -0.196 & &  0.674 \\ 
C & & 0.069 & & -0.076 & &  1.410 \\ 
C & & -1.142 & & 0.049 & &  0.723 \\ 
H & & -2.116 & & 0.140 & &  -1.185 \\ 
H & & 2.149 & & -0.293 & &  -1.294 \\ 
H & & -2.058 & & 0.146 & &  1.291 \\ 
O & & 0.047 & & -0.076 & &  2.768 \\ 
O & & 0.044 & & -0.097 & &  -2.768 \\ 
C & & 2.541 & & -0.329 & &  1.400 \\ 
O & & 2.613 & & -0.388 & &  2.636 \\ 
O & & 3.619 & & -0.375 & &  0.621 \\ 
C & & 4.891 & & -0.596 & &  1.268 \\ 
H & & 5.015 & & 0.082 & &  2.110 \\ 
H & & 5.637 & & -0.403 & &  0.502 \\ 
H & & 4.949 & & -1.629 & &  1.611 \\ 
H & & 0.983 & & -0.194 & &  3.062 \\ 
H & & -0.850 & & -0.008 & &  -3.124 \\ 
\hline
\end{tabular}
\end{table}
 \newpage
\subsection*{For R-BQ (left) and R-HQ (right) with R = CF$_3$}
\begin{table}[H]
\centering
\begin{tabular}{ccccccc}
\hline
\textbf{Atom} & & \textbf{x} & & \textbf{y} & & \textbf{z} \\ 
\hline
C & & -1.271 & & 0.003 & &  -0.672 \\ 
C & & -0.008 & & -0.002 & &  -1.435 \\ 
C & & 1.270 & & -0.004 & &  -0.673 \\ 
C & & 1.274 & & -0.002 & &  0.668 \\ 
C & & -0.003 & & 0.001 & &  1.441 \\ 
C & & -1.263 & & 0.004 & &  0.671 \\ 
H & & -2.187 & & 0.005 & &  -1.249 \\ 
H & & 2.181 & & -0.006 & &  -1.258 \\ 
H & & -2.174 & & 0.008 & &  1.257 \\ 
O & & 0.000 & & 0.000 & &  2.668 \\ 
O & & 0.011 & & -0.005 & &  -2.661 \\ 
C & & 2.561 & & -0.005 & &  1.448 \\ 
F & & 2.666 & & -1.094 & &  2.242 \\ 
F & & 2.669 & & 1.082 & &  2.245 \\ 
F & & 3.644 & & -0.005 & &  0.637 \\ 
\hline
\end{tabular}
\hspace{2.5cm}
\begin{tabular}{ccccccc}
\hline
\textbf{Atom} & & \textbf{x} & & \textbf{y} & & \textbf{z} \\ 
\hline
C & & -1.195 & & 0.010 & &  -0.714 \\ 
C & & 0.024 & & 0.001 & &  -1.397 \\ 
C & & 1.217 & & -0.010 & &  -0.682 \\ 
C & & 1.196 & & -0.011 & &  0.716 \\ 
C & & -0.026 & & -0.000 & &  1.403 \\ 
C & & -1.217 & & 0.010 & &  0.673 \\ 
H & & -2.120 & & 0.018 & &  -1.275 \\ 
H & & 2.168 & & -0.018 & &  -1.200 \\ 
H & & -2.166 & & 0.017 & &  1.196 \\ 
O & & 0.000 & & -0.000 & &  2.766 \\ 
O & & -0.021 & & 0.003 & &  -2.762 \\ 
C & & 2.481 & & -0.023 & &  1.478 \\ 
F & & 2.607 & & -1.108 & &  2.288 \\ 
F & & 2.630 & & 1.063 & &  2.285 \\ 
F & & 3.568 & & -0.036 & &  0.665 \\ 
H & & -0.903 & & 0.005 & &  3.112 \\ 
H & & 0.871 & & -0.003 & &  -3.131 \\ 
\hline
\end{tabular}
\end{table}
 \newpage
\subsection*{For R-BQ (left) and R-HQ (right) with R = NH$_2$}
\begin{table}[H]
\centering
\begin{tabular}{ccccccc}
\hline
\textbf{Atom} & & \textbf{x} & & \textbf{y} & & \textbf{z} \\ 
\hline
C & & -1.257 & & 0.001 & &  -0.721 \\ 
C & & -0.019 & & -0.000 & &  -1.443 \\ 
C & & 1.261 & & -0.000 & &  -0.673 \\ 
C & & 1.285 & & -0.000 & &  0.667 \\ 
C & & 0.028 & & 0.000 & &  1.429 \\ 
C & & -1.273 & & 0.001 & &  0.653 \\ 
H & & -2.178 & & 0.001 & &  -1.291 \\ 
H & & 2.169 & & -0.000 & &  -1.265 \\ 
H & & 2.200 & & 0.000 & &  1.245 \\ 
O & & 0.004 & & 0.000 & &  2.654 \\ 
O & & 0.046 & & -0.002 & &  -2.689 \\ 
N & & -2.363 & & 0.001 & &  1.428 \\ 
H & & -2.249 & & 0.003 & &  2.432 \\ 
H & & -3.294 & & 0.004 & &  1.042 \\ 
\hline
\end{tabular}
\hspace{2.5cm}
\begin{tabular}{ccccccc}
\hline
\textbf{Atom} & & \textbf{x} & & \textbf{y} & & \textbf{z} \\ 
\hline
C & & -1.226 & & 0.027 & &  -0.661 \\ 
C & & -0.035 & & -0.001 & &  -1.390 \\ 
C & & 1.197 & & -0.027 & &  -0.739 \\ 
C & & 1.223 & & -0.026 & &  0.659 \\ 
C & & 0.044 & & 0.000 & &  1.392 \\ 
C & & -1.203 & & 0.026 & &  0.738 \\ 
H & & -2.182 & & 0.045 & &  -1.175 \\ 
H & & 2.114 & & -0.052 & &  -1.313 \\ 
H & & 2.174 & & -0.051 & &  1.180 \\ 
O & & -0.000 & & -0.000 & &  2.771 \\ 
O & & -0.028 & & -0.004 & &  -2.766 \\ 
N & & -2.371 & & 0.107 & &  1.494 \\ 
H & & -2.271 & & -0.233 & &  2.441 \\ 
H & & -3.198 & & -0.263 & &  1.046 \\ 
H & & 0.889 & & -0.118 & &  3.130 \\ 
H & & -0.936 & & 0.004 & &  -3.098 \\ 
\hline
\end{tabular}
\end{table}
 \newpage
\subsection*{For R-BQ (left) and R-HQ (right) with R = N(CH$_3$)$_2$}
\begin{table}[H]
\centering
\begin{tabular}{ccccccc}
\hline
\textbf{Atom} & & \textbf{x} & & \textbf{y} & & \textbf{z} \\ 
\hline
C & & -1.242 & & -0.670 & &  -0.740 \\ 
C & & -0.062 & & -0.275 & &  -1.431 \\ 
C & & 1.067 & & 0.282 & &  -0.643 \\ 
C & & 1.004 & & 0.379 & &  0.691 \\ 
C & & -0.170 & & -0.105 & &  1.440 \\ 
C & & -1.373 & & -0.578 & &  0.643 \\ 
H & & -2.062 & & -1.016 & &  -1.353 \\ 
H & & 1.938 & & 0.602 & &  -1.203 \\ 
H & & 1.812 & & 0.771 & &  1.296 \\ 
O & & -0.129 & & -0.161 & &  2.661 \\ 
O & & 0.073 & & -0.357 & &  -2.679 \\ 
N & & -2.486 & & -0.951 & &  1.298 \\ 
C & & -2.759 & & -0.696 & &  2.720 \\ 
H & & -2.253 & & -1.422 & &  3.357 \\ 
H & & -3.834 & & -0.775 & &  2.865 \\ 
H & & -2.432 & & 0.300 & &  3.008 \\ 
C & & -3.558 & & -1.602 & &  0.547 \\ 
H & & -4.106 & & -0.880 & &  -0.067 \\ 
H & & -4.247 & & -2.063 & &  1.249 \\ 
H & & -3.146 & & -2.377 & &  -0.101 \\ 
\hline
\end{tabular}
\hspace{2.5cm}
\begin{tabular}{ccccccc}
\hline
\textbf{Atom} & & \textbf{x} & & \textbf{y} & & \textbf{z} \\ 
\hline
C & & -1.245 & & 0.035 & &  -0.619 \\ 
C & & -0.036 & & -0.041 & &  -1.316 \\ 
C & & 1.176 & & -0.062 & &  -0.628 \\ 
C & & 1.180 & & -0.036 & &  0.770 \\ 
C & & -0.020 & & 0.021 & &  1.465 \\ 
C & & -1.246 & & 0.078 & &  0.774 \\ 
H & & -2.174 & & 0.063 & &  -1.175 \\ 
H & & 2.113 & & -0.117 & &  -1.170 \\ 
H & & 2.114 & & -0.083 & &  1.317 \\ 
O & & -0.032 & & -0.001 & &  2.836 \\ 
O & & -0.107 & & -0.087 & &  -2.688 \\ 
N & & -2.421 & & 0.148 & &  1.582 \\ 
C & & -2.715 & & 1.515 & &  2.044 \\ 
H & & -3.521 & & 1.476 & &  2.778 \\ 
H & & -3.025 & & 2.168 & &  1.216 \\ 
H & & -1.834 & & 1.950 & &  2.516 \\ 
C & & -3.613 & & -0.475 & &  1.008 \\ 
H & & -4.025 & & 0.098 & &  0.165 \\ 
H & & -4.383 & & -0.535 & &  1.779 \\ 
H & & -3.378 & & -1.483 & &  0.668 \\ 
H & & -0.966 & & -0.156 & &  3.077 \\ 
H & & 0.783 & & -0.116 & &  -3.062 \\ 
\hline
\end{tabular}
\end{table}
 \newpage
\subsection*{For R-BQ (left) and R-HQ (right) with R = NO$_2$}
\begin{table}[H]
\centering
\begin{tabular}{ccccccc}
\hline
\textbf{Atom} & & \textbf{x} & & \textbf{y} & & \textbf{z} \\ 
\hline
C & & -1.205 & & 0.032 & &  -0.688 \\ 
C & & 0.051 & & -0.063 & &  -1.457 \\ 
C & & 1.333 & & -0.075 & &  -0.699 \\ 
C & & 1.323 & & 0.006 & &  0.637 \\ 
C & & 0.058 & & 0.045 & &  1.438 \\ 
C & & -1.194 & & 0.082 & &  0.655 \\ 
H & & -2.124 & & 0.047 & &  -1.261 \\ 
H & & 2.251 & & -0.133 & &  -1.270 \\ 
H & & -2.105 & & 0.130 & &  1.238 \\ 
O & & 0.061 & & 0.005 & &  2.659 \\ 
O & & 0.072 & & -0.134 & &  -2.679 \\ 
N & & 2.603 & & 0.059 & &  1.350 \\ 
O & & 2.699 & & 0.857 & &  2.277 \\ 
O & & 3.507 & & -0.682 & &  0.962 \\ 
\hline
\end{tabular}
\hspace{2.5cm}
\begin{tabular}{ccccccc}
\hline
\textbf{Atom} & & \textbf{x} & & \textbf{y} & & \textbf{z} \\ 
\hline
C & & -1.180 & & 0.031 & &  -0.686 \\ 
C & & 0.015 & & 0.004 & &  -1.439 \\ 
C & & 1.232 & & -0.023 & &  -0.787 \\ 
C & & 1.263 & & -0.023 & &  0.621 \\ 
C & & 0.070 & & 0.003 & &  1.389 \\ 
C & & -1.148 & & 0.031 & &  0.695 \\ 
H & & -2.137 & & 0.052 & &  -1.195 \\ 
H & & 2.157 & & -0.044 & &  -1.342 \\ 
H & & -2.065 & & 0.051 & &  1.269 \\ 
O & & 0.042 & & 0.004 & &  2.729 \\ 
O & & 0.022 & & 0.003 & &  -2.798 \\ 
N & & 2.532 & & -0.050 & &  1.260 \\ 
O & & 2.583 & & -0.053 & &  2.522 \\ 
O & & 3.568 & & -0.070 & &  0.584 \\ 
H & & 0.979 & & -0.018 & &  3.034 \\ 
H & & -0.878 & & 0.021 & &  -3.151 \\ 
\hline
\end{tabular}
\end{table}
 \newpage
\subsection*{For R-BQ (left) and R-HQ (right) with R = OH}
\begin{table}[H]
\centering
\begin{tabular}{ccccccc}
\hline
\textbf{Atom} & & \textbf{x} & & \textbf{y} & & \textbf{z} \\ 
\hline
C & & -1.266 & & -0.002 & &  -0.690 \\ 
C & & -0.008 & & -0.000 & &  -1.430 \\ 
C & & 1.266 & & 0.002 & &  -0.664 \\ 
C & & 1.271 & & 0.002 & &  0.677 \\ 
C & & 0.011 & & 0.000 & &  1.445 \\ 
C & & -1.273 & & -0.002 & &  0.663 \\ 
H & & -2.182 & & -0.004 & &  -1.269 \\ 
H & & 2.177 & & 0.004 & &  -1.249 \\ 
H & & 2.183 & & 0.004 & &  1.261 \\ 
O & & 0.000 & & 0.000 & &  2.669 \\ 
O & & 0.019 & & -0.001 & &  -2.664 \\ 
O & & -2.361 & & -0.004 & &  1.443 \\ 
H & & -3.168 & & -0.005 & &  0.906 \\ 
\hline
\end{tabular}
\hspace{2.5cm}
\begin{tabular}{ccccccc}
\hline
\textbf{Atom} & & \textbf{x} & & \textbf{y} & & \textbf{z} \\ 
\hline
C & & -1.223 & & 0.011 & &  -0.669 \\ 
C & & -0.028 & & -0.002 & &  -1.395 \\ 
C & & 1.196 & & -0.011 & &  -0.730 \\ 
C & & 1.215 & & -0.010 & &  0.666 \\ 
C & & 0.034 & & 0.001 & &  1.403 \\ 
C & & -1.194 & & 0.012 & &  0.725 \\ 
H & & -2.179 & & 0.019 & &  -1.184 \\ 
H & & 2.118 & & -0.020 & &  -1.297 \\ 
H & & 2.165 & & -0.018 & &  1.189 \\ 
O & & 0.000 & & -0.000 & &  2.778 \\ 
O & & -0.008 & & -0.004 & &  -2.766 \\ 
O & & -2.331 & & 0.023 & &  1.485 \\ 
H & & -3.107 & & 0.025 & &  0.910 \\ 
H & & 0.904 & & -0.009 & &  3.117 \\ 
H & & -0.909 & & -0.001 & &  -3.117 \\ 
\hline
\end{tabular}
\end{table}
 \newpage
\subsection*{For R-BQ (left) and R-HQ (right) with R = OCH$_3$}
\begin{table}[H]
\centering
\begin{tabular}{ccccccc}
\hline
\textbf{Atom} & & \textbf{x} & & \textbf{y} & & \textbf{z} \\ 
\hline
C & & -1.264 & & 0.015 & &  -0.698 \\ 
C & & -0.004 & & 0.002 & &  -1.429 \\ 
C & & 1.269 & & -0.014 & &  -0.659 \\ 
C & & 1.270 & & -0.015 & &  0.681 \\ 
C & & 0.008 & & -0.001 & &  1.445 \\ 
C & & -1.278 & & 0.014 & &  0.659 \\ 
H & & -2.169 & & 0.026 & &  -1.290 \\ 
H & & 2.181 & & -0.024 & &  -1.242 \\ 
H & & 2.180 & & -0.026 & &  1.269 \\ 
O & & 0.000 & & 0.000 & &  2.669 \\ 
O & & 0.035 & & 0.003 & &  -2.664 \\ 
O & & -2.353 & & 0.025 & &  1.443 \\ 
C & & -3.645 & & 0.041 & &  0.816 \\ 
H & & -4.365 & & 0.046 & &  1.628 \\ 
H & & -3.777 & & -0.850 & &  0.199 \\ 
H & & -3.757 & & 0.940 & &  0.206 \\ 
\hline
\end{tabular}
\hspace{2.5cm}
\begin{tabular}{ccccccc}
\hline
\textbf{Atom} & & \textbf{x} & & \textbf{y} & & \textbf{z} \\ 
\hline
C & & -1.223 & & 0.002 & &  -0.673 \\ 
C & & -0.022 & & 0.000 & &  -1.392 \\ 
C & & 1.200 & & -0.002 & &  -0.728 \\ 
C & & 1.214 & & -0.002 & &  0.670 \\ 
C & & 0.033 & & -0.000 & &  1.401 \\ 
C & & -1.202 & & 0.002 & &  0.723 \\ 
H & & -2.165 & & 0.004 & &  -1.205 \\ 
H & & 2.123 & & -0.004 & &  -1.293 \\ 
H & & 2.163 & & -0.004 & &  1.196 \\ 
O & & -0.000 & & 0.000 & &  2.776 \\ 
O & & -0.002 & & 0.000 & &  -2.765 \\ 
O & & -2.317 & & 0.004 & &  1.505 \\ 
C & & -3.588 & & 0.005 & &  0.855 \\ 
H & & -4.328 & & 0.007 & &  1.652 \\ 
H & & -3.715 & & -0.889 & &  0.239 \\ 
H & & -3.713 & & 0.898 & &  0.238 \\ 
H & & 0.905 & & -0.002 & &  3.113 \\ 
H & & -0.903 & & 0.001 & &  -3.113 \\ 
\hline
\end{tabular}
\end{table}
 \newpage
\subsection*{For R-BQ (left) and R-HQ (right) with R = 2OCH$_3$}
\begin{table}[H]
\centering
\begin{tabular}{ccccccc}
\hline
\textbf{Atom} & & \textbf{x} & & \textbf{y} & & \textbf{z} \\ 
\hline
C & & -1.266 & & 0.017 & &  -0.676 \\ 
C & & 0.001 & & -0.008 & &  -1.413 \\ 
C & & 1.267 & & -0.042 & &  -0.676 \\ 
C & & 1.280 & & -0.052 & &  0.676 \\ 
C & & 0.001 & & -0.024 & &  1.459 \\ 
C & & -1.279 & & -0.001 & &  0.676 \\ 
H & & -2.168 & & 0.045 & &  -1.271 \\ 
H & & 2.170 & & -0.059 & &  -1.271 \\ 
O & & 0.001 & & -0.018 & &  2.678 \\ 
O & & 0.001 & & -0.004 & &  -2.652 \\ 
O & & -2.353 & & -0.001 & &  1.468 \\ 
C & & -3.642 & & -0.001 & &  0.838 \\ 
H & & -4.365 & & -0.066 & &  1.646 \\ 
H & & -3.740 & & -0.863 & &  0.176 \\ 
H & & -3.787 & & 0.923 & &  0.275 \\ 
O & & 2.354 & & -0.085 & &  1.468 \\ 
C & & 3.641 & & -0.117 & &  0.838 \\ 
H & & 4.365 & & -0.152 & &  1.647 \\ 
H & & 3.735 & & -1.006 & &  0.211 \\ 
H & & 3.791 & & 0.782 & &  0.236 \\ 
\hline
\end{tabular}
\hspace{2.5cm}
\begin{tabular}{ccccccc}
\hline
\textbf{Atom} & & \textbf{x} & & \textbf{y} & & \textbf{z} \\ 
\hline
C & & -1.212 & & 0.006 & &  -0.701 \\ 
C & & 0.004 & & 0.001 & &  -1.384 \\ 
C & & 1.223 & & -0.007 & &  -0.705 \\ 
C & & 1.203 & & -0.006 & &  0.689 \\ 
C & & -0.003 & & 0.000 & &  1.400 \\ 
C & & -1.214 & & 0.005 & &  0.701 \\ 
H & & -2.142 & & 0.012 & &  -1.254 \\ 
H & & 2.147 & & -0.012 & &  -1.264 \\ 
O & & 0.000 & & 0.000 & &  2.775 \\ 
O & & 0.064 & & 0.002 & &  -2.754 \\ 
O & & -2.345 & & 0.008 & &  1.464 \\ 
C & & -3.601 & & 0.005 & &  0.789 \\ 
H & & -4.358 & & 0.003 & &  1.569 \\ 
H & & -3.712 & & -0.888 & &  0.168 \\ 
H & & -3.718 & & 0.898 & &  0.169 \\ 
O & & 2.312 & & -0.009 & &  1.490 \\ 
C & & 3.594 & & -0.042 & &  0.862 \\ 
H & & 4.323 & & -0.034 & &  1.669 \\ 
H & & 3.710 & & -0.952 & &  0.268 \\ 
H & & 3.739 & & 0.835 & &  0.226 \\ 
H & & 0.930 & & -0.006 & &  3.050 \\ 
H & & -0.827 & & 0.008 & &  -3.128 \\ 
\hline
\end{tabular}
\end{table}
 \newpage
\subsection*{For R-BQ (left) and R-HQ (right) with R = F}
\begin{table}[H]
\centering
\begin{tabular}{ccccccc}
\hline
\textbf{Atom} & & \textbf{x} & & \textbf{y} & & \textbf{z} \\ 
\hline
C & & -1.278 & & -0.002 & &  -0.669 \\ 
C & & -0.012 & & -0.010 & &  -1.428 \\ 
C & & 1.265 & & -0.019 & &  -0.670 \\ 
C & & 1.282 & & -0.020 & &  0.672 \\ 
C & & 0.029 & & -0.011 & &  1.453 \\ 
C & & -1.238 & & -0.003 & &  0.667 \\ 
H & & -2.208 & & 0.005 & &  -1.222 \\ 
H & & 2.171 & & -0.024 & &  -1.262 \\ 
H & & 2.201 & & -0.026 & &  1.244 \\ 
O & & 0.008 & & -0.011 & &  2.678 \\ 
O & & -0.016 & & -0.010 & &  -2.657 \\ 
F & & -2.355 & & 0.003 & &  1.405 \\ 
\hline
\end{tabular}
\hspace{2.5cm}
\begin{tabular}{ccccccc}
\hline
\textbf{Atom} & & \textbf{x} & & \textbf{y} & & \textbf{z} \\ 
\hline
C & & -1.256 & & 0.002 & &  -0.633 \\ 
C & & -0.066 & & -0.002 & &  -1.363 \\ 
C & & 1.161 & & -0.007 & &  -0.697 \\ 
C & & 1.198 & & -0.007 & &  0.696 \\ 
C & & 0.021 & & -0.003 & &  1.447 \\ 
C & & -1.186 & & 0.002 & &  0.751 \\ 
H & & -2.226 & & 0.006 & &  -1.116 \\ 
H & & 2.079 & & -0.010 & &  -1.270 \\ 
H & & 2.154 & & -0.010 & &  1.207 \\ 
O & & -0.021 & & -0.002 & &  2.815 \\ 
O & & -0.046 & & -0.002 & &  -2.732 \\ 
F & & -2.346 & & 0.006 & &  1.464 \\ 
H & & 0.879 & & -0.004 & &  3.166 \\ 
H & & -0.946 & & -0.001 & &  -3.083 \\ 
\hline
\end{tabular}
\end{table}
 \newpage
\subsection*{For R-BQ (left) and R-HQ (right) with R = 4F}
\begin{table}[H]
\centering
\begin{tabular}{ccccccc}
\hline
\textbf{Atom} & & \textbf{x} & & \textbf{y} & & \textbf{z} \\ 
\hline
C & & -1.261 & & -0.007 & &  -0.671 \\ 
C & & 0.000 & & -0.000 & &  -1.454 \\ 
C & & 1.261 & & 0.007 & &  -0.671 \\ 
C & & 1.261 & & 0.007 & &  0.671 \\ 
C & & -0.000 & & 0.000 & &  1.454 \\ 
C & & -1.261 & & -0.007 & &  0.671 \\ 
O & & 0.000 & & -0.000 & &  2.671 \\ 
O & & -0.000 & & -0.000 & &  -2.671 \\ 
F & & -2.384 & & -0.013 & &  1.378 \\ 
F & & -2.384 & & -0.013 & &  -1.378 \\ 
F & & 2.384 & & 0.013 & &  -1.378 \\ 
F & & 2.384 & & 0.013 & &  1.379 \\ 
\hline
\end{tabular}
\hspace{2.5cm}
\begin{tabular}{ccccccc}
\hline
\textbf{Atom} & & \textbf{x} & & \textbf{y} & & \textbf{z} \\ 
\hline
C & & -1.192 & & -0.000 & &  -0.673 \\ 
C & & -0.016 & & 0.000 & &  -1.421 \\ 
C & & 1.186 & & 0.000 & &  -0.715 \\ 
C & & 1.192 & & 0.000 & &  0.673 \\ 
C & & 0.016 & & -0.000 & &  1.421 \\ 
C & & -1.186 & & -0.000 & &  0.715 \\ 
O & & 0.000 & & 0.000 & &  2.776 \\ 
O & & 0.000 & & -0.000 & &  -2.776 \\ 
F & & -2.353 & & -0.000 & &  1.384 \\ 
F & & -2.367 & & -0.000 & &  -1.343 \\ 
F & & 2.353 & & 0.000 & &  -1.384 \\ 
F & & 2.367 & & 0.000 & &  1.343 \\ 
H & & 0.908 & & -0.000 & &  3.113 \\ 
H & & -0.908 & & 0.000 & &  -3.113 \\ 
\hline
\end{tabular}
\end{table}
 \newpage
\subsection*{For R-BQ (left) and R-HQ (right) with R = SH}
\begin{table}[H]
\centering
\begin{tabular}{ccccccc}
\hline
\textbf{Atom} & & \textbf{x} & & \textbf{y} & & \textbf{z} \\ 
\hline
C & & -1.264 & & -0.004 & &  -0.685 \\ 
C & & -0.002 & & -0.001 & &  -1.429 \\ 
C & & 1.270 & & 0.004 & &  -0.666 \\ 
C & & 1.270 & & 0.005 & &  0.675 \\ 
C & & 0.007 & & 0.000 & &  1.437 \\ 
C & & -1.281 & & -0.004 & &  0.669 \\ 
H & & -2.170 & & -0.008 & &  -1.278 \\ 
H & & 2.183 & & 0.007 & &  -1.249 \\ 
H & & 2.180 & & 0.008 & &  1.262 \\ 
O & & 0.000 & & 0.000 & &  2.664 \\ 
O & & 0.015 & & -0.001 & &  -2.663 \\ 
S & & -2.703 & & -0.008 & &  1.694 \\ 
H & & -3.589 & & -0.011 & &  0.682 \\ 
\hline
\end{tabular}
\hspace{2.5cm}
\begin{tabular}{ccccccc}
\hline
\textbf{Atom} & & \textbf{x} & & \textbf{y} & & \textbf{z} \\ 
\hline
C & & -1.223 & & 0.004 & &  -0.670 \\ 
C & & -0.029 & & 0.000 & &  -1.393 \\ 
C & & 1.197 & & -0.004 & &  -0.727 \\ 
C & & 1.220 & & -0.004 & &  0.665 \\ 
C & & 0.033 & & -0.000 & &  1.396 \\ 
C & & -1.198 & & 0.004 & &  0.728 \\ 
H & & -2.172 & & 0.007 & &  -1.195 \\ 
H & & 2.117 & & -0.007 & &  -1.296 \\ 
H & & 2.170 & & -0.008 & &  1.188 \\ 
O & & -0.000 & & 0.000 & &  2.770 \\ 
O & & -0.004 & & 0.000 & &  -2.764 \\ 
S & & -2.675 & & 0.009 & &  1.722 \\ 
H & & -3.536 & & 0.012 & &  0.691 \\ 
H & & 0.900 & & -0.006 & &  3.122 \\ 
H & & -0.903 & & 0.003 & &  -3.118 \\ 
\hline
\end{tabular}
\end{table}
 \newpage
\subsection*{For R-BQ (left) and R-HQ (right) with R = Cl}
\begin{table}[H]
\centering
\begin{tabular}{ccccccc}
\hline
\textbf{Atom} & & \textbf{x} & & \textbf{y} & & \textbf{z} \\ 
\hline
C & & -1.269 & & -0.004 & &  -0.684 \\ 
C & & 0.006 & & -0.002 & &  -1.433 \\ 
C & & 1.271 & & 0.004 & &  -0.666 \\ 
C & & 1.263 & & 0.005 & &  0.676 \\ 
C & & 0.001 & & 0.000 & &  1.447 \\ 
C & & -1.273 & & -0.003 & &  0.659 \\ 
H & & -2.179 & & -0.006 & &  -1.268 \\ 
H & & 2.188 & & 0.006 & &  -1.242 \\ 
H & & 2.172 & & 0.008 & &  1.265 \\ 
O & & 0.000 & & 0.000 & &  2.671 \\ 
O & & 0.006 & & -0.007 & &  -2.662 \\ 
Cl & & -2.745 & & -0.005 & &  1.570 \\ 
\hline
\end{tabular}
\hspace{2.5cm}
\begin{tabular}{ccccccc}
\hline
\textbf{Atom} & & \textbf{x} & & \textbf{y} & & \textbf{z} \\ 
\hline
C & & -1.222 & & 0.001 & &  -0.680 \\ 
C & & -0.026 & & 0.000 & &  -1.397 \\ 
C & & 1.194 & & -0.001 & &  -0.718 \\ 
C & & 1.213 & & -0.001 & &  0.672 \\ 
C & & 0.026 & & -0.000 & &  1.410 \\ 
C & & -1.185 & & 0.001 & &  0.713 \\ 
H & & -2.178 & & 0.001 & &  -1.189 \\ 
H & & 2.119 & & -0.001 & &  -1.280 \\ 
H & & 2.163 & & -0.001 & &  1.195 \\ 
O & & 0.000 & & -0.000 & &  2.776 \\ 
O & & 0.010 & & 0.000 & &  -2.766 \\ 
Cl & & -2.695 & & 0.001 & &  1.604 \\ 
H & & 0.905 & & -0.001 & &  3.117 \\ 
H & & -0.885 & & 0.000 & &  -3.128 \\ 
\hline
\end{tabular}
\end{table}
 \newpage
\subsection*{For R-BQ (left) and R-HQ (right) with R = 2Cl}
\begin{table}[H]
\centering
\begin{tabular}{ccccccc}
\hline
\textbf{Atom} & & \textbf{x} & & \textbf{y} & & \textbf{z} \\ 
\hline
C & & -1.270 & & -0.005 & &  -0.678 \\ 
C & & 0.000 & & -0.001 & &  -1.431 \\ 
C & & 1.270 & & 0.005 & &  -0.678 \\ 
C & & 1.269 & & 0.005 & &  0.664 \\ 
C & & -0.000 & & 0.000 & &  1.459 \\ 
C & & -1.269 & & -0.005 & &  0.664 \\ 
H & & -2.184 & & -0.009 & &  -1.258 \\ 
H & & 2.184 & & 0.009 & &  -1.258 \\ 
O & & 0.000 & & 0.000 & &  2.676 \\ 
O & & -0.000 & & -0.002 & &  -2.660 \\ 
Cl & & -2.732 & & -0.010 & &  1.582 \\ 
Cl & & 2.732 & & 0.012 & &  1.582 \\ 
\hline
\end{tabular}
\hspace{2.5cm}
\begin{tabular}{ccccccc}
\hline
\textbf{Atom} & & \textbf{x} & & \textbf{y} & & \textbf{z} \\ 
\hline
C & & -1.215 & & 0.000 & &  -0.677 \\ 
C & & -0.027 & & -0.000 & &  -1.410 \\ 
C & & 1.186 & & -0.000 & &  -0.713 \\ 
C & & 1.215 & & -0.000 & &  0.677 \\ 
C & & 0.027 & & 0.000 & &  1.410 \\ 
C & & -1.186 & & 0.000 & &  0.713 \\ 
H & & -2.169 & & 0.000 & &  -1.190 \\ 
H & & 2.169 & & -0.000 & &  1.190 \\ 
O & & -0.000 & & -0.000 & &  2.772 \\ 
O & & 0.000 & & -0.000 & &  -2.772 \\ 
Cl & & -2.691 & & 0.000 & &  1.600 \\ 
Cl & & 2.691 & & -0.000 & &  -1.600 \\ 
H & & 0.901 & & -0.000 & &  3.123 \\ 
H & & -0.901 & & -0.000 & &  -3.123 \\ 
\hline
\end{tabular}
\end{table}
 \newpage
\subsection*{For R-BQ (left) and R-HQ (right) with R = 4Cl}
\begin{table}[H]
\centering
\begin{tabular}{ccccccc}
\hline
\textbf{Atom} & & \textbf{x} & & \textbf{y} & & \textbf{z} \\ 
\hline
C & & -1.277 & & -0.004 & &  -0.675 \\ 
C & & 0.000 & & 0.000 & &  -1.450 \\ 
C & & 1.277 & & 0.003 & &  -0.675 \\ 
C & & 1.277 & & 0.003 & &  0.675 \\ 
C & & -0.000 & & 0.000 & &  1.450 \\ 
C & & -1.277 & & -0.003 & &  0.675 \\ 
O & & 0.000 & & 0.000 & &  2.667 \\ 
O & & 0.000 & & 0.002 & &  -2.667 \\ 
Cl & & -2.715 & & -0.006 & &  1.610 \\ 
Cl & & -2.715 & & -0.008 & &  -1.610 \\ 
Cl & & 2.715 & & 0.005 & &  -1.610 \\ 
Cl & & 2.715 & & 0.007 & &  1.610 \\ 
\hline
\end{tabular}
\hspace{2.5cm}
\begin{tabular}{ccccccc}
\hline
\textbf{Atom} & & \textbf{x} & & \textbf{y} & & \textbf{z} \\ 
\hline
C & & -1.216 & & -0.013 & &  -0.678 \\ 
C & & -0.026 & & -0.010 & &  -1.418 \\ 
C & & 1.193 & & -0.009 & &  -0.730 \\ 
C & & 1.209 & & -0.009 & &  0.667 \\ 
C & & 0.019 & & -0.011 & &  1.407 \\ 
C & & -1.200 & & -0.013 & &  0.718 \\ 
O & & -0.004 & & -0.011 & &  2.760 \\ 
O & & -0.003 & & -0.008 & &  -2.771 \\ 
Cl & & -2.672 & & -0.016 & &  1.632 \\ 
Cl & & -2.715 & & -0.019 & &  -1.563 \\ 
Cl & & 2.665 & & -0.009 & &  -1.644 \\ 
Cl & & 2.708 & & -0.006 & &  1.552 \\ 
H & & 0.906 & & -0.008 & &  3.099 \\ 
H & & -0.913 & & -0.009 & &  -3.111 \\ 
\hline
\end{tabular}
\end{table}
 \newpage
\subsection*{For R-BQ (left) and R-HQ (right) with R = 2Cl2CN}
\begin{table}[H]
\centering
\begin{tabular}{ccccccc}
\hline
\textbf{Atom} & & \textbf{x} & & \textbf{y} & & \textbf{z} \\ 
\hline
C & & -1.296 & & 0.012 & &  -0.690 \\ 
C & & -0.018 & & 0.015 & &  -1.448 \\ 
C & & 1.257 & & 0.019 & &  -0.656 \\ 
C & & 1.241 & & 0.020 & &  0.698 \\ 
C & & -0.054 & & 0.016 & &  1.459 \\ 
C & & -1.312 & & 0.012 & &  0.669 \\ 
O & & -0.030 & & 0.016 & &  2.675 \\ 
O & & 0.035 & & 0.014 & &  -2.664 \\ 
Cl & & -2.717 & & 0.007 & &  -1.634 \\ 
Cl & & -2.756 & & 0.007 & &  1.578 \\ 
C & & 2.440 & & 0.023 & &  1.471 \\ 
C & & 2.476 & & 0.022 & &  -1.397 \\ 
N & & 3.427 & & 0.026 & &  2.081 \\ 
N & & 3.482 & & 0.025 & &  -1.979 \\ 
\hline
\end{tabular}
\hspace{2.5cm}
\begin{tabular}{ccccccc}
\hline
\textbf{Atom} & & \textbf{x} & & \textbf{y} & & \textbf{z} \\ 
\hline
C & & -1.228 & & -0.002 & &  -0.660 \\ 
C & & -0.033 & & -0.000 & &  -1.417 \\ 
C & & 1.192 & & 0.002 & &  -0.748 \\ 
C & & 1.226 & & 0.002 & &  0.676 \\ 
C & & 0.040 & & 0.000 & &  1.417 \\ 
C & & -1.198 & & -0.002 & &  0.732 \\ 
O & & 0.000 & & -0.000 & &  2.760 \\ 
O & & -0.025 & & 0.000 & &  -2.759 \\ 
Cl & & -2.717 & & -0.004 & &  -1.528 \\ 
Cl & & -2.645 & & -0.004 & &  1.652 \\ 
C & & 2.469 & & 0.004 & &  1.364 \\ 
C & & 2.407 & & 0.003 & &  -1.484 \\ 
N & & 3.474 & & 0.005 & &  1.953 \\ 
N & & 3.424 & & 0.005 & &  -2.052 \\ 
H & & 0.888 & & 0.002 & &  3.147 \\ 
H & & -0.936 & & -0.001 & &  -3.101 \\ 
\hline
\end{tabular}
\end{table}
 \newpage

 \bibliography{reference_SI}